\documentstyle[aaspp4,emulateapj5,apjfonts]{article}

\lefthead{van der Wel et al.}
\righthead{The FP at z=1 for Field Galaxies}
\slugcomment{Submitted to ApJ}

\begin{document}

\title{Mass-to-Light Ratios of Field Early-Type Galaxies at $z\sim1$ from Ultra-Deep Spectroscopy: Evidence for Mass-dependent Evolution\altaffilmark{1}}

\author{A.~van~der~Wel\altaffilmark{2}, M.~Franx\altaffilmark{2}, P.G.~van~Dokkum\altaffilmark{3}, H.-W.~Rix\altaffilmark{4}
G.D. Illingworth\altaffilmark{5}, and P. Rosati\altaffilmark{6}}
\altaffiltext{1}{Based on observations collected at the European Southern Observatory, Chile (169.A-0458).}
\altaffiltext{2}{Leiden Observatory, P.O.Box 9513, NL-2300 AA, Leiden, The Netherlands}
\altaffiltext{3}{Department of Astronomy, Yale University, P.O. Box 208101, New Haven, CT 06520-8101}
\altaffiltext{4}{Max-Planck-Institut f\"ur Astronomie, K\"onigstuhl 17, D-69117 Heidelberg, Germany}
\altaffiltext{5}{University of California Observatories/Lick Observatory, University of California, Santa Cruz, CA 95064}
\altaffiltext{6}{European Southern Observatory, Karl-Schwarzschild-Strasse 2, D-85748 Garching, Germany}
\begin{abstract}
We present an analysis of the Fundamental Plane for a sample of 27  
field early-type galaxies in the redshift range $0.6<z<1.15$
in the Chandra Deep Field-South and the 
field of the background cluster RDCS 1252.9-2927. 
Sixteen of the galaxies are at $z>0.95$.
The galaxies in this sample have high signal-to-noise spectra 
obtained at the Very Large Telescope and high resolution imaging from the 
HST Advanced Camera for Surveys.
From comparison with lower redshift data,
we find that the mean evolution of the mass-to-light ratio ($M/L$) of our sample
is $\Delta \ln {(M/L_B)} = (-1.75\pm0.16)z$, with a large galaxy-to-galaxy scatter.
This value can be too low by 0.3 due to selection effects, resulting in
$\Delta \ln {(M/L_B)} = (-1.43\pm0.16)z$.
The strong correlation between $M/L$ and rest-frame color indicates that the observed scatter 
is not due to measurement errors, but due to intrinsic differences between the stellar populations
of the galaxies,
such that our results can be used as a calibration for converting luminosities of 
high redshift galaxies into masses.
This pace of evolution is much faster than the evolution of cluster galaxies.
However, we find that the measured $M/L$ evolution strongly depends on galaxy mass.
For galaxies with masses $M>2\times10^{11} M_{\odot}$, we find no significant 
difference between the evolution of field and cluster galaxies: 
$\Delta \ln {(M/L_B)} = (-1.20\pm0.17)z$ for field galaxies and 
$\Delta \ln {(M/L_B)} = (-1.12\pm0.06)z$ for cluster galaxies.
The relation between the measured $M/L$ evolution and mass is partially
due to selection effects, as the galaxies are selected by luminosity, not mass.
However, even when taking selection effects into account, we still find a relation
between $M/L$ evolution and mass, which is most likely caused by a lower mean age and 
a larger intrinsic scatter for low mass galaxies.
Results from lensing early-type galaxies, which are mass-selected, show a very similar 
trend with mass. This, combined with our findings, provides evidence
for down-sizing, i.e., for the proposition that low mass galaxies are younger
than high mass galaxies.
Previous studies of the rate of evolution of field early-type galaxies found a large
range of mutually exclusive values. We show that these differences are largely
caused by the differences between fitting methods: 
most literature studies are consistent with our result and with one another
when using the same method.
Finally, five of the early-type galaxies in our sample have AGN. 
There is tentative evidence that the stellar
populations in these galaxies are younger than those of galaxies without AGN.
\end{abstract}


\section{Introduction}
Understanding the formation and evolution of early-type galaxies is a key issue
when addressing the mass assembly and star formation history of the galaxy 
population as a whole and the formation of structure in the universe,
as 50\% or more of all stars in the present day universe are in early-type 
galaxies and bulges (see, e.g., Bell et al. 2003).

In hierarchical galaxy formation theories (e.g., Cole et al. 2000), 
massive galaxies assemble late, such 
that strong evolution of the mass density from $z=1$ to the present day is 
expected (see, e.g., Kauffmann \& Charlot 1998).
Measuring the mass density requires a measurement of the luminosity density,
and an accurate determination of the $M/L$. $M/L$ can be estimated from models
(see, e.g., Bell et al. 2004),
but these estimates are uncertain due to the age/metallicity degeneracy and the unknown 
IMF of the stellar populations of the galaxies (Bruzual \& Charlot 2003). 

The Fundamental Plane (Djorgovski \& Davis 1987; Dressler et al. 1987)
provides a tool to measure the evolution of  
$M/L$ without model uncertainties.
The $M/L$ offset of high redshift galaxies from the local FP can be used to calibrate 
high redshift galaxy masses and to estimate
the age of their stellar populations (Franx 1993).
This technique has been used successfully to measure the luminosity weighted ages
of massive cluster galaxies, which have formed most of their stars at redshifts
$z\ge2$ (see, e.g., van Dokkum \& Franx 1996; van Dokkum \& Stanford 2003, Holden et al. 2005).
However, it is not clear whether galaxies in the general field evolve in the same way.
In fact, in the hierarchical picture the formation redshift of galaxies
with a given mass depends on environment (Diaferio et al. 2001). This 
would lead to substantial age differences between field and cluster galaxies
at any redshift (van Dokkum et al. 2001). Since this is a generic property of all hierarchical
formation models, measuring this difference is a critical test for those theories.

Various authors have measured the $M/L$ evolution of field early-
type galaxies through deep spectroscopy of magnitude limited samples.
The results are much less conclusive than the results from cluster studies 
and the comparison between field and cluster has proved to be very hard.
Some authors claim much faster evolution for  
field galaxies than for cluster galaxies (Treu et al. 2002; Gebhardt et al. 2003),
but others find that field and cluster galaxies evolve at comparable rates
(van Dokkum et al. 2001, van Dokkum \& Ellis 2003, van der Wel et al. 2005).
Studies involving lensing galaxies 
(Kochanek et al. 2000, Rusin et al. 2003, van de Ven et al. 2003),
indicate the presence of a mix of fast and 
slowly evolving galaxies. 
It is unclear whether the differences between the various results are caused by 
selection 
effects, measurement errors due to low signal-to-noise spectra, 
low number statistics, or contamination by late-type galaxies.

This paper describes a study of early-type galaxies at $z\sim 1$ using
much higher quality data than in previous studies.
The substantially large number of objects with very high signal-to-noise spectra 
enables 
us to accurately measure the $M/L$ evolution of the field early-type galaxy population, 
to compare the cluster and field populations, to study correlations 
between $M/L$, $M$, and rest-frame color, and to describe the possible 
effects of biases.
Also, we carefully
compare the samples and results from previous studies and this 
study in order to verify previous claims about the evolution of
field galaxies, and to see whether previous results are in fact 
consistent with each other and these new results.

In Section 2 we describe the sample selection, the spectroscopic observations
and data reduction, and the measurement of velocity dispersions. Section 3 
describes the measurement of the structural parameters, colors, 
morphologies, and the available X-ray data. 
In Section 4 we present the results
Throughout this paper we use Vega magnitudes, and assume 
$(\Omega_M,\Omega_{\Lambda})=(0.3,0.7)$, with a Hubble constant of
$H_0=70$~km~s$^{-1}$~Mpc$^{-1}$.

\begin{figure*}
\figurenum{1}
\leavevmode
\hbox{
\epsfxsize=5.8cm
\epsffile{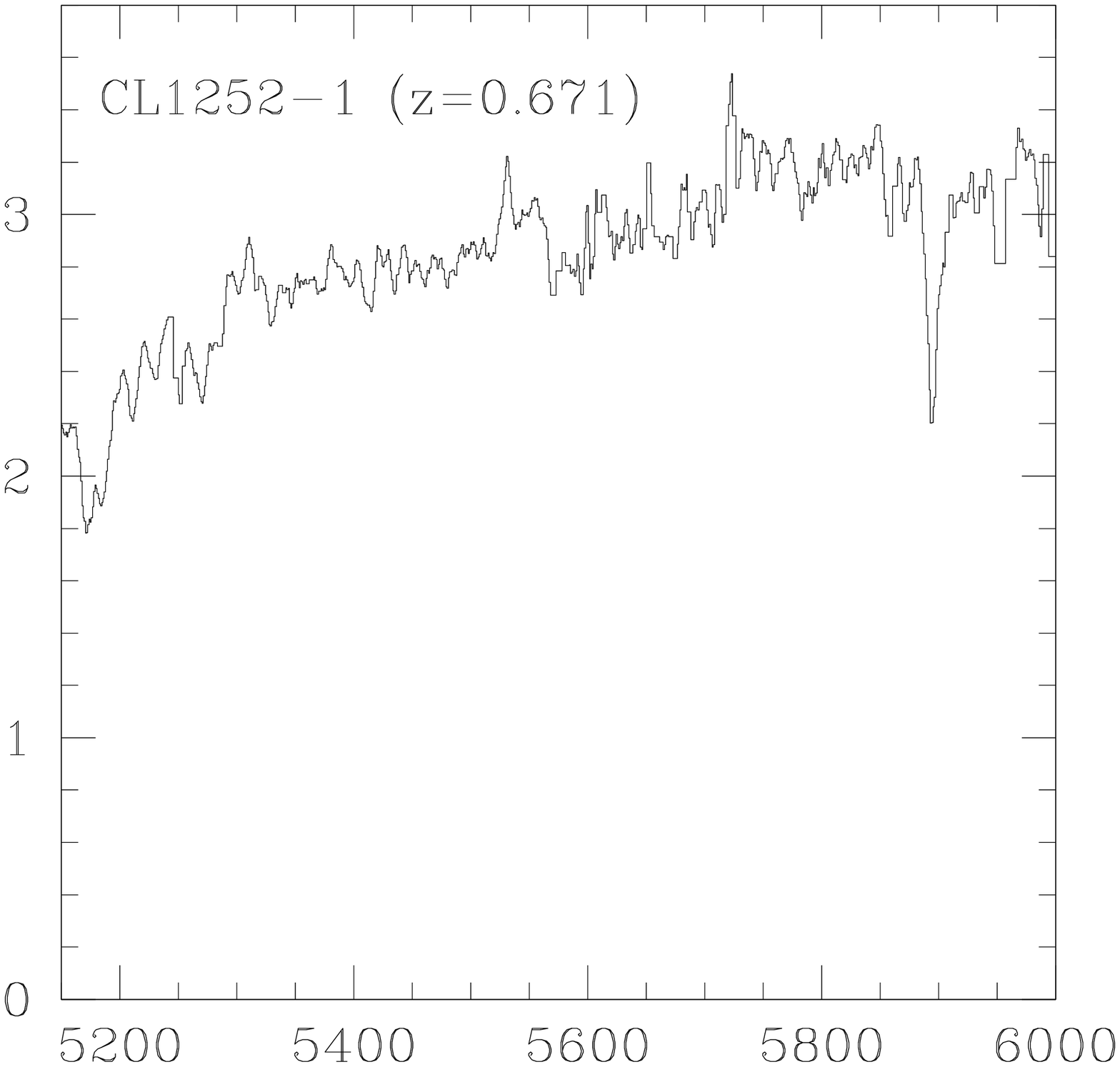}
\epsfxsize=5.8cm
\epsffile{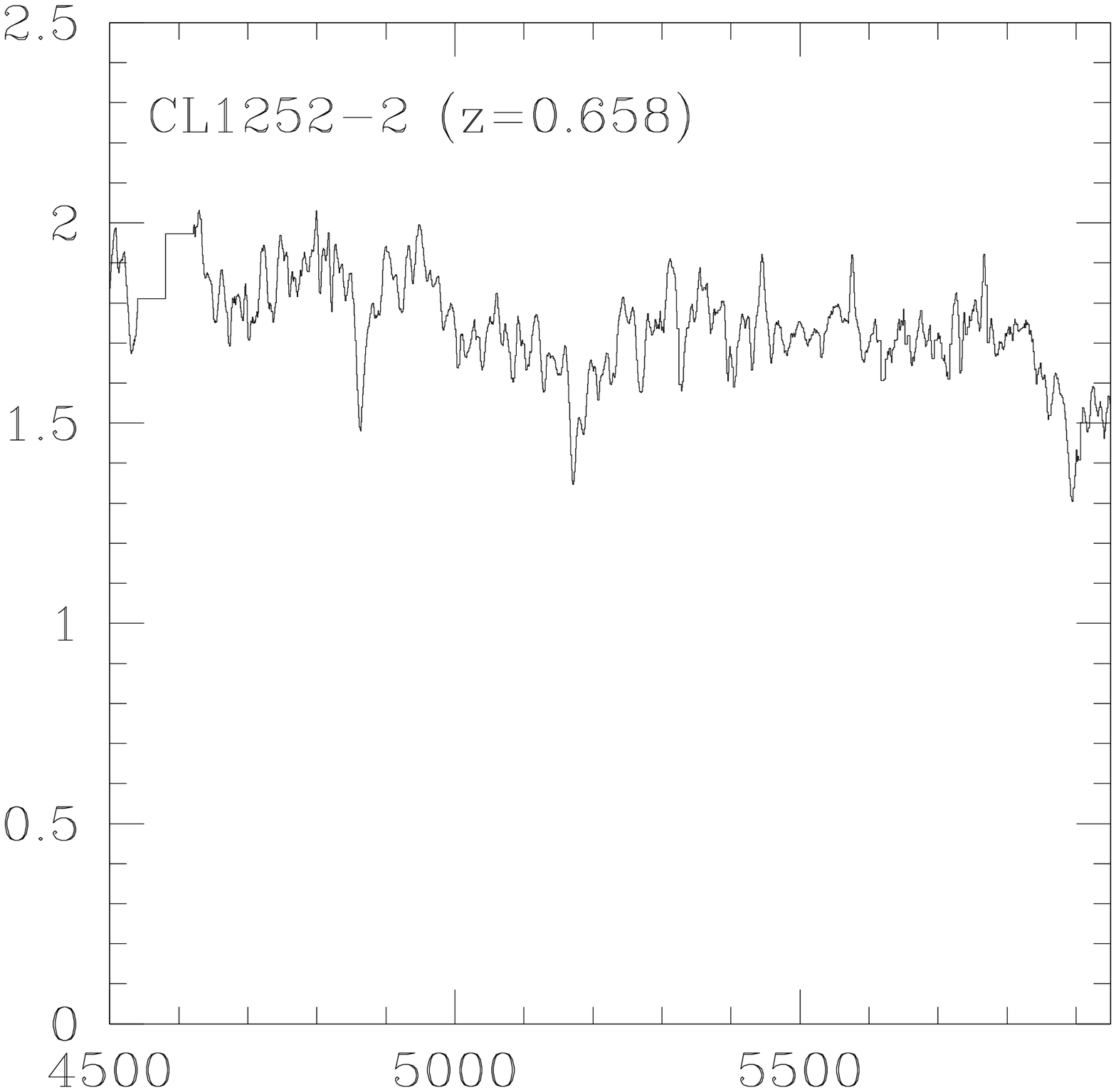}
\epsfxsize=5.8cm
\epsffile{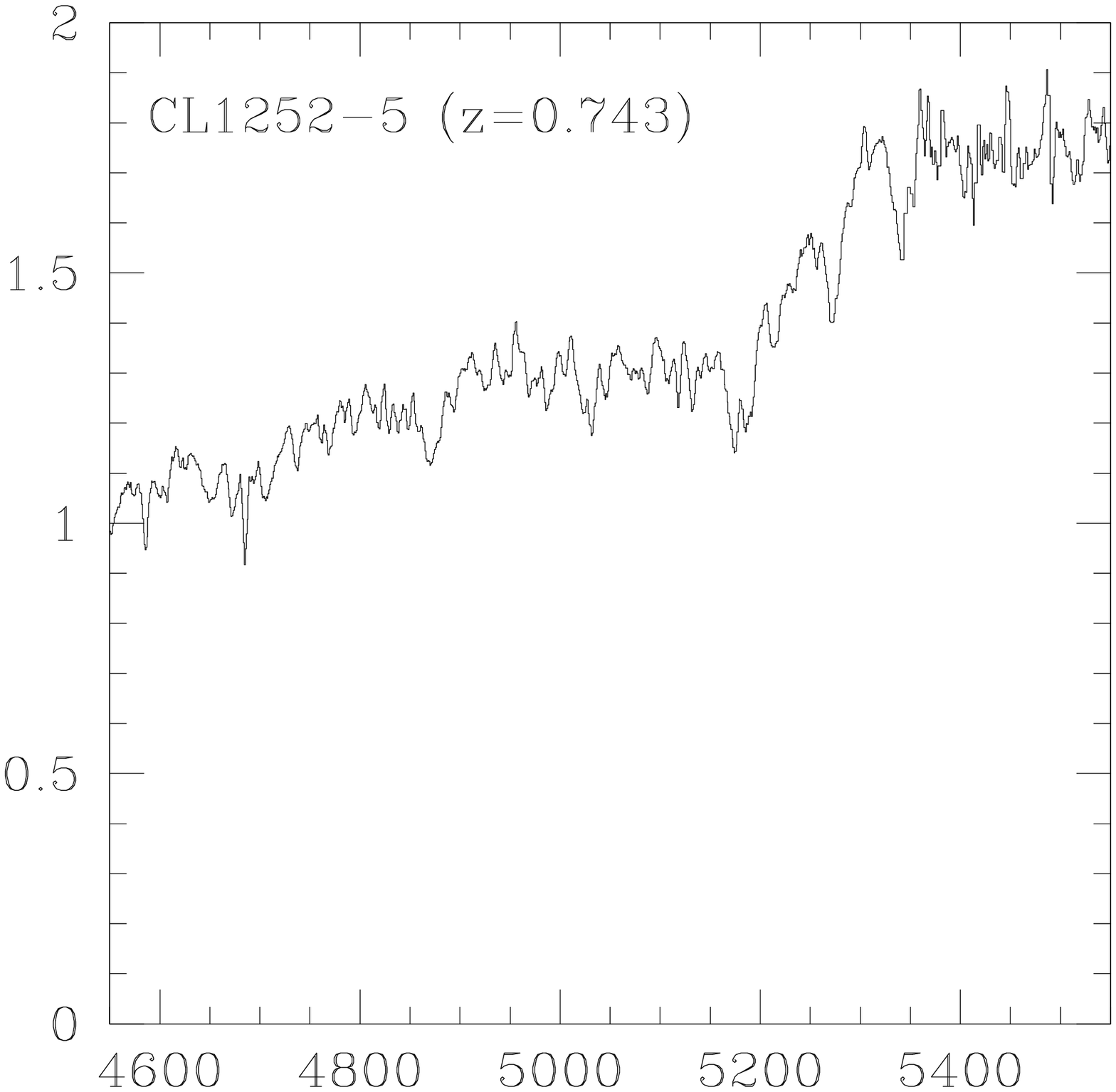}
}
\hbox{
\epsfxsize=5.8cm
\epsffile{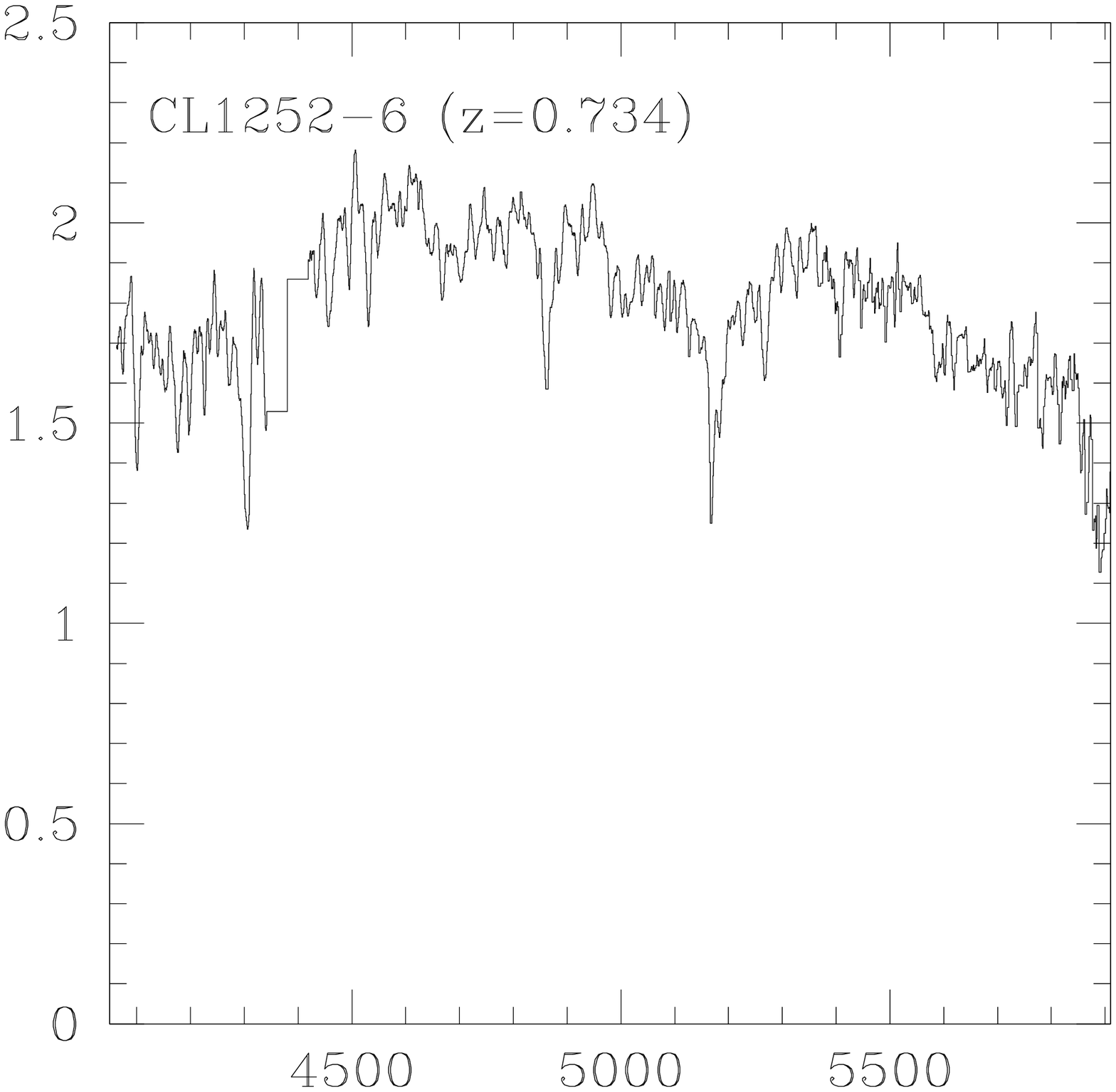}
\epsfxsize=5.8cm
\epsffile{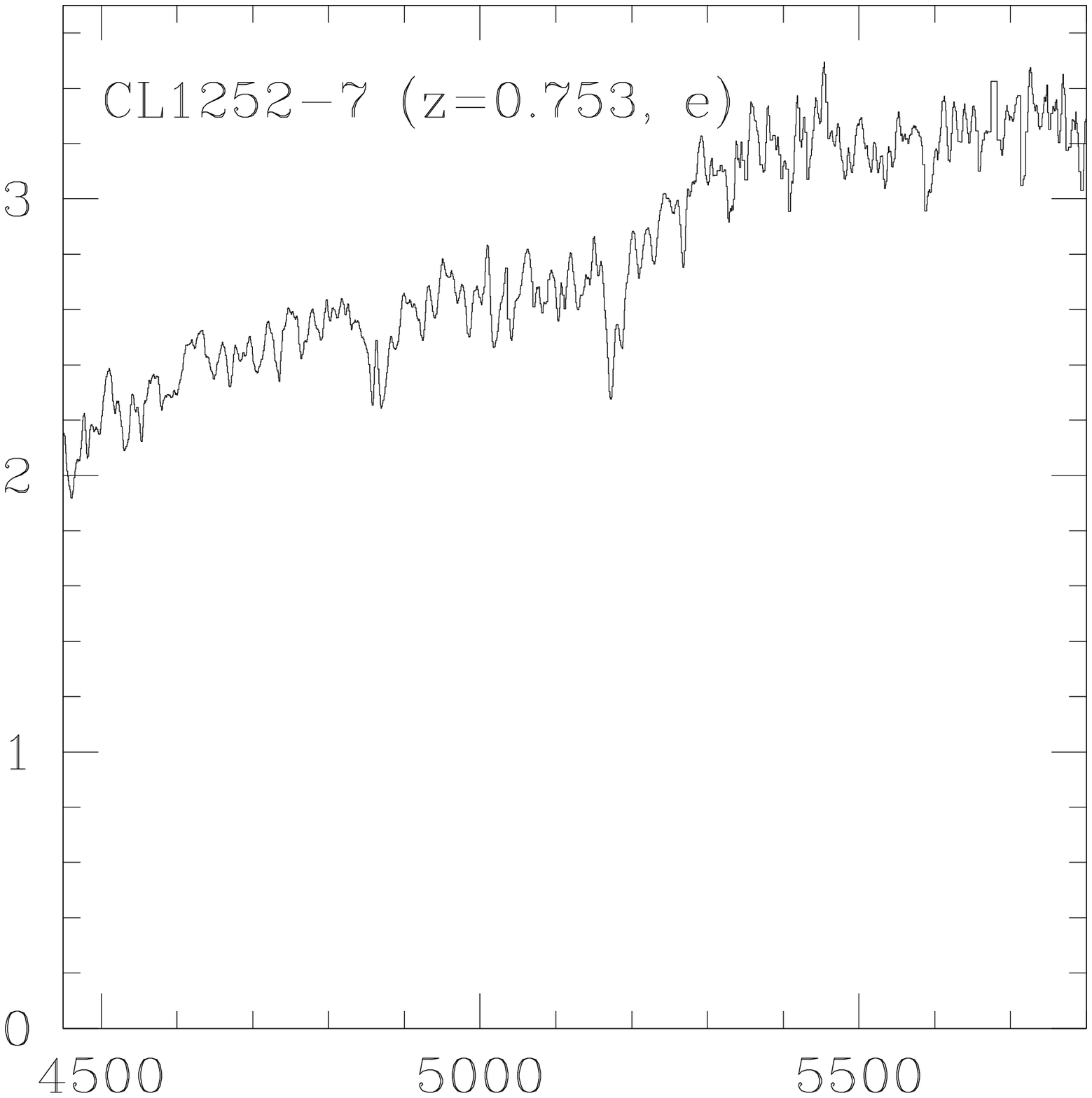}
\epsfxsize=5.8cm
\epsffile{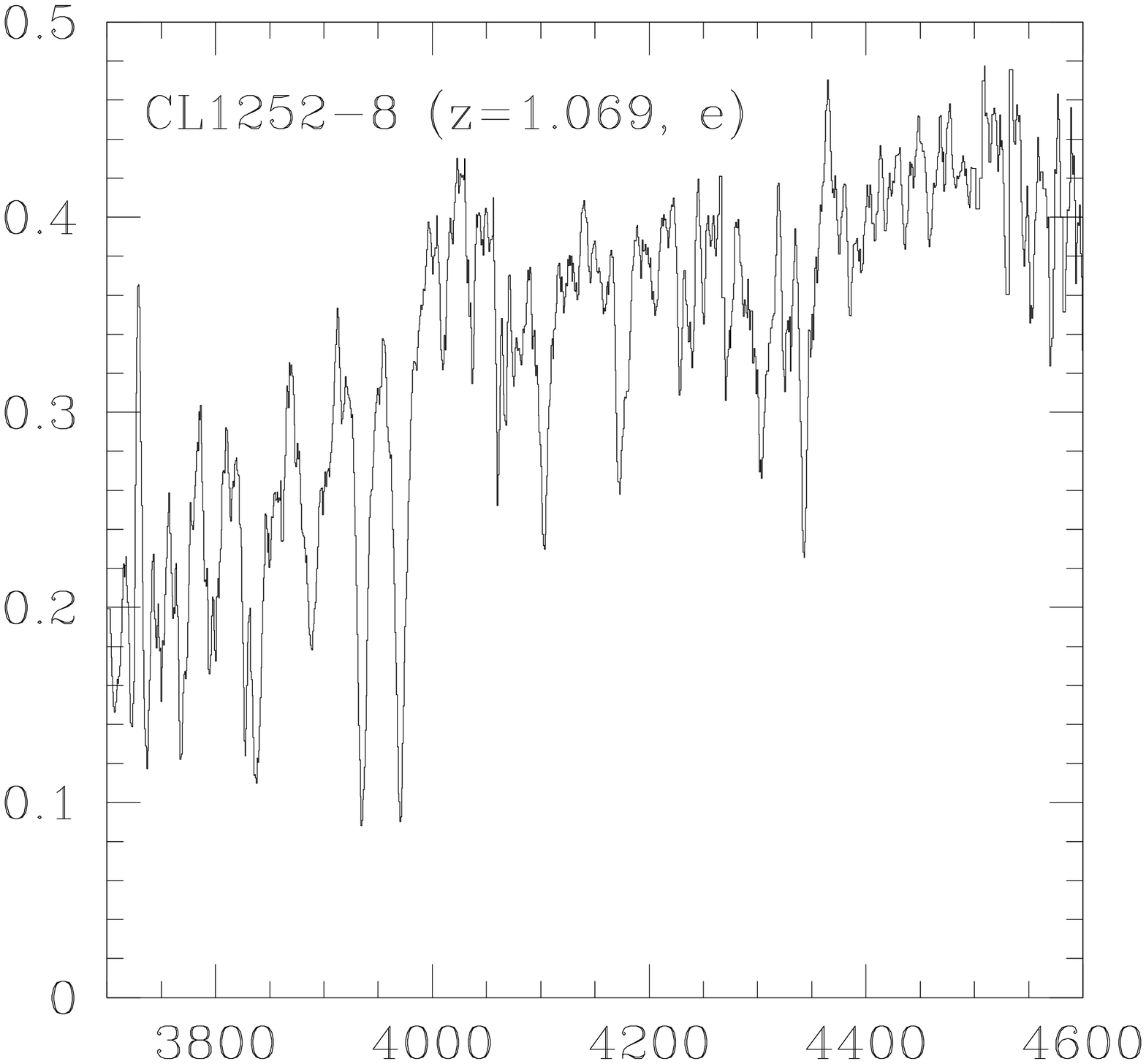}
}
\hbox{
\epsfxsize=5.8cm
\epsffile{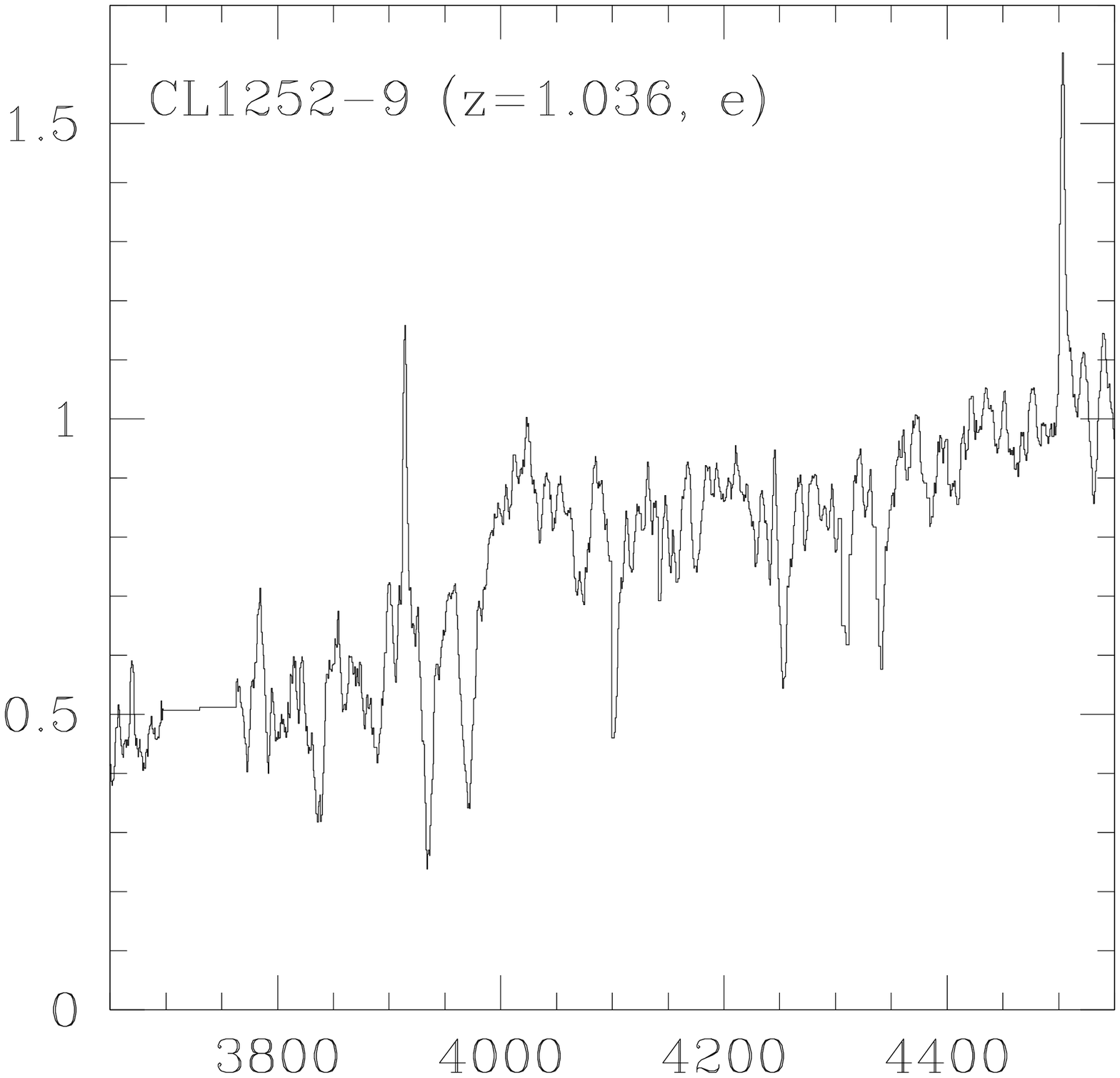}
\epsfxsize=5.8cm
\epsffile{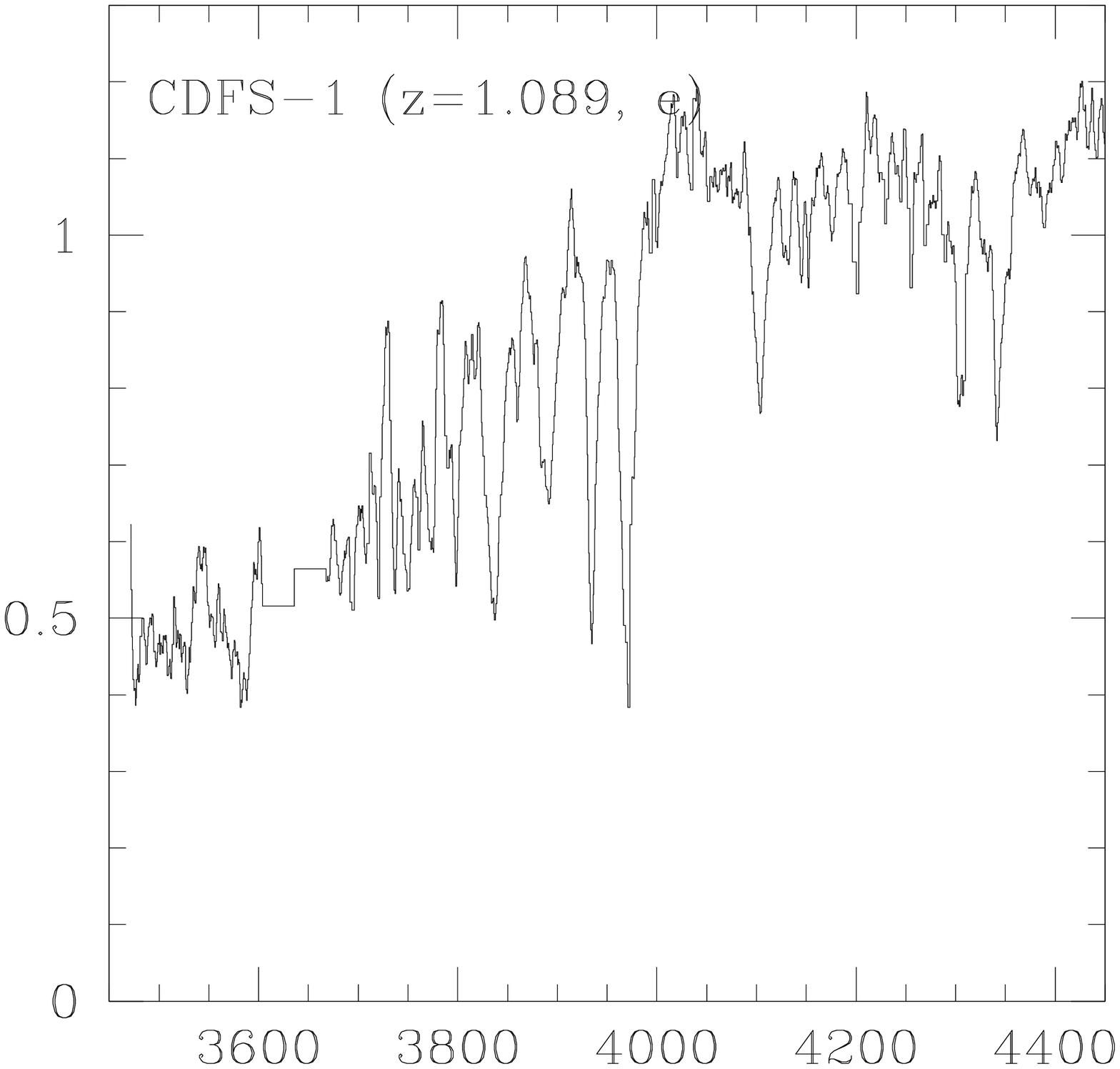}
\epsfxsize=5.8cm
\epsffile{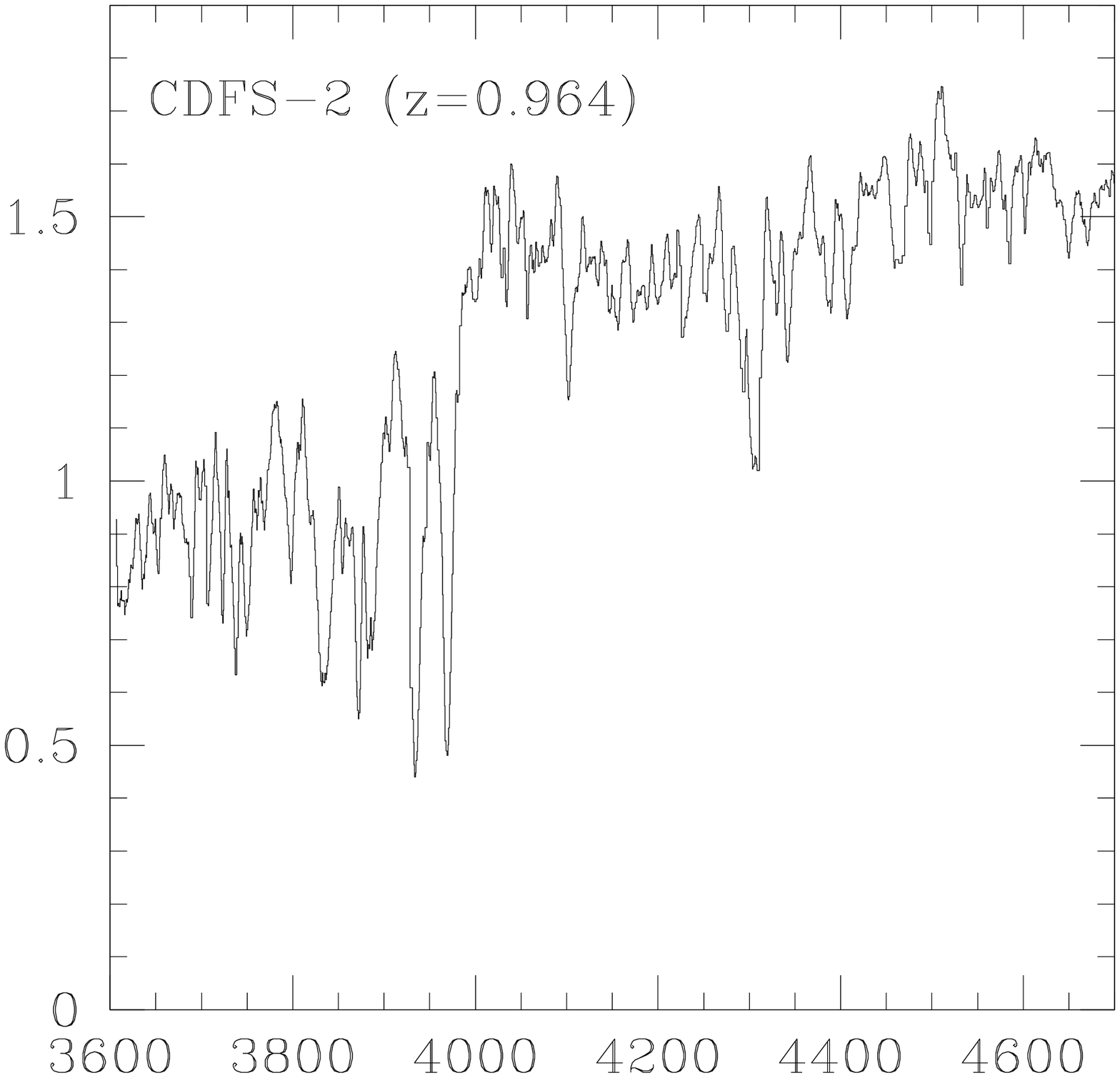}
}
\hbox{
\epsfxsize=5.8cm
\epsffile{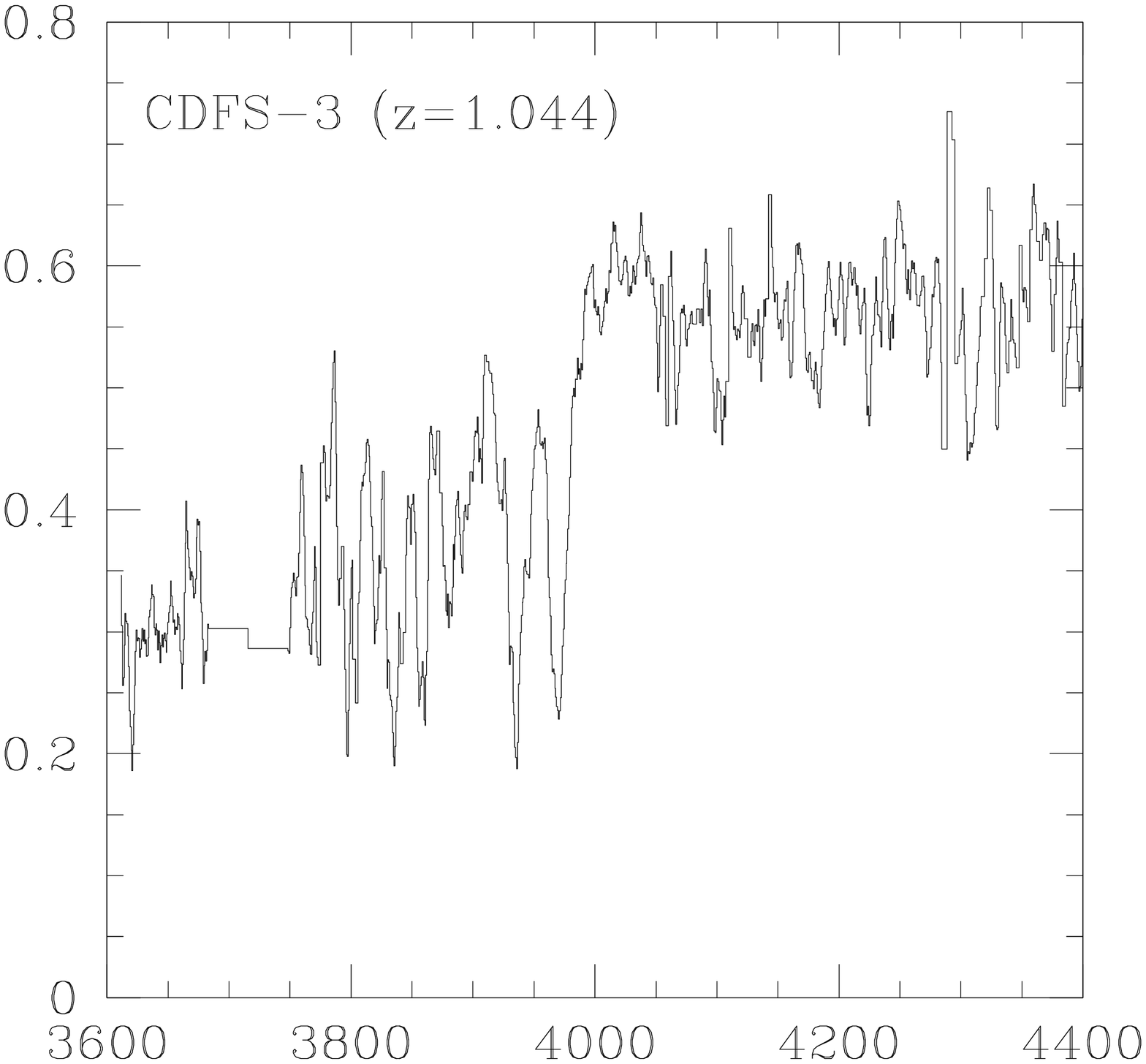}
\epsfxsize=5.8cm
\epsffile{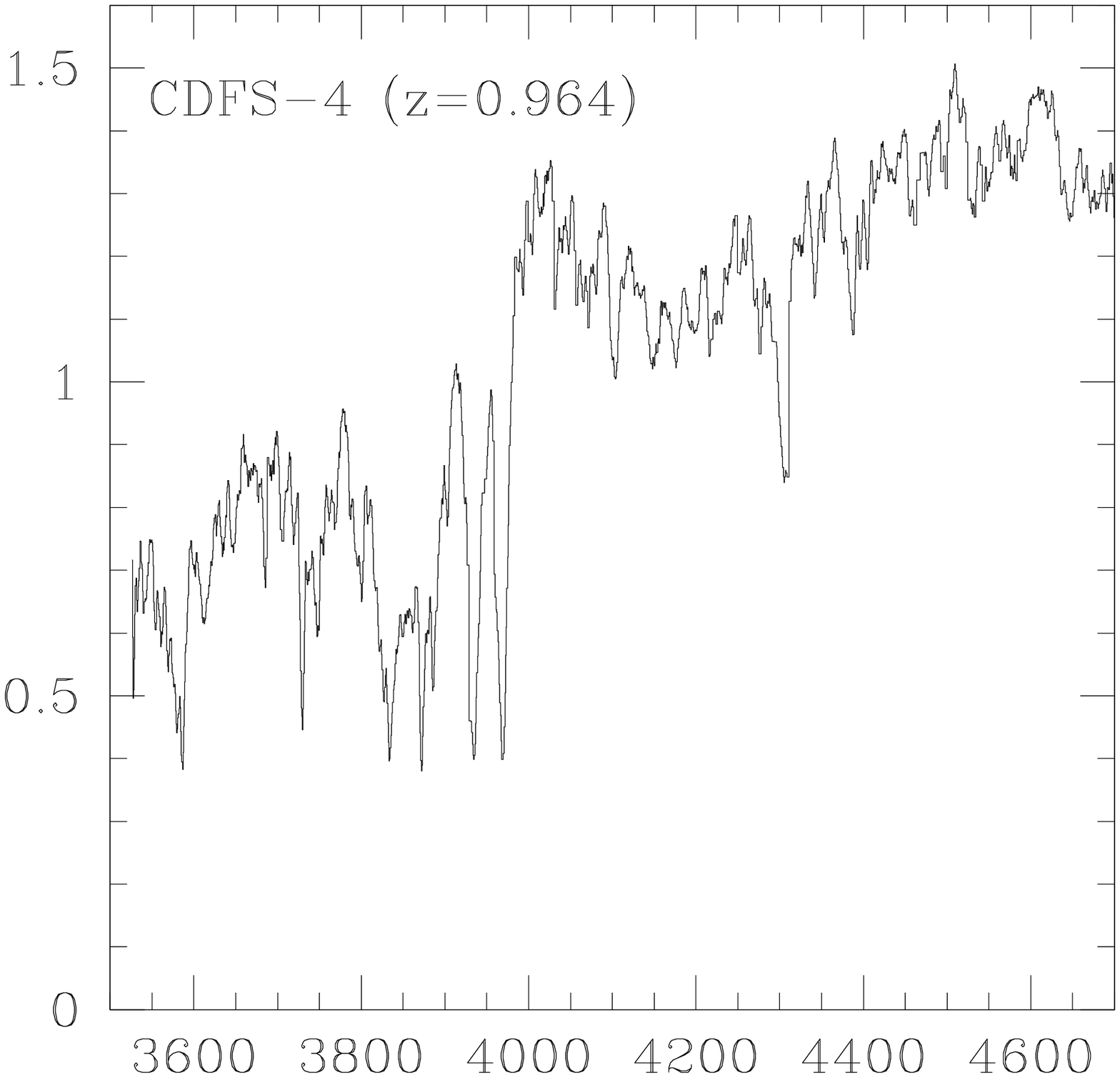}
\epsfxsize=5.8cm
\epsffile{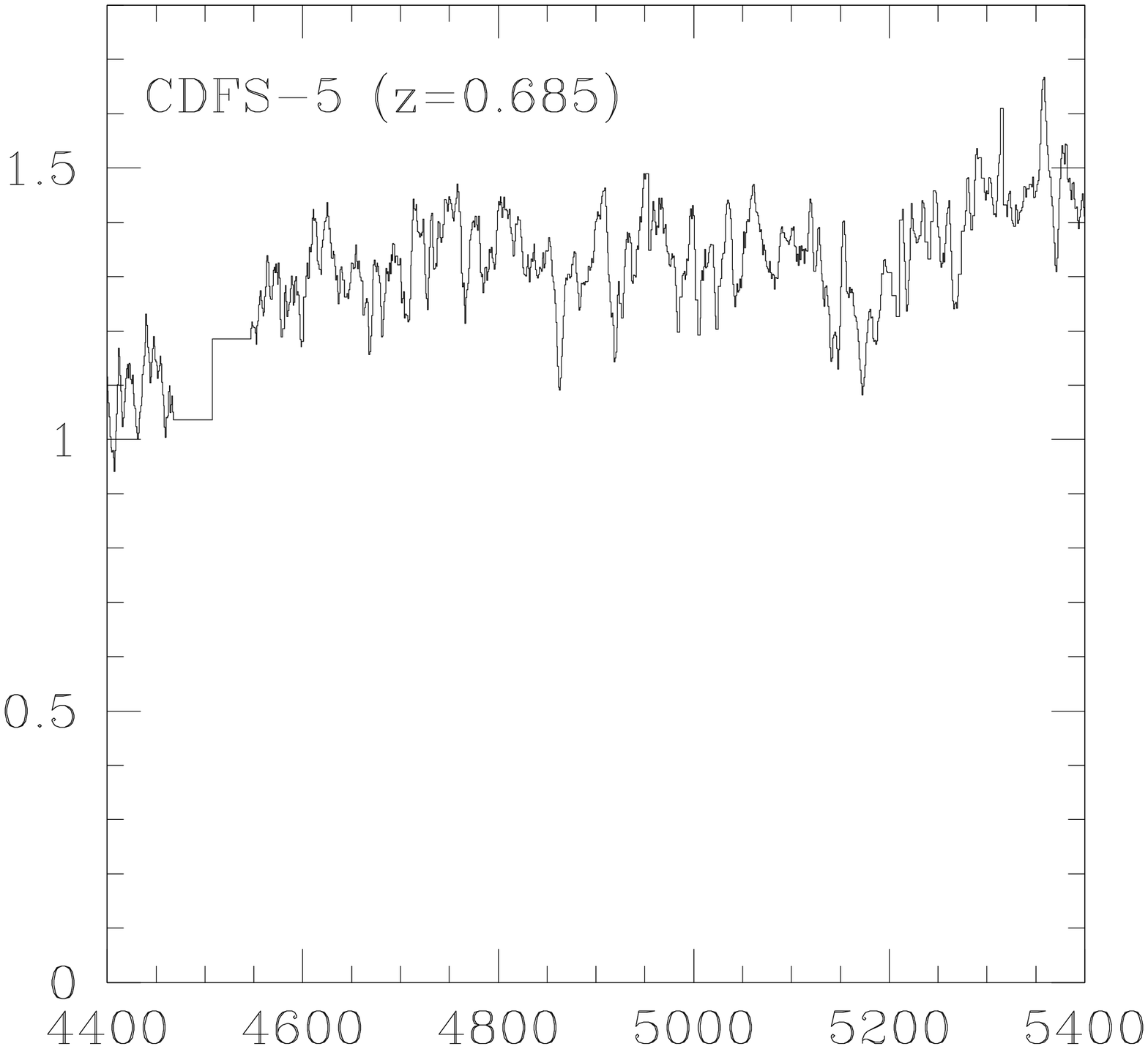}
}
\hbox{
\figcaption{\small 
Rest-frame spectra in 8\AA~bins of the galaxies in our sample with 
early-type morphologies, used in the analysis in the subsequent sections. 
The area around 7600\AA~(observed wavelength) and the positions of bright 
skylines are excluded from the binning.
The wavelength range differs from object to object due to differences in 
redshift and slit position. Every spectrum is labeled with the redshifts 
and an 'e' if the spectrum shows one or more emission lines.}\label{fig:spec}}
\end{figure*}

\begin{figure*}
\figurenum{1}
\leavevmode
\hbox{
\epsfxsize=5.8cm
\epsffile{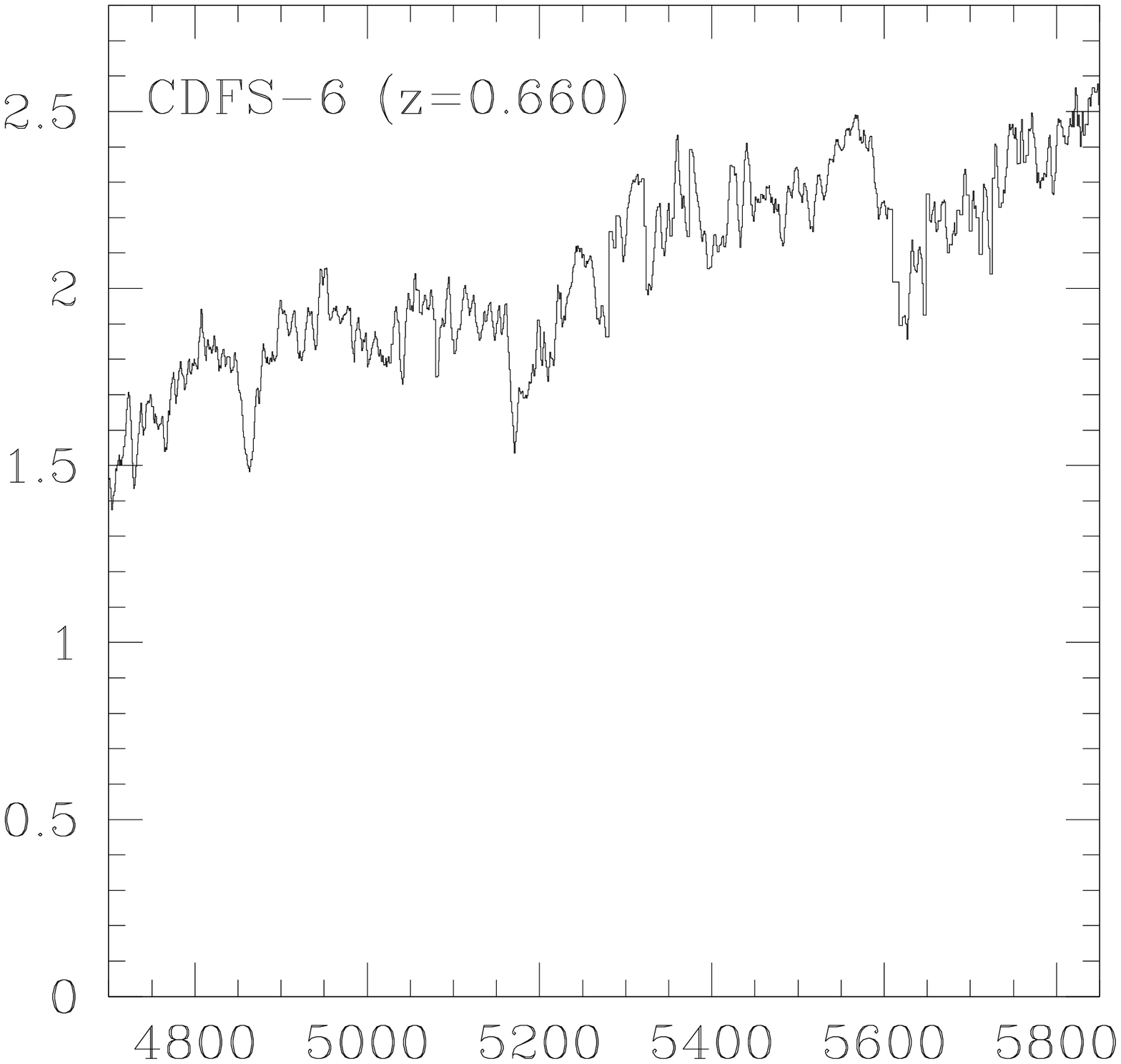}
\epsfxsize=5.8cm
\epsffile{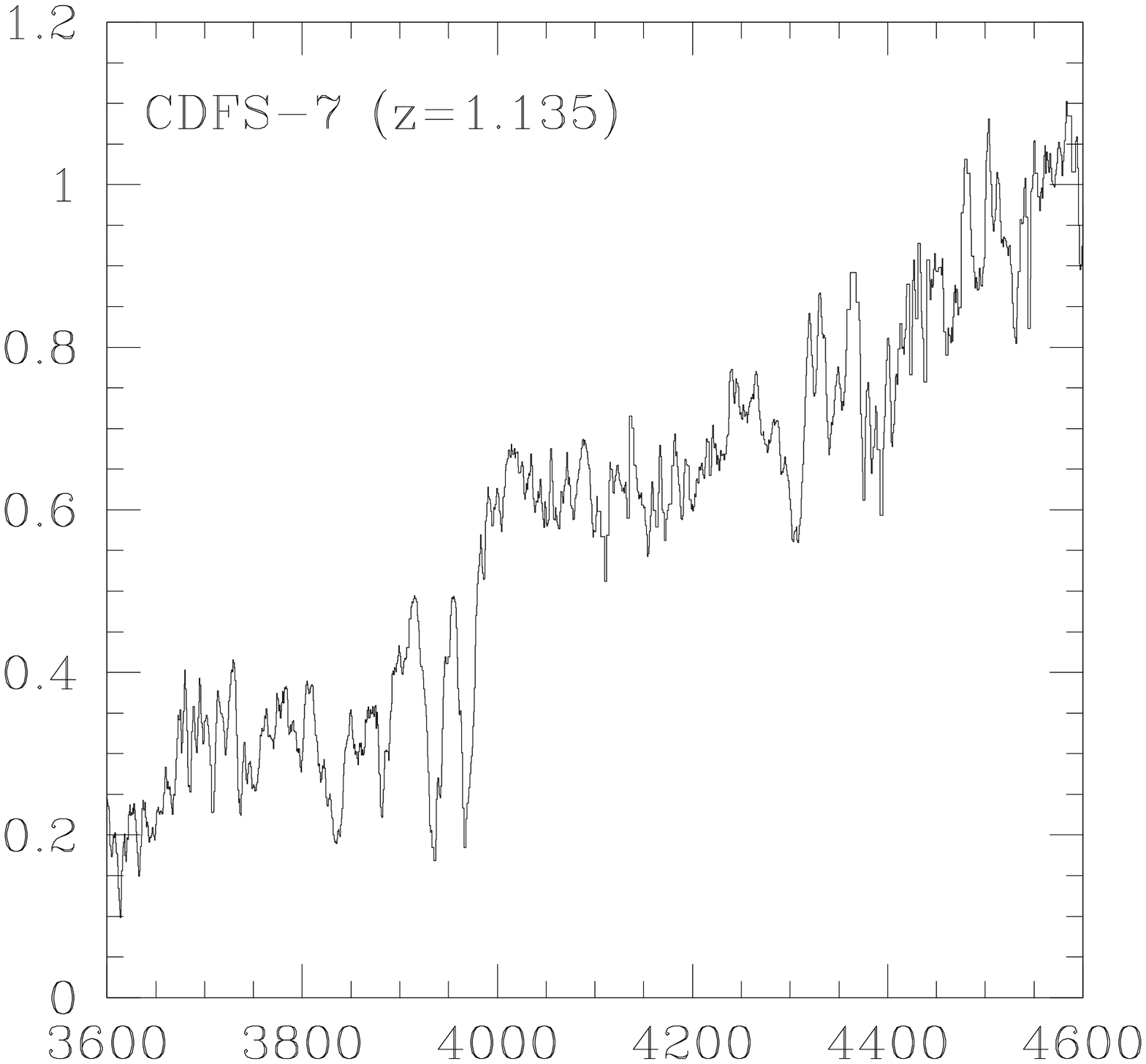}
\epsfxsize=5.8cm
\epsffile{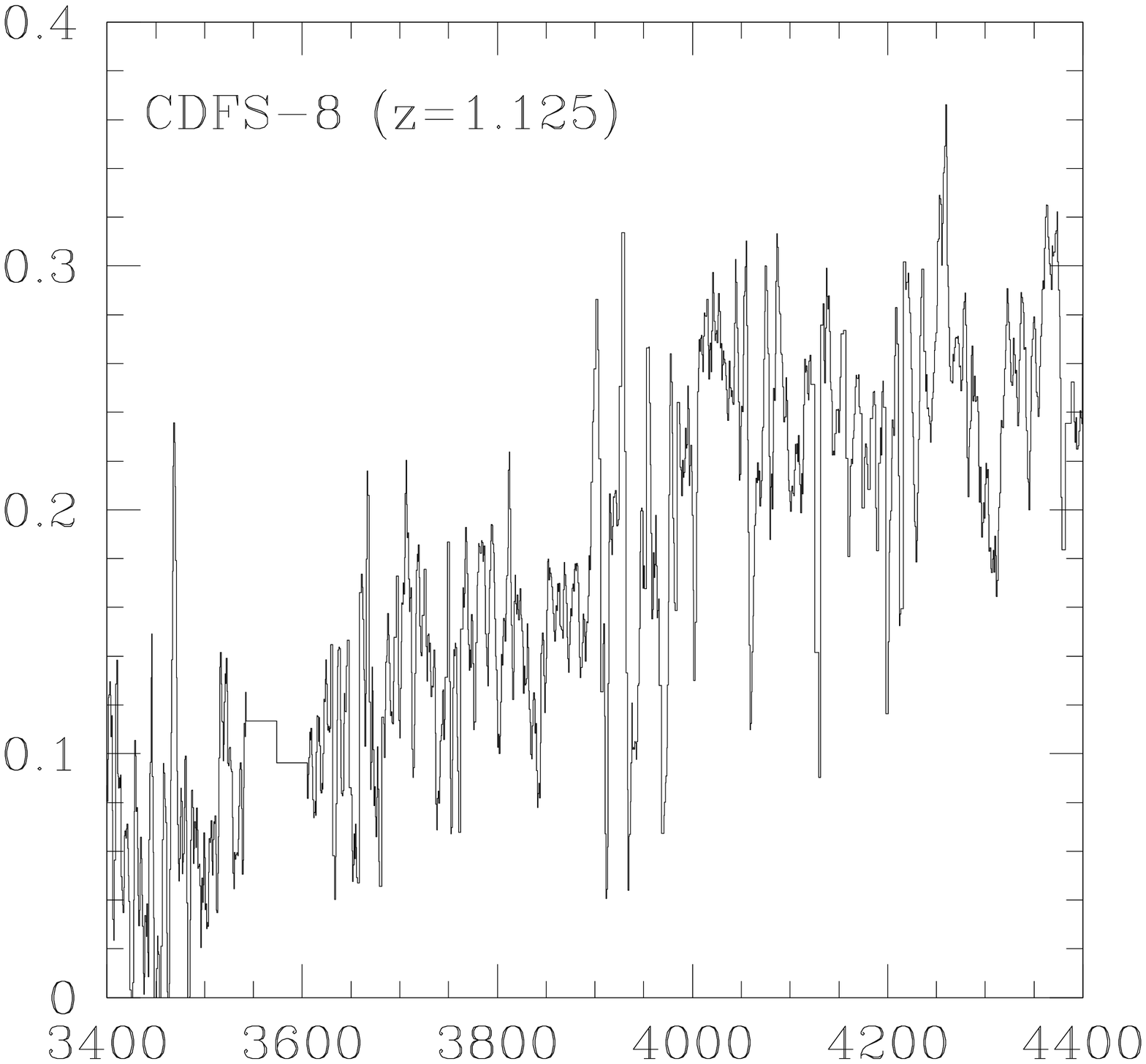}
}
\hbox{
\epsfxsize=5.8cm
\epsffile{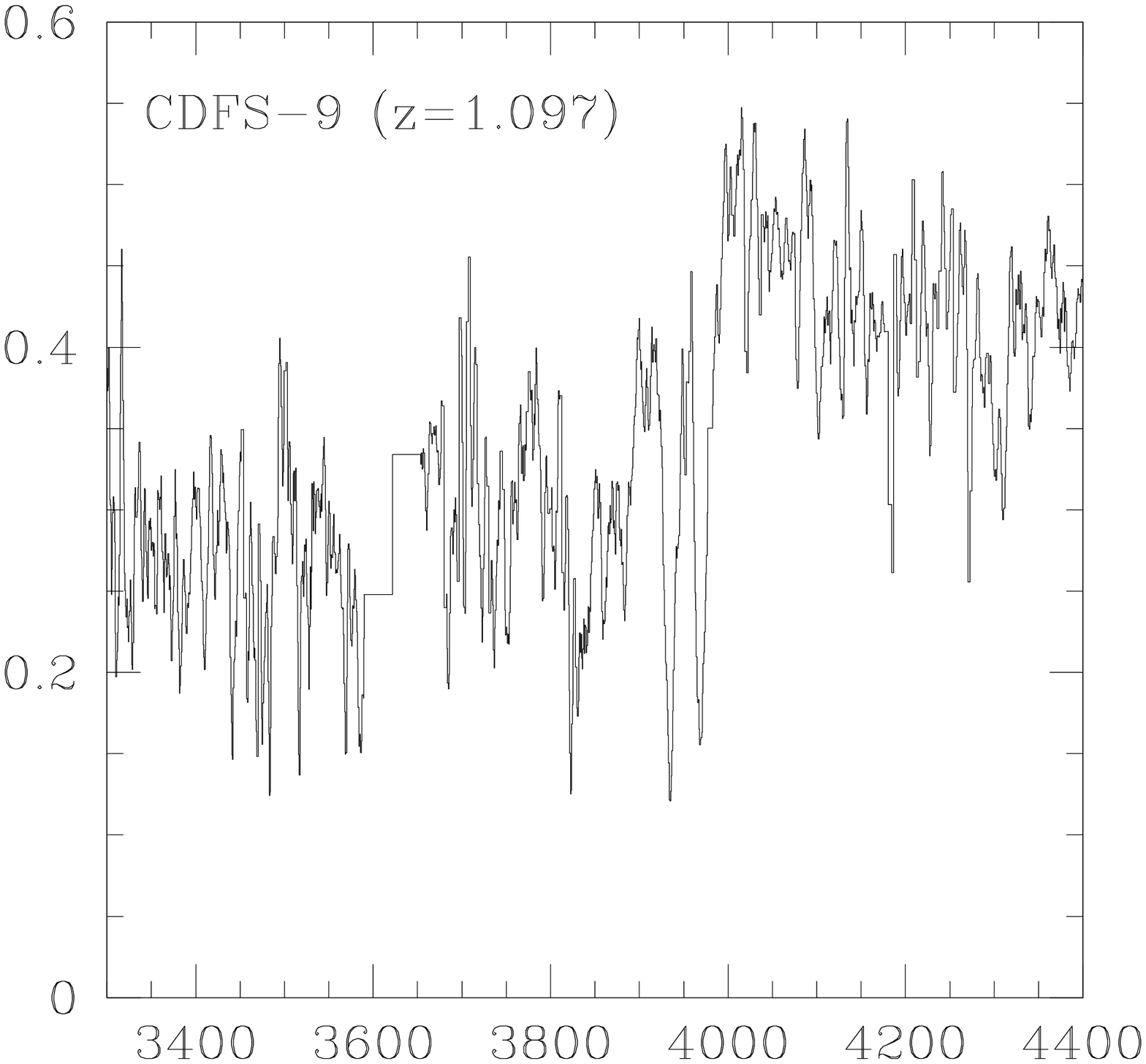}
\epsfxsize=5.8cm
\epsffile{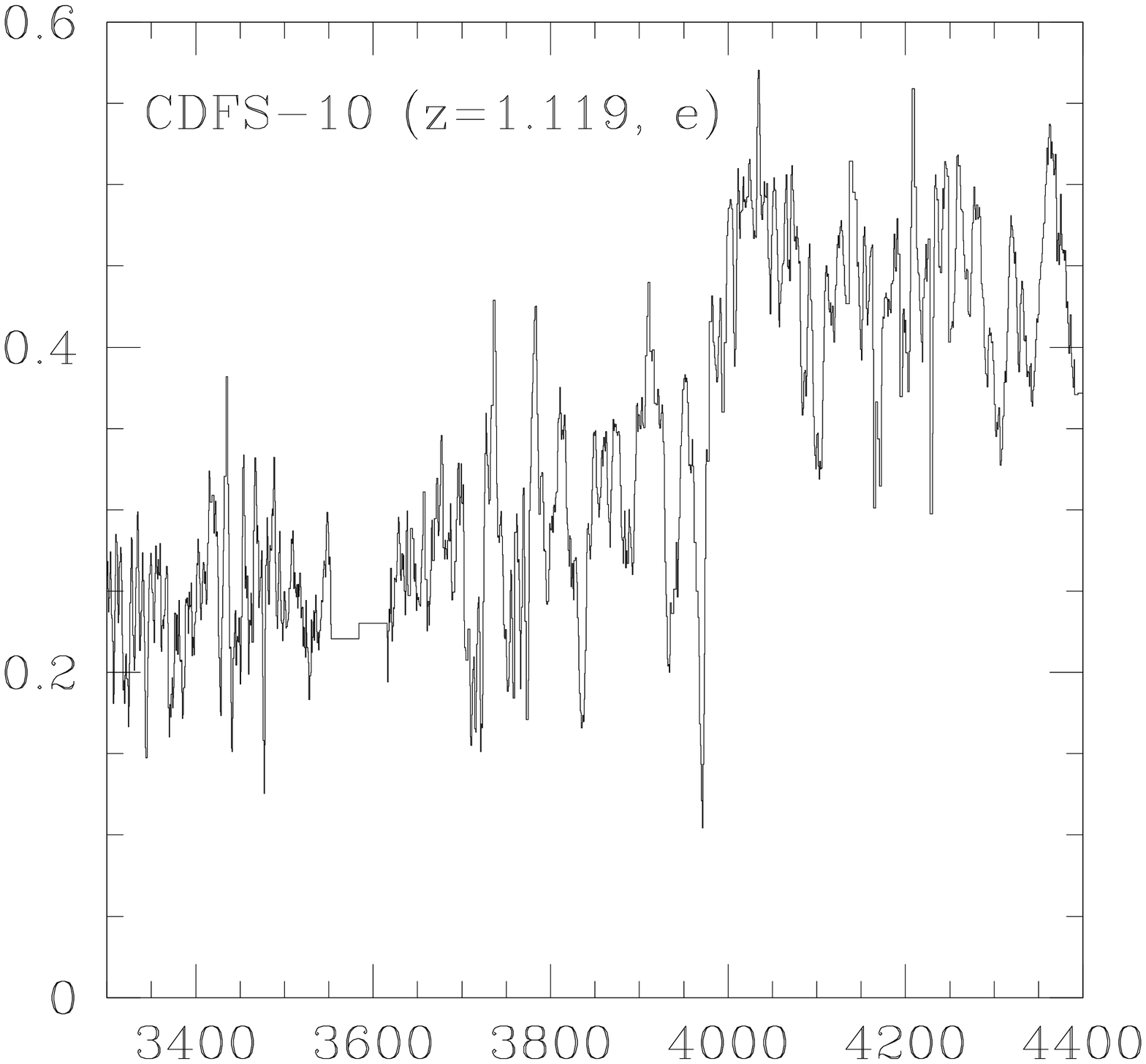}
\epsfxsize=5.8cm
\epsffile{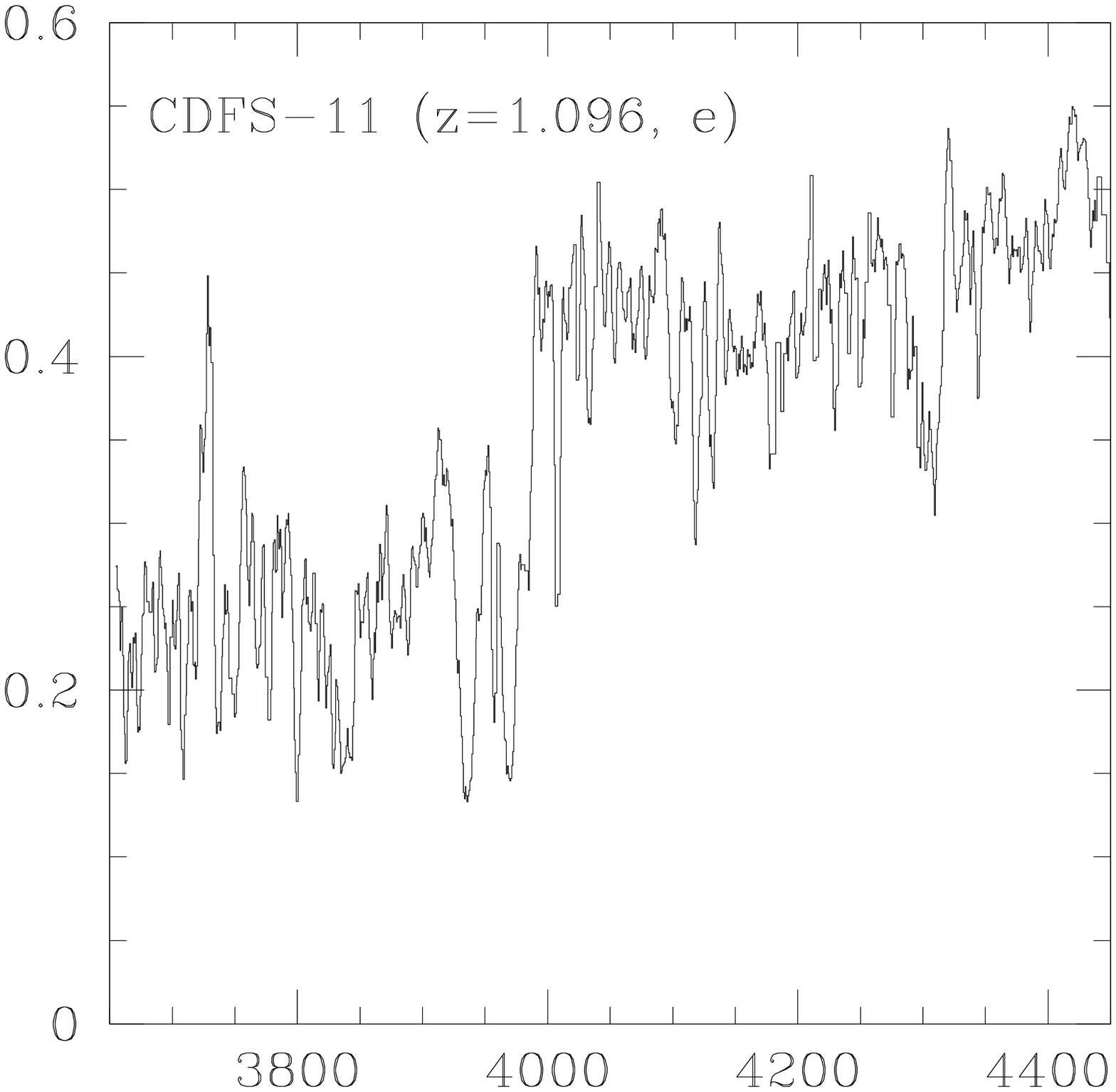}
}
\hbox{
\epsfxsize=5.8cm
\epsffile{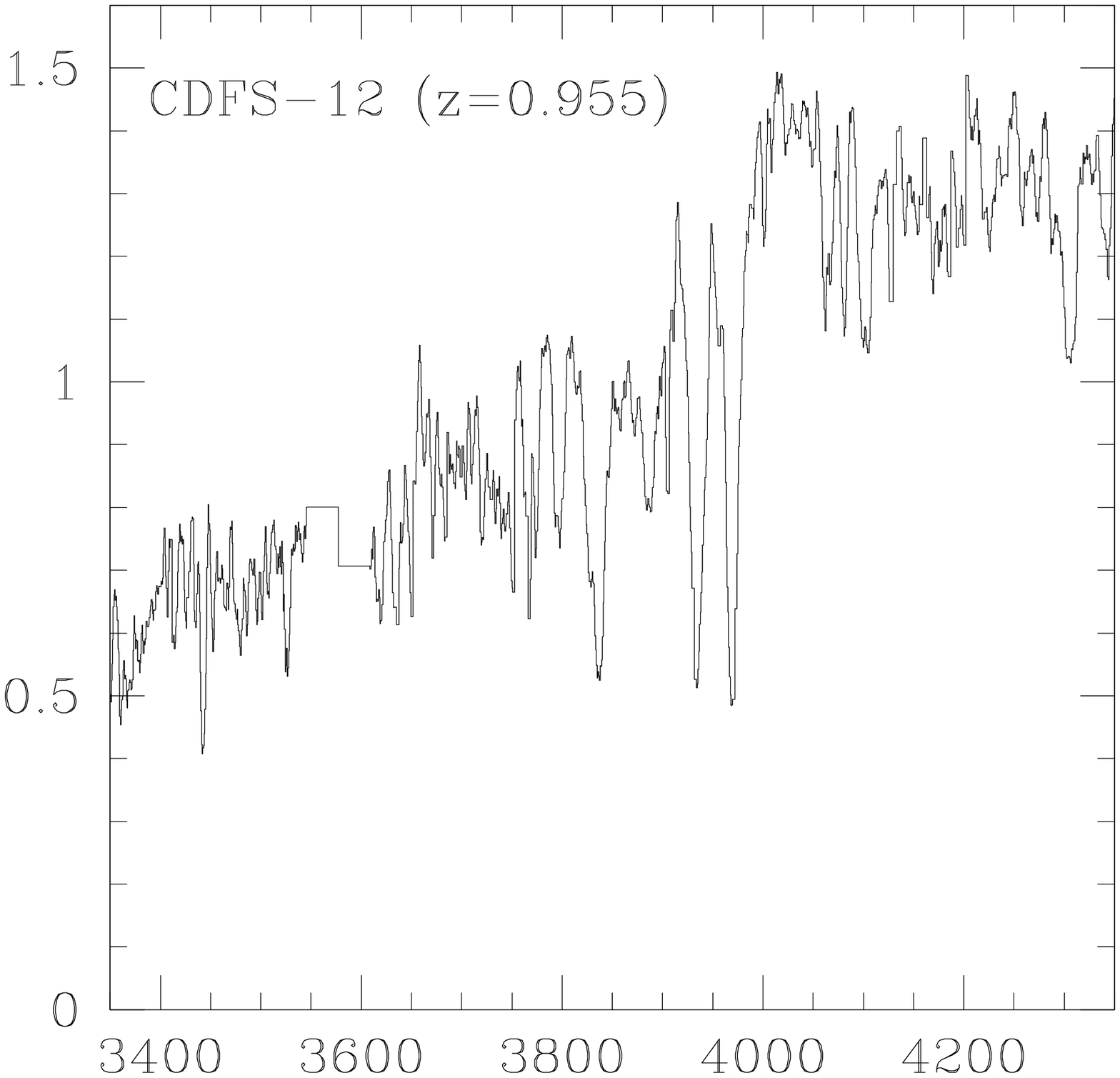}
\epsfxsize=5.8cm
\epsffile{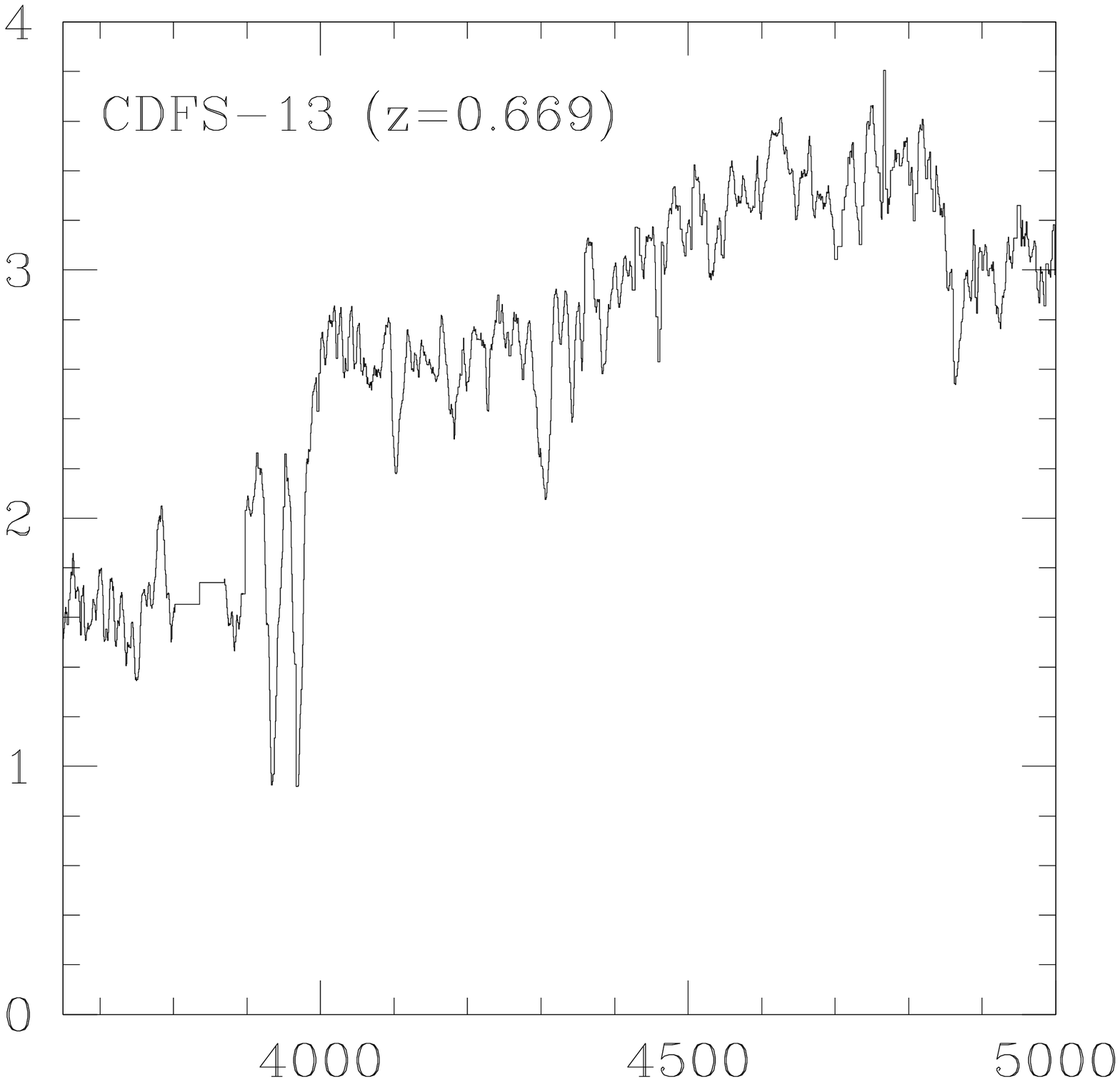}
\epsfxsize=5.8cm
\epsffile{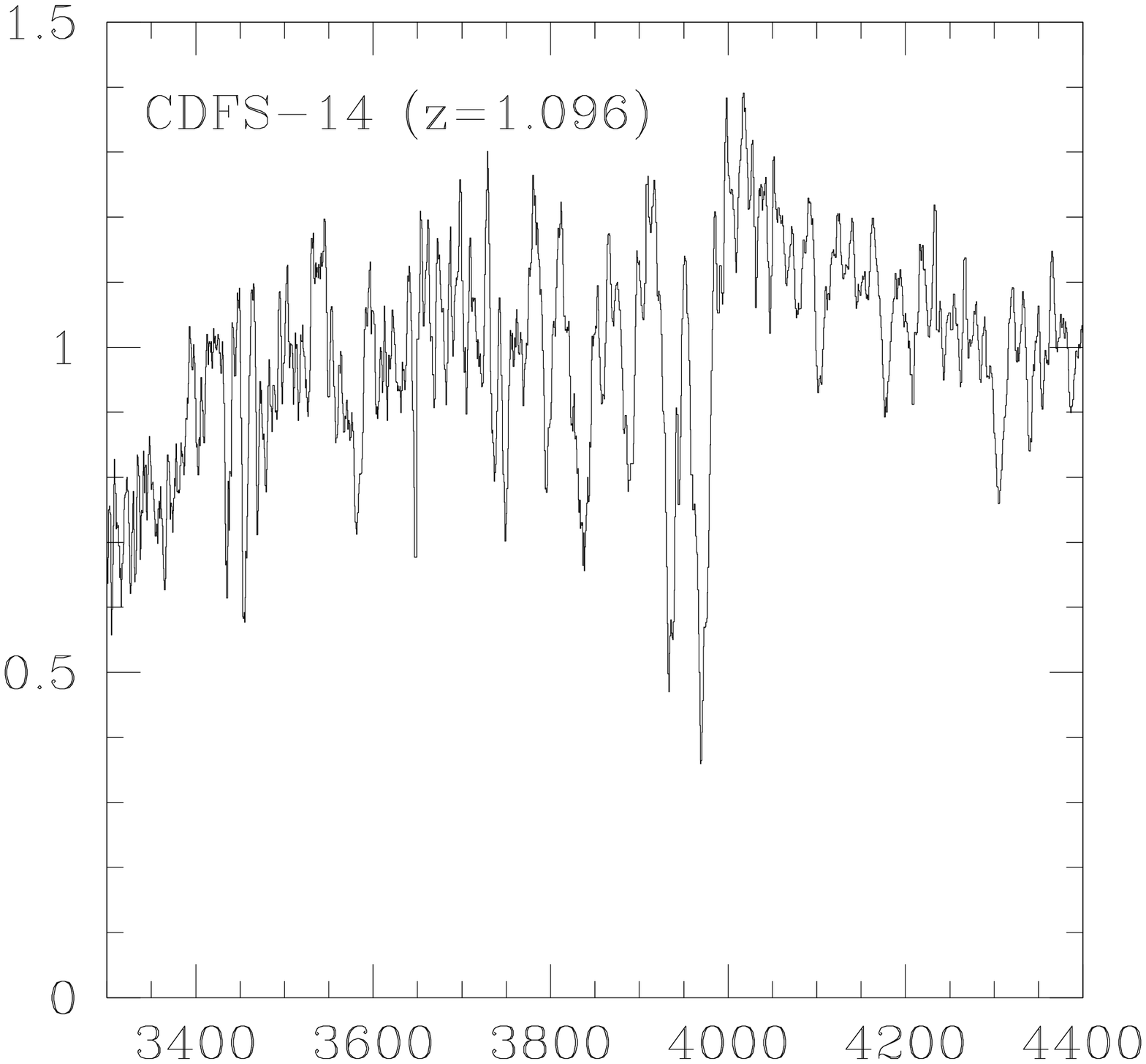}
}
\hbox{
\epsfxsize=5.8cm
\epsffile{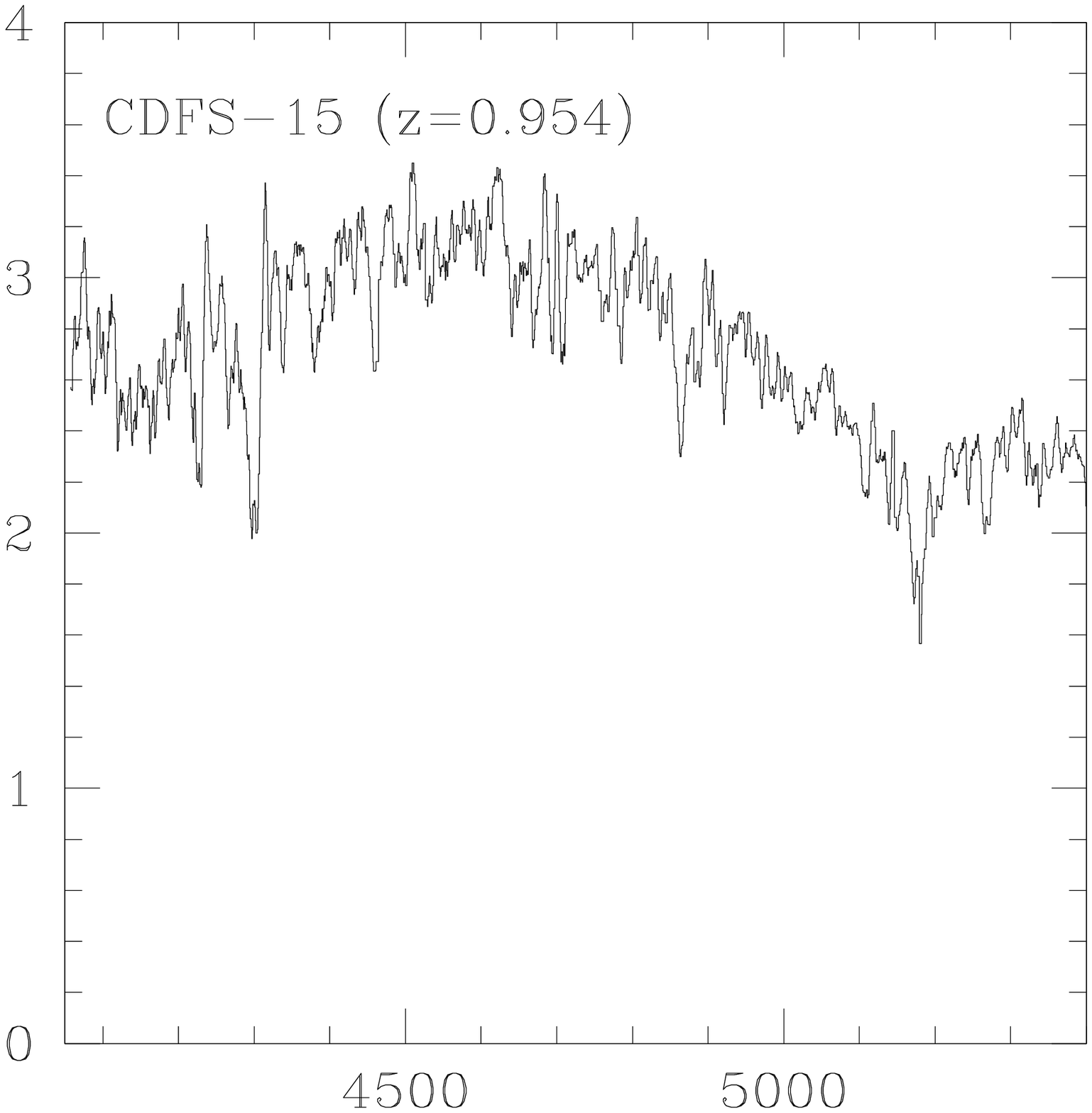}
\epsfxsize=5.8cm
\epsffile{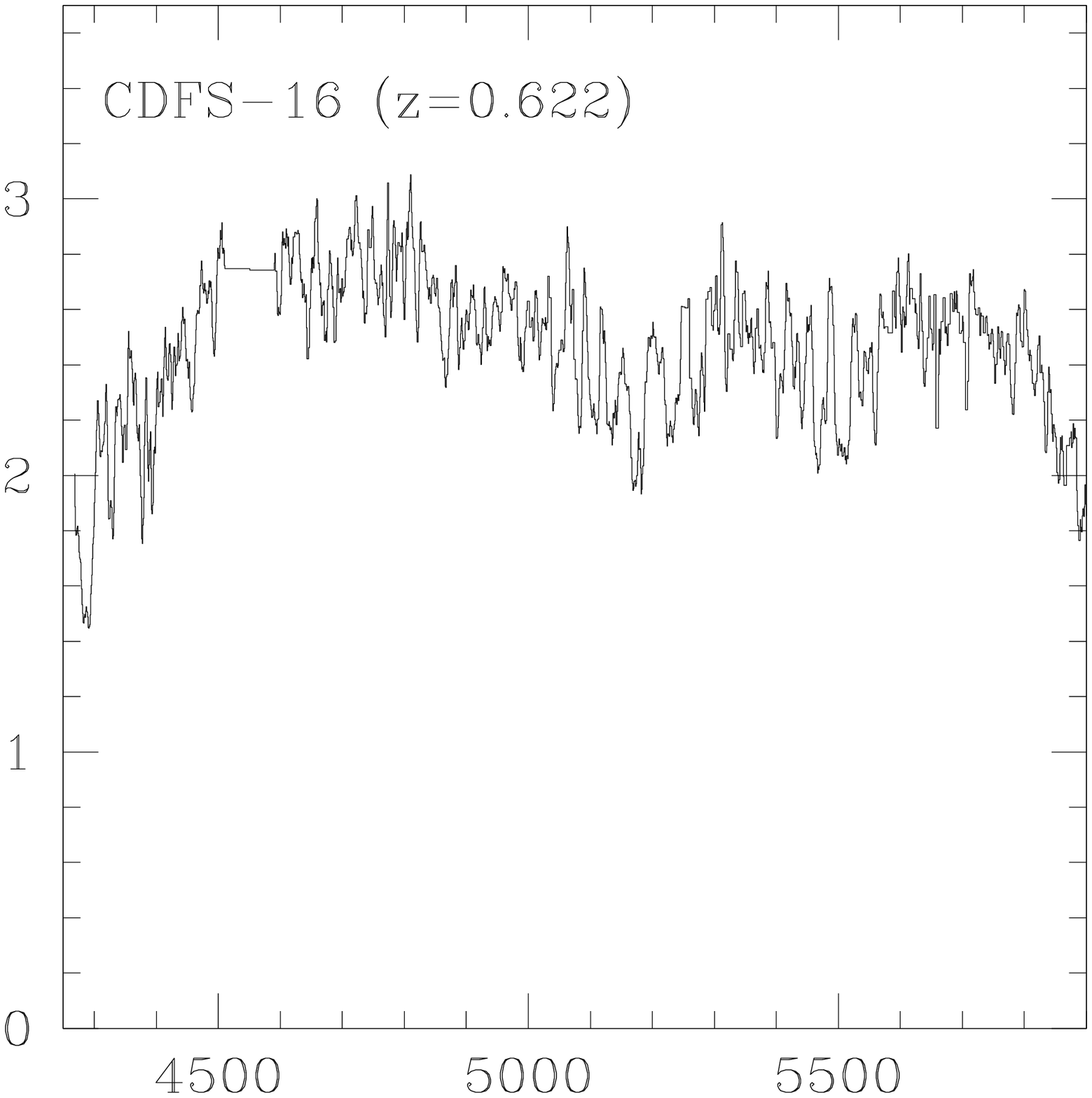}
\epsfxsize=5.8cm
\epsffile{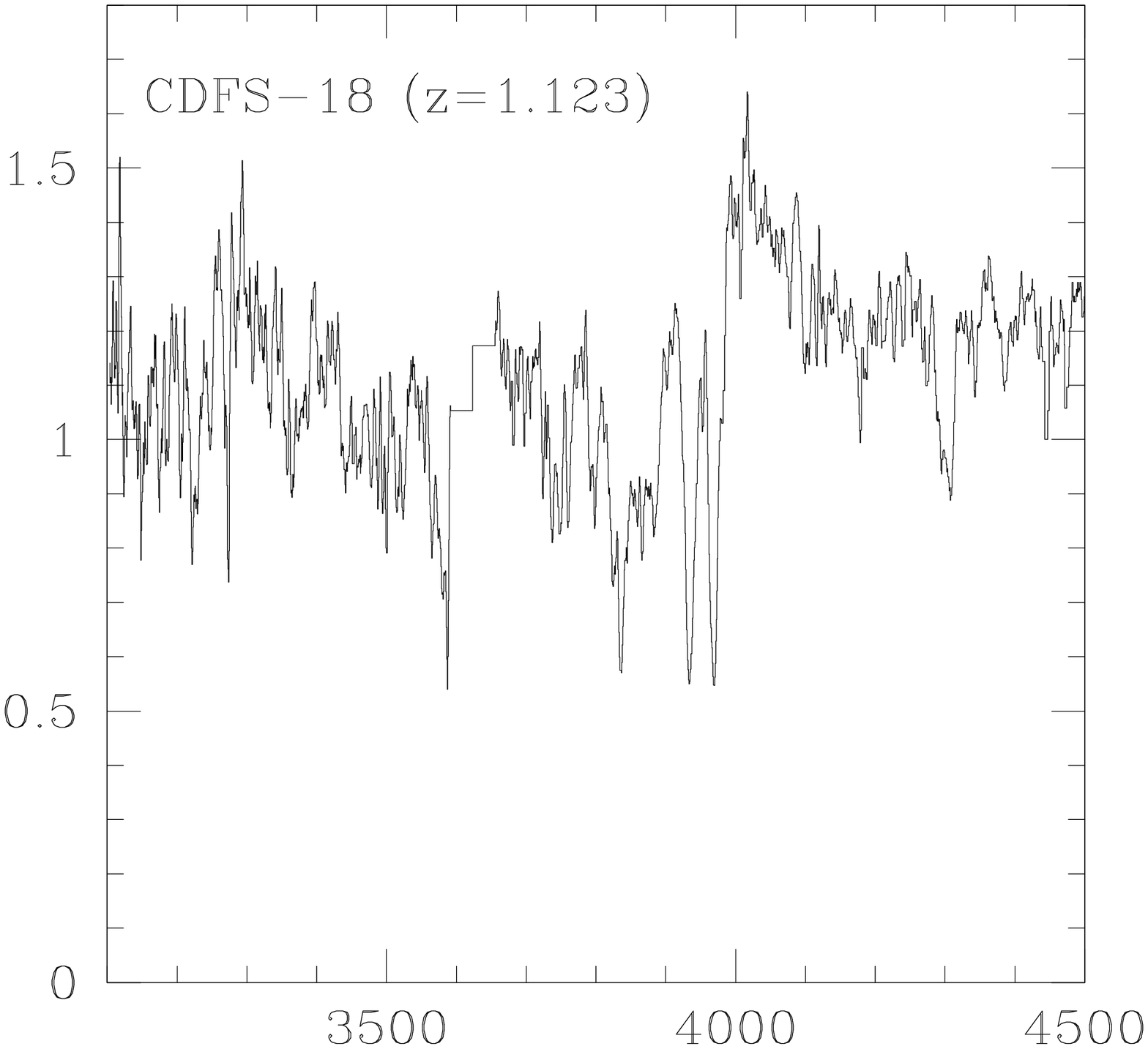}
}
\hbox{
\figcaption{\small Continued}
\label{fig:spec}}
\end{figure*}

\begin{figure*}
\figurenum{1}
\leavevmode
\hbox{
\epsfxsize=5.8cm
\epsffile{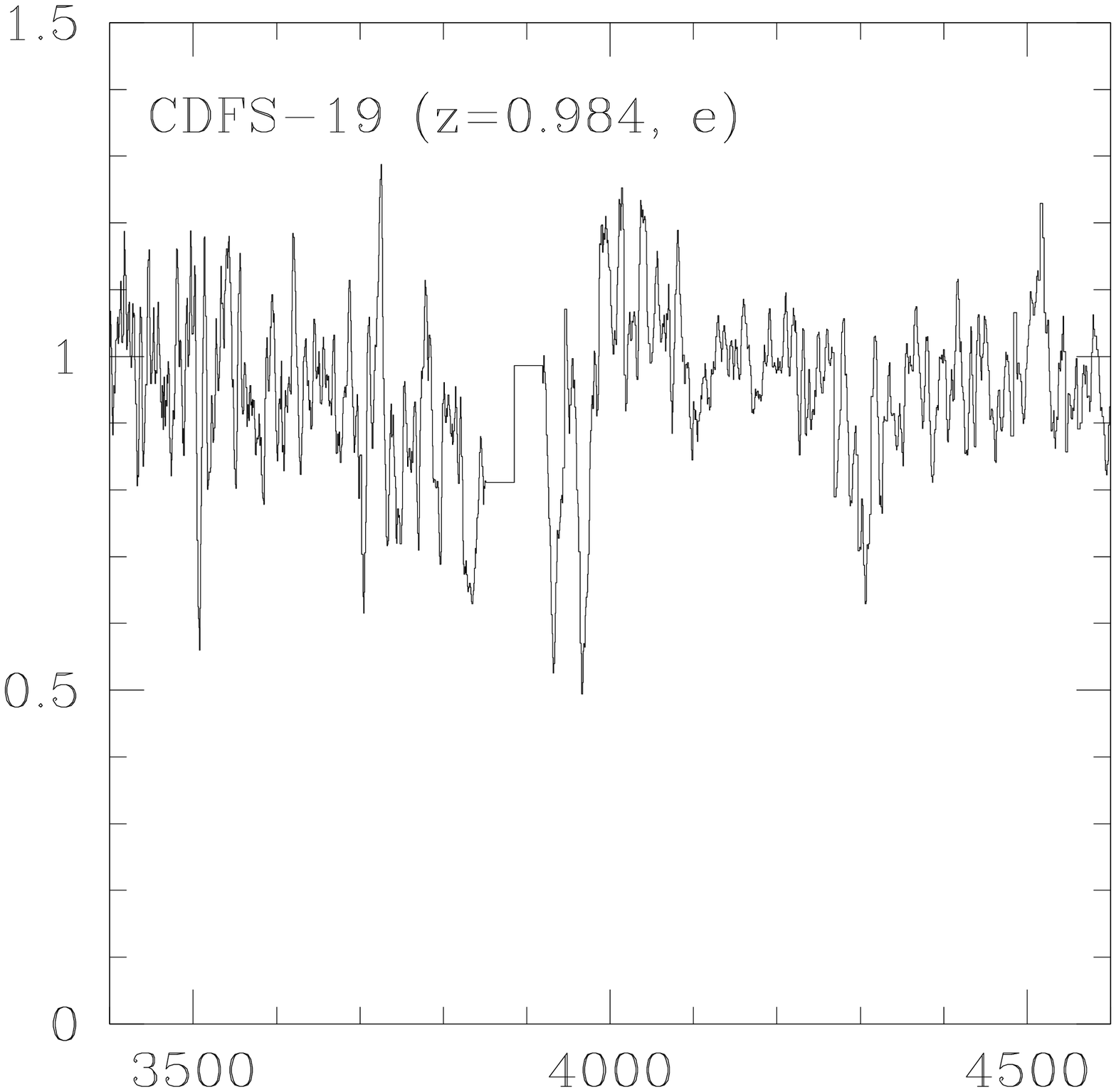}
\epsfxsize=5.8cm
\epsffile{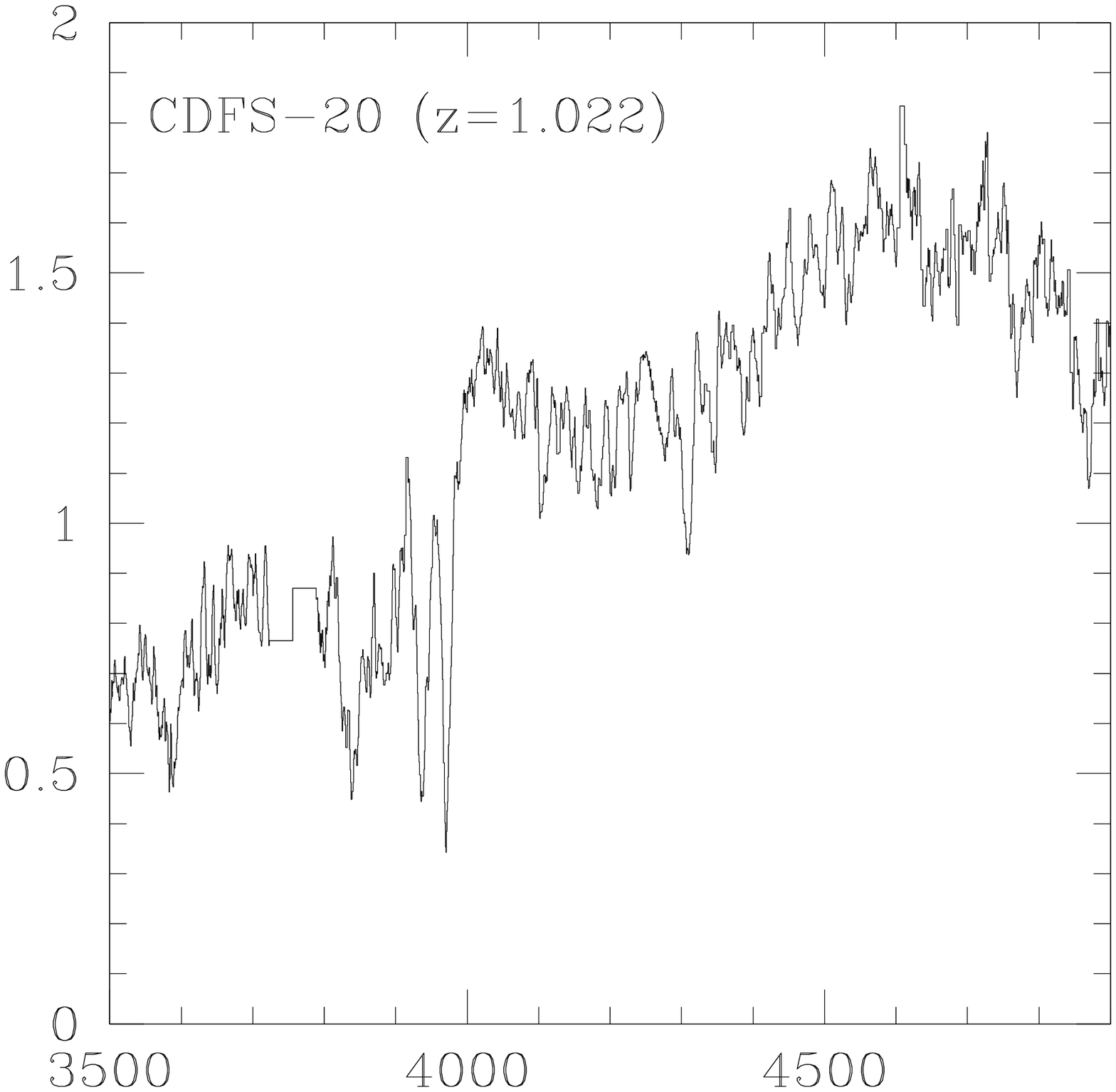}
\epsfxsize=5.8cm
\epsffile{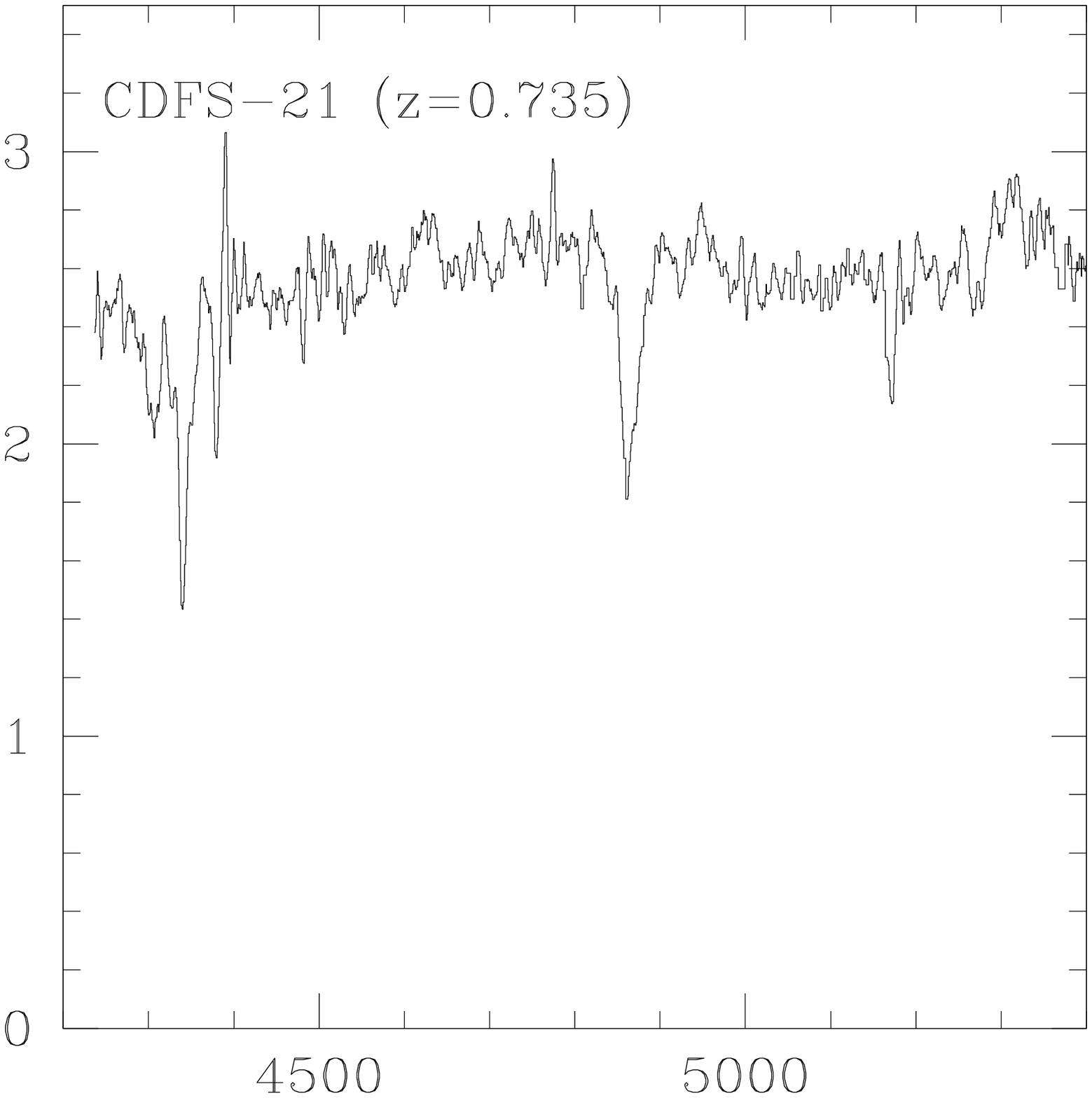}
}
\hbox{
\epsfxsize=5.8cm
\epsffile{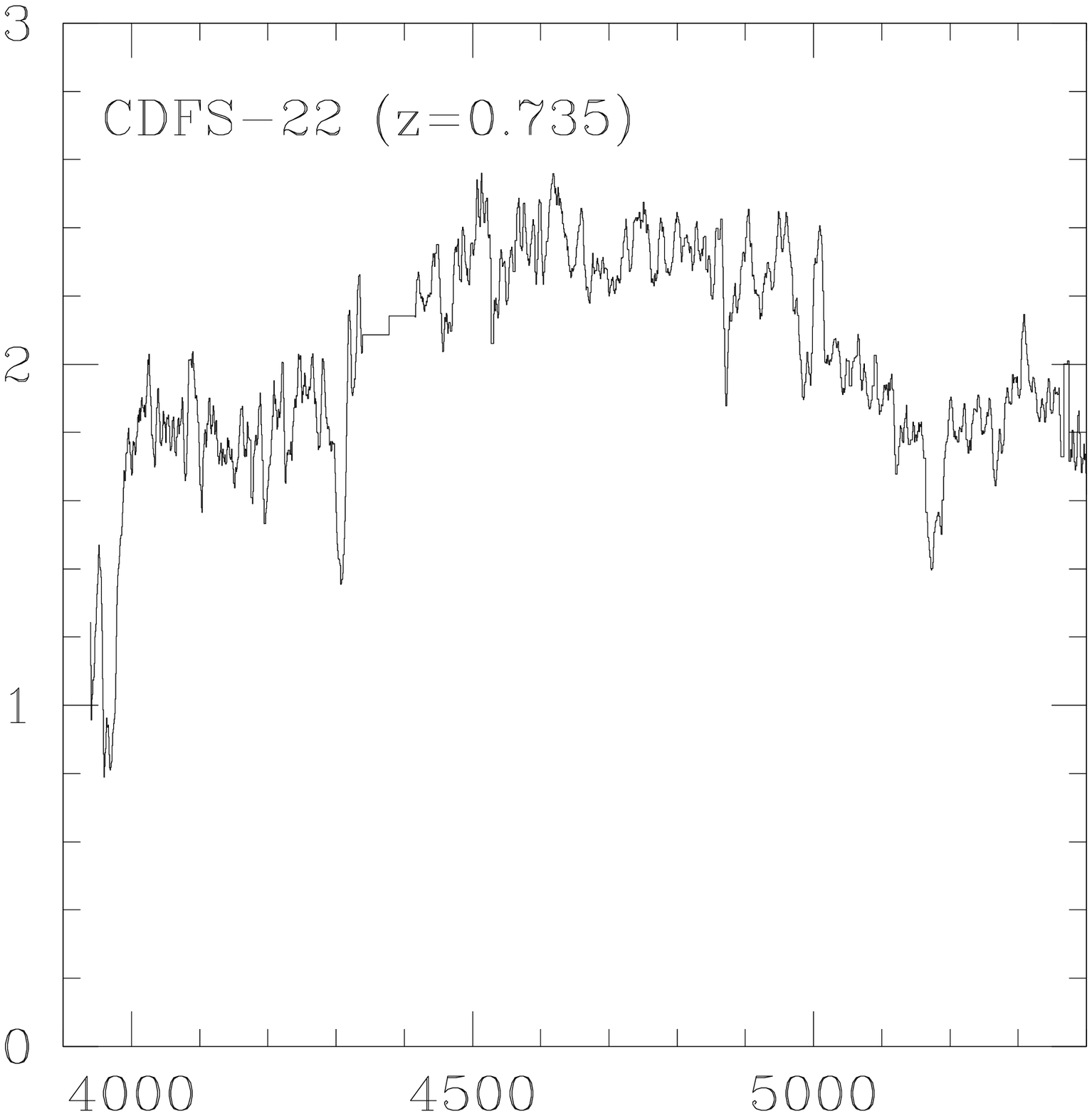}
\epsfxsize=5.8cm
\epsffile{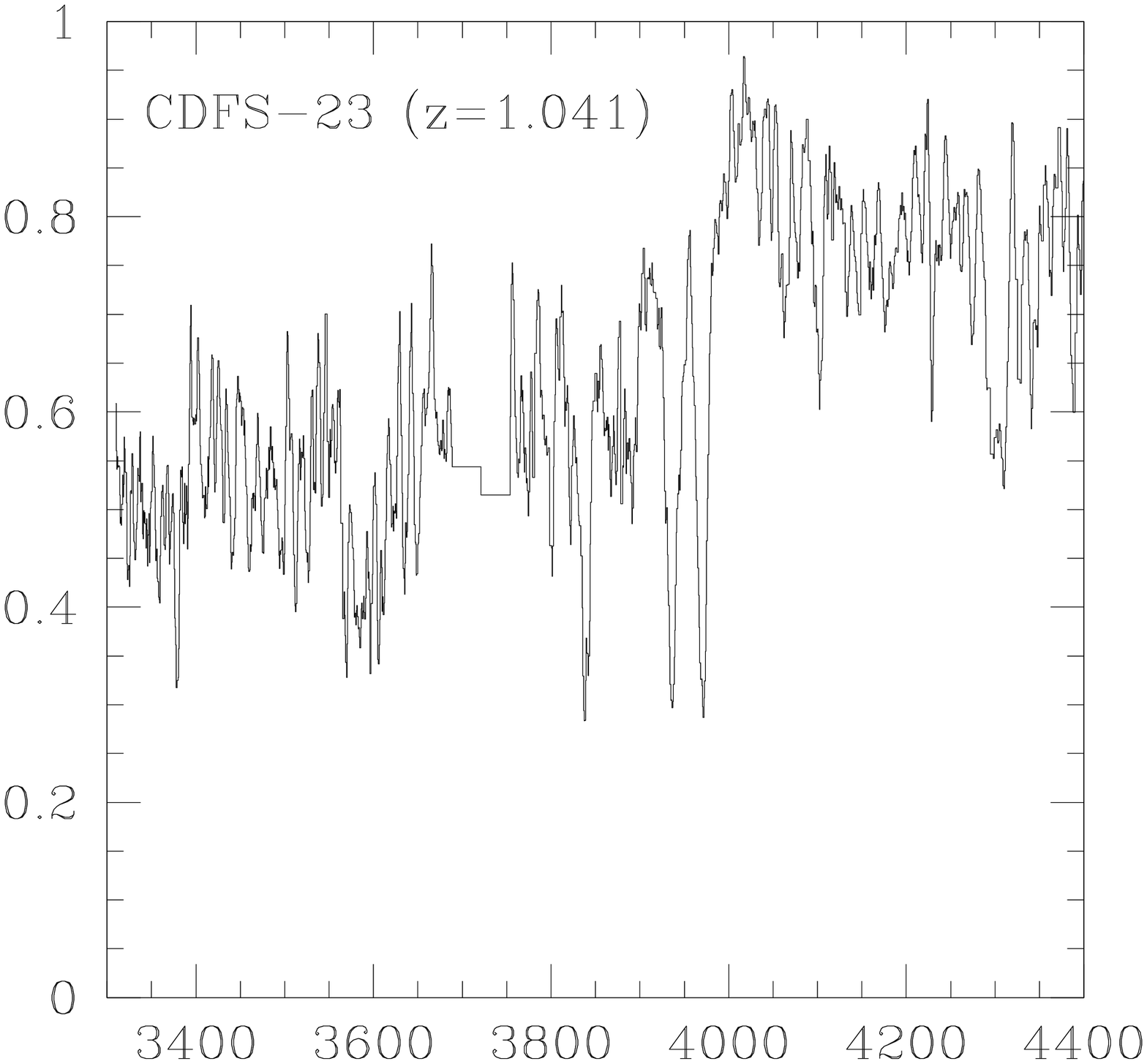}
\epsfxsize=5.8cm
\epsffile{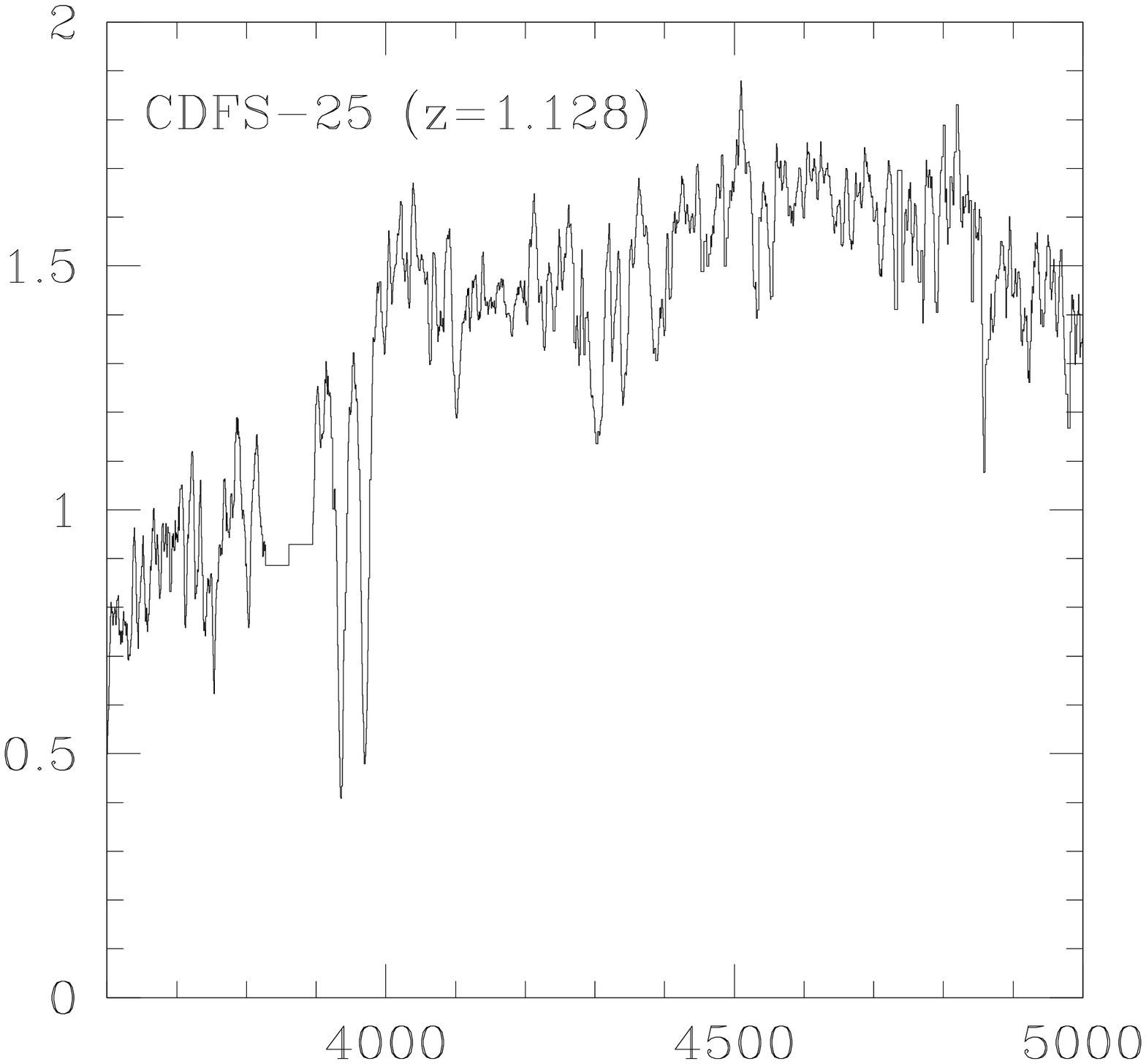}
}
\hbox{
\epsfxsize=5.8cm
\epsffile{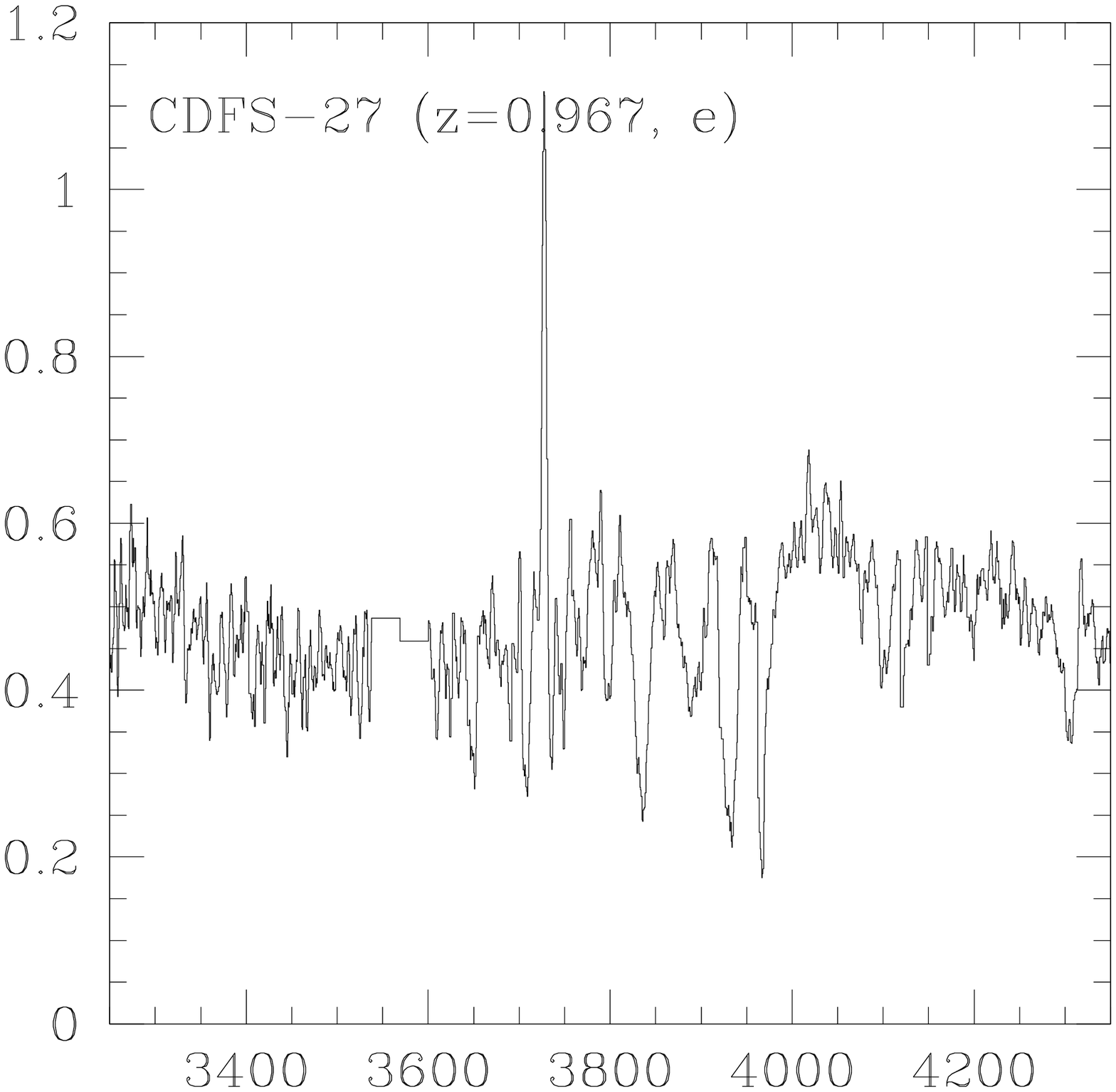}
\epsfxsize=5.8cm
\epsffile{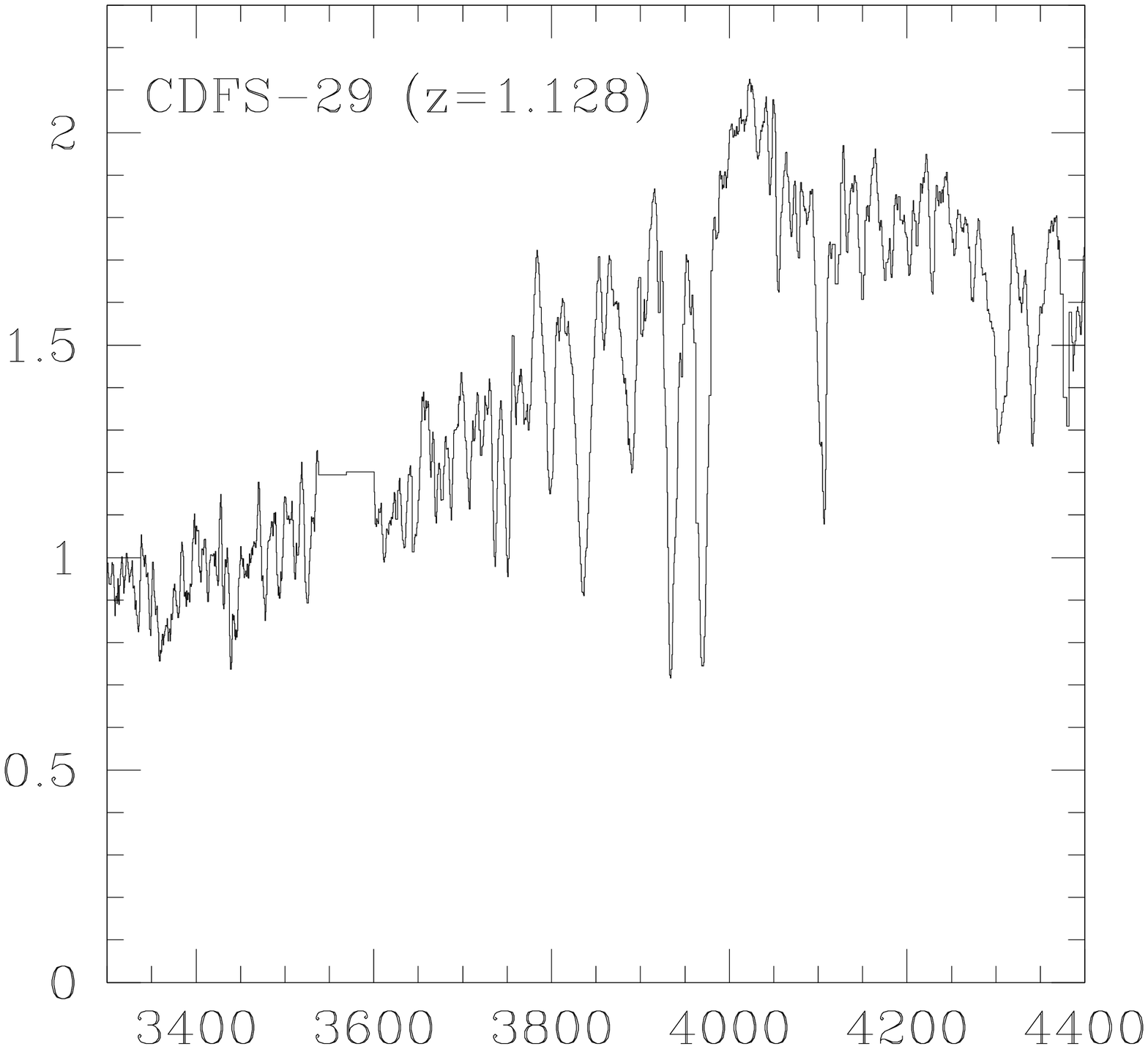}
}
\hbox{
\figcaption{\small Continued}
\label{fig:spec}}
\end{figure*}

\begin{figure*}[t]
\figurenum{2}
\begin{center}
\leavevmode
\hbox{
\epsfxsize=5.8cm
\epsffile{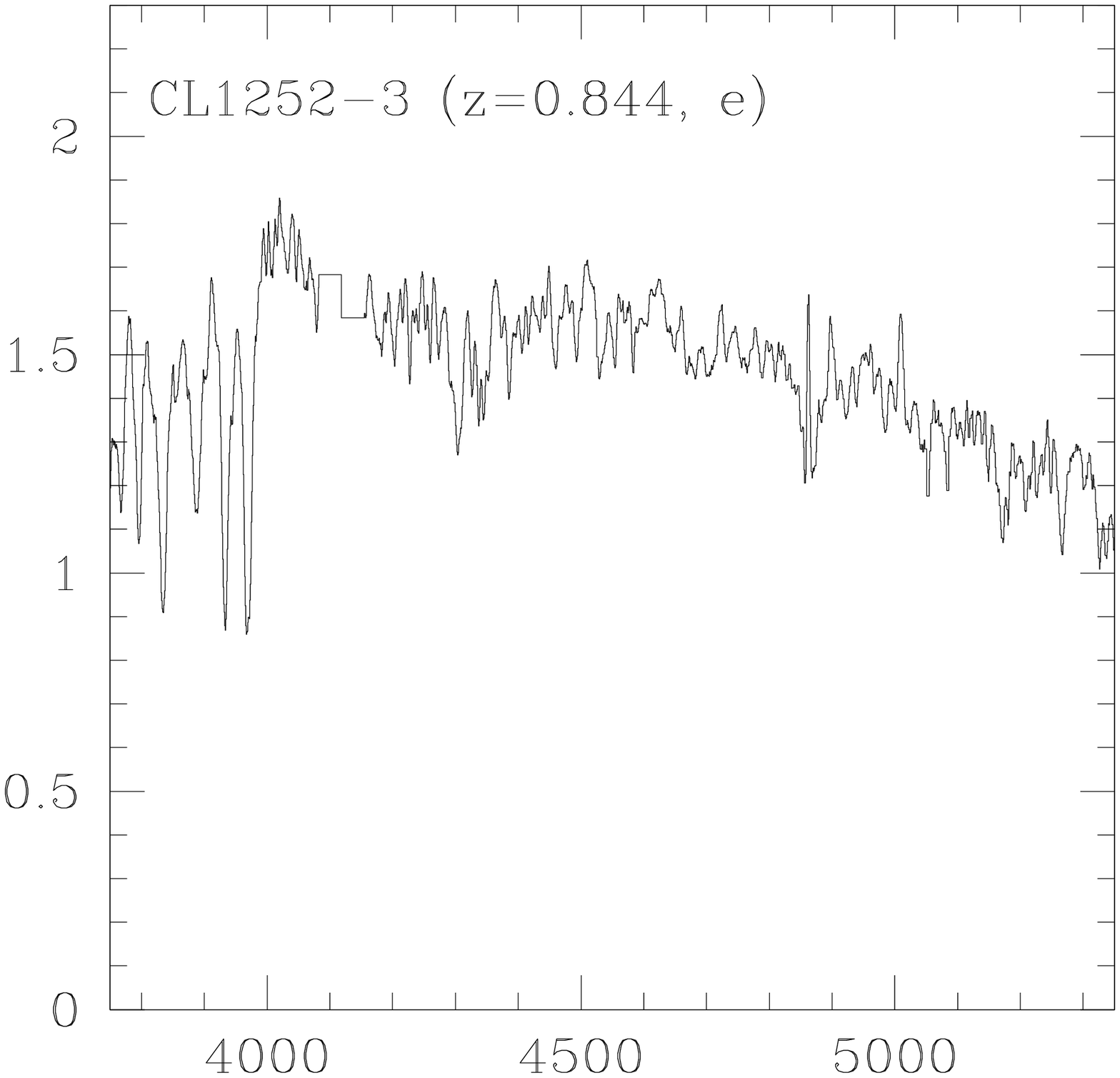}
\epsfxsize=5.8cm
\epsffile{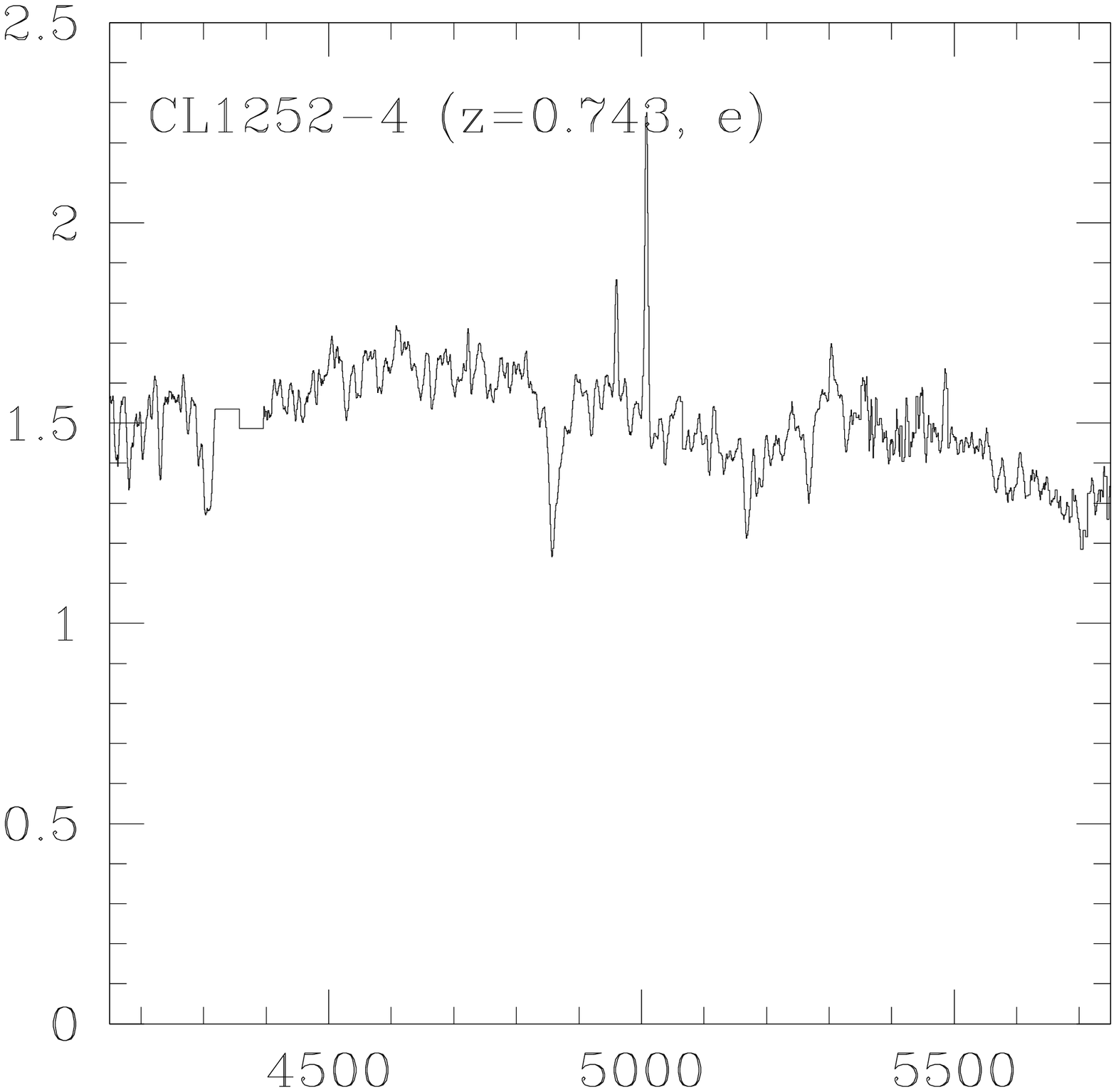}
\epsfxsize=5.8cm
\epsffile{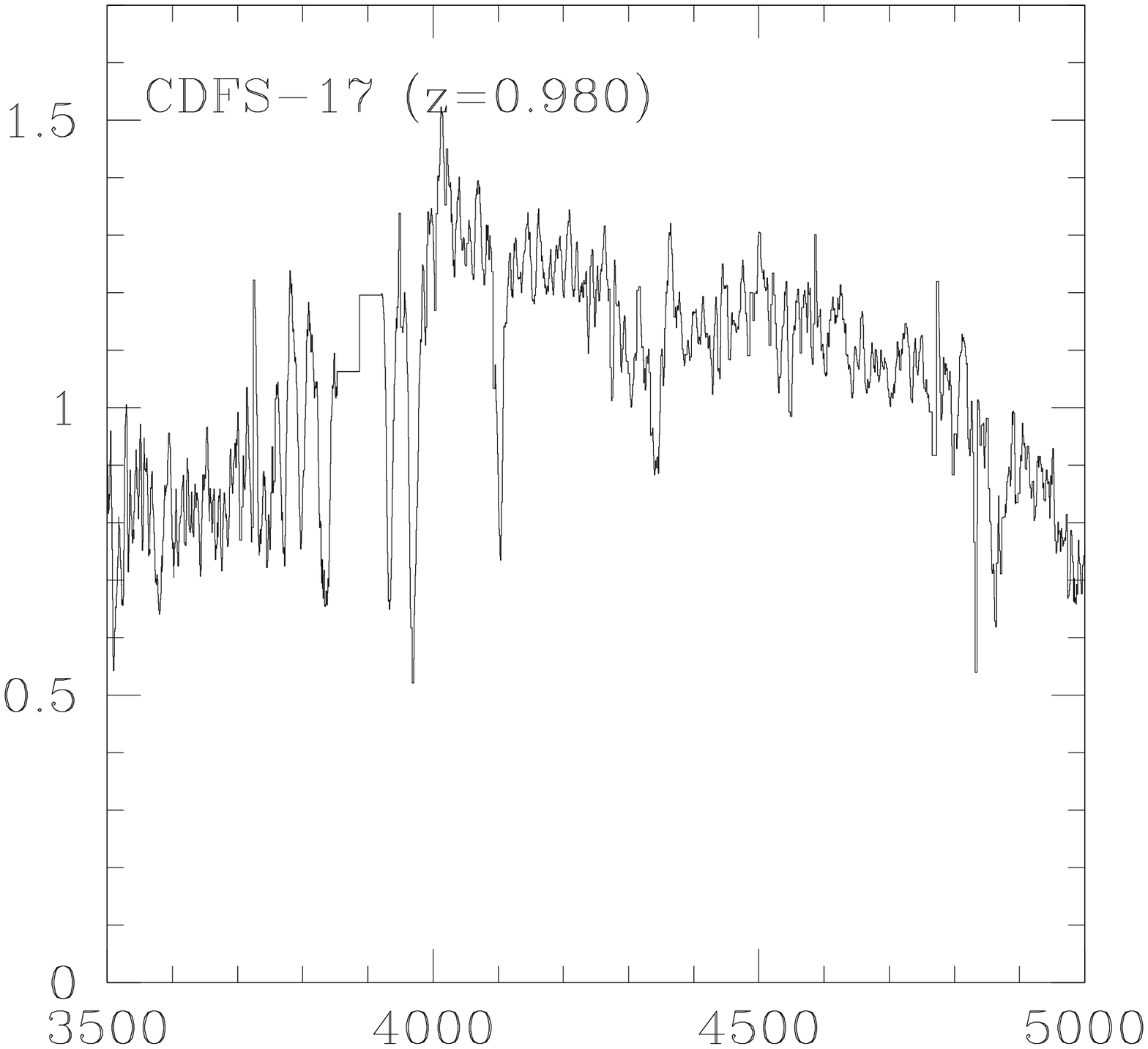}
}
\hbox{
\epsfxsize=5.8cm
\epsffile{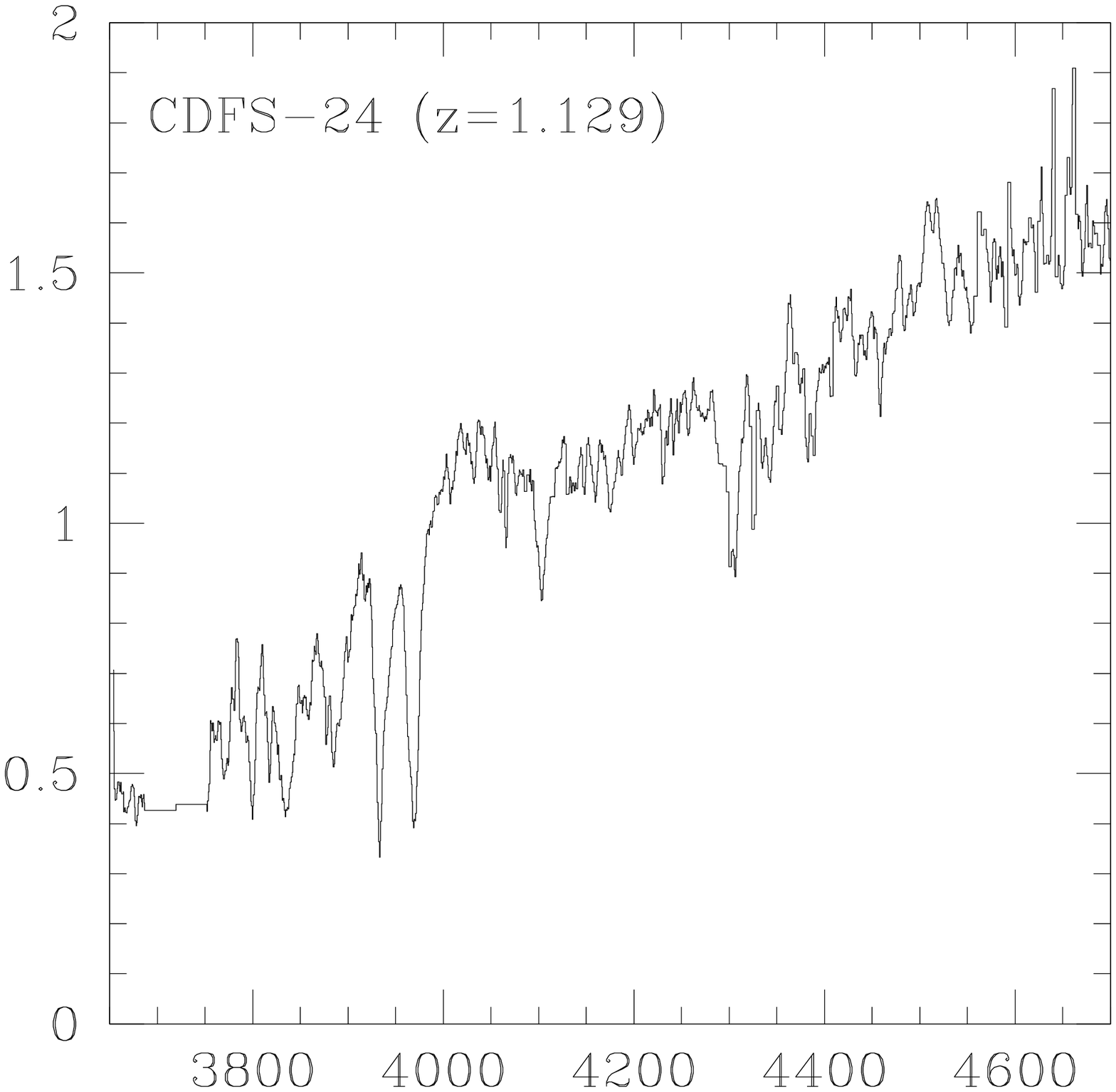}
\epsfxsize=5.8cm
\epsffile{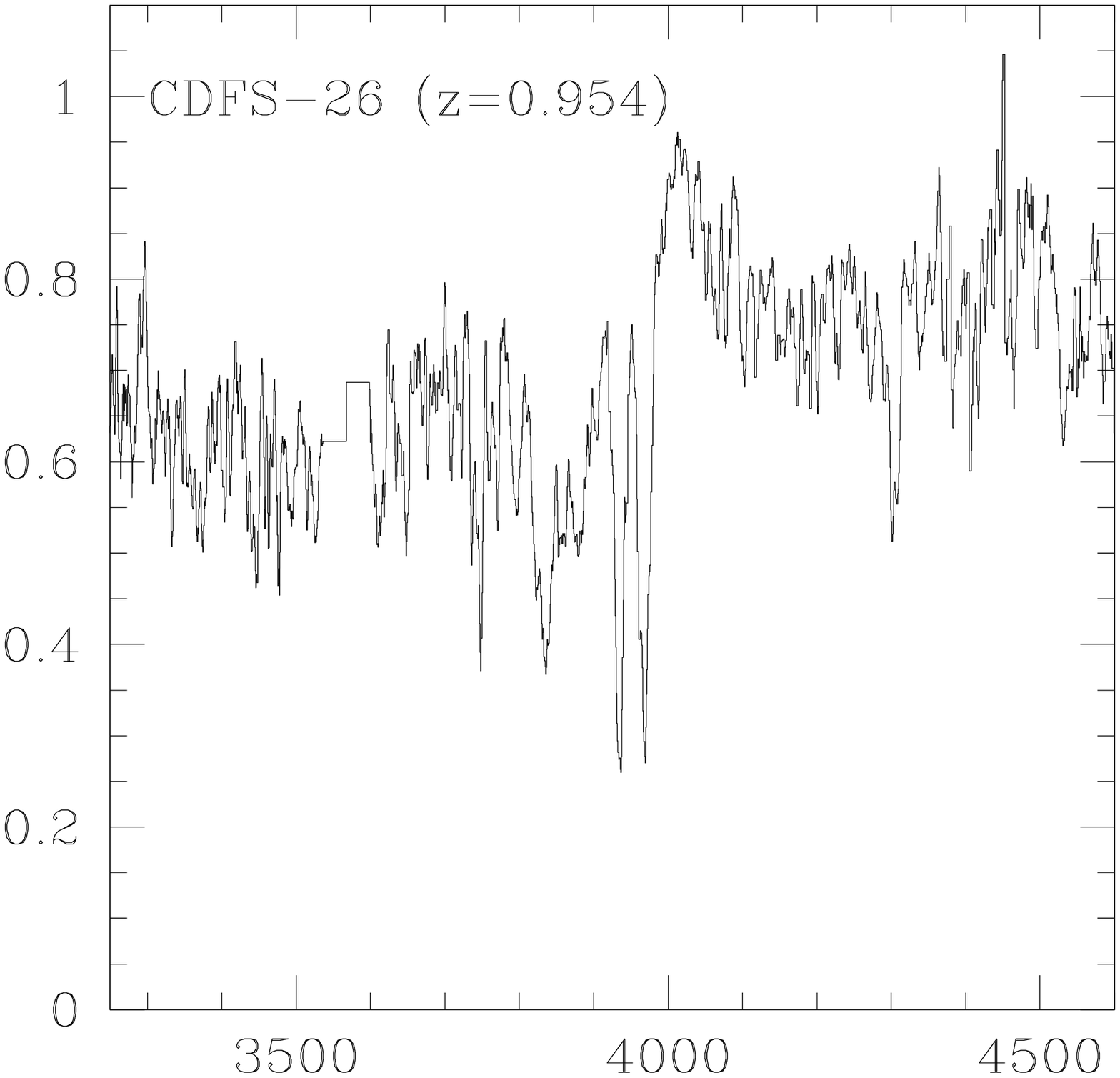}
\epsfxsize=5.8cm
\epsffile{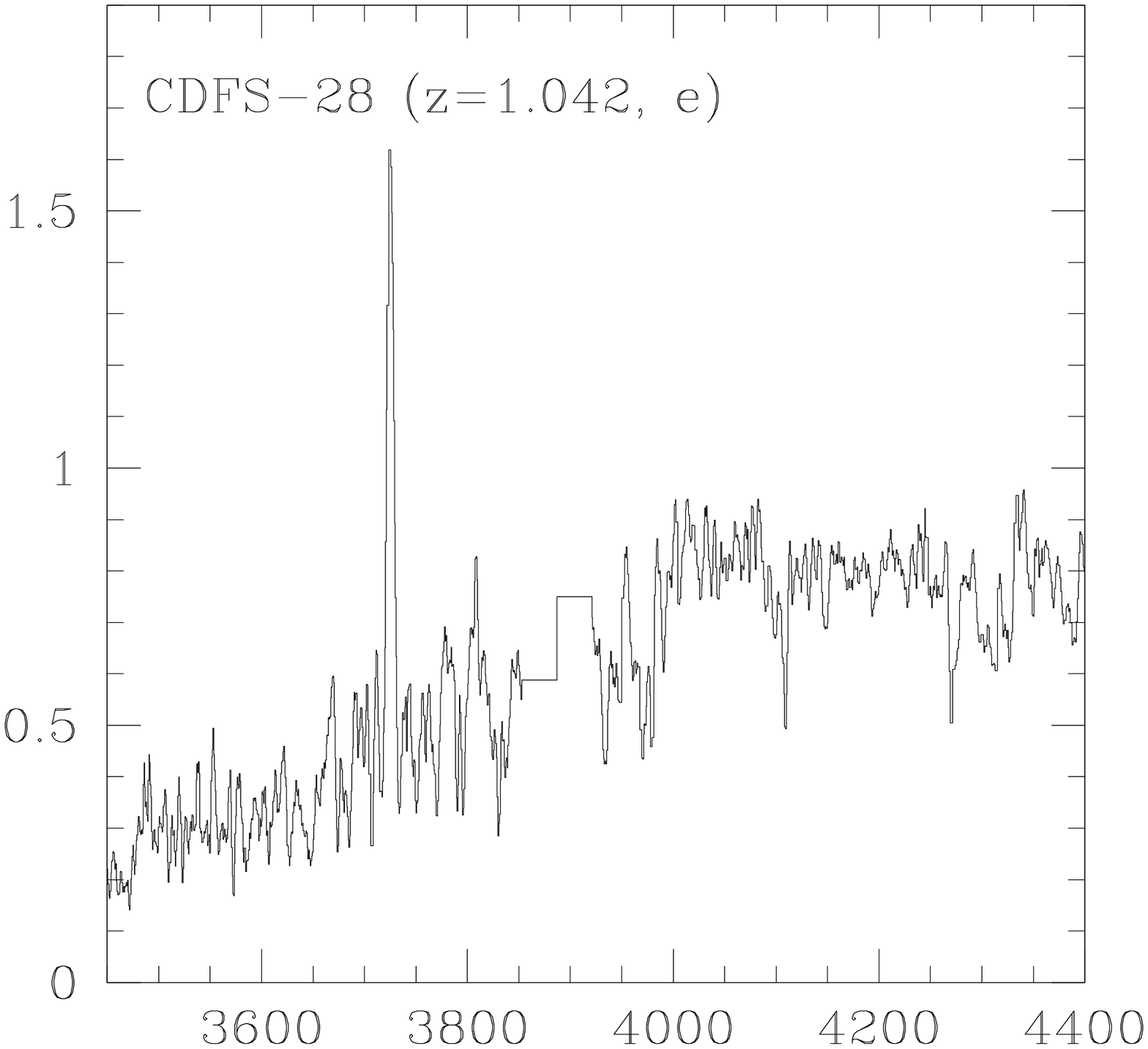}
}
\end{center}
\hbox{
\figcaption{\small Rest-frame spectra in 8\AA~bins of the galaxies in our sample with 
late-type and irregular morphologies. These are not included in the analysis 
in the subsequent sections. For further explanation of the spectra, see Figure
1.
}
\label{fig:spec}}
\end{figure*}

\section{Spectroscopy}
\subsection{Sample selection and Observations}\label{sec:obs}
We selected galaxies in the Chandra Deep Field-South (CDFS) and the 
RDCS1252.9-2927 cluster field (CL1252, Rosati et al. 2004),
which both have deep optical 
imaging from the Advanced Camera for Surveys (ACS) on the HST. 
GOODS\footnote{http://www.stsci.edu/science/goods/} provides
publicly available imaging in four filters
(Giavalisco et al. 2004):
F475W, F606W, F775W, and F850LP (hereafter $b$, $v$, $i$,
 and $z$). As these data were not yet available
when we started this project, 
we used ground-based COMBO-17 photometry (Wolf et al. 2004) to select our 
sample for the first observing run. 
For subsequent runs  
version 0.5 of the ACS GOODS data were available, and for the last run we used the 
version 1.0 data release.
(Blakeslee et al. 2003)
provide ACS imaging on the CL1252 field in the $i$ and $z$ bands.

In order to construct a sample of early-type galaxies at $z\sim 1$ in the CDFS, we 
selected objects with $i-z>0.86$ and COMBO-17 photometric redshifts
in the range
$0.8<z_{phot}<1.4$. 
(We use $z$ when we mean 'redshift', and $z_{mag}$ if we mean $z$-band magnitude, 
but when indicating a color we omit the $mag$ subscript for clearity.)
This color cut selects galaxies redder than a local Sbc galaxy 
at $z=1$.
Therefore, this study only includes galaxies that are on the red sequence
at $z\sim 1$.
We morphologically classified all galaxies satisfying these criteria and brighter than $z_{mag}=21.5$,
distinguishing between early- and late-type galaxies using the ACS imaging.
The classification was based on compactness, regularity and the presence of spiral arms.
26 out of the 52 galaxies satisfying our selection criteria we classified as early-type galaxies.
We designed multi-slit masks
for three different pointings, selected by the number of primary targets that could 
be included. 
Open spaces in the masks were filled with early-type galaxies satisfying the 
color and redshift criteria but fainter than $z_{mag}=21.5$, early-type galaxies 
with lower photometric redshifts and late-type galaxies with $i-z>0.85$ and redshifts
$0.8<z_{phot}<1.4$.
Switching from using ground-based $i-z$ colors from COMBO-17 to ACS $i-z$ colors from GOODS
did not lead to large differences between the selected samples,
although several objects changed priority.

The same selection criteria were used for the CL1252 field.
However,
the mask design for the CL1252 field was geometrically constrained 
because the 
primary targets were cluster galaxies at $z=1.24$.
The two brightest 
galaxies in the cluster are the two central galaxies, which had to 
be included in a single slit because of their small angular separation of 
$1\farcs 5$. Therefore, not only  the positions, but also the position angles 
of the designed masks were fixed. Unfortunately, only two galaxies brighter than $z_{mag}=21.5$,
redder than $i-z=0.85$ and early-type morphologies could be included, additionally 
to the cluster galaxies. Similarly to the CDFS masks, fillers were included.

We carried out the observations with FORS2 in MXU mode on ESO's Kueyen, 
one of the VLT unit telescopes. 
We used the 600z grism together with the OG590 order 
separation filter to obtain 
a sufficiently high spectral resolution ($\sigma\approx 80$~km~s$^{-1}$) and to 
cover the wavelength range around the Balmer/4000\AA~break for galaxies at $z\sim 1$.
The observations were carried out in series of four 
dithered exposures with spatial offsets of $1\farcs 5$ or $2''$ and equal 
exposure times ranging from 14 to 30 minutes each. 

In total, 51 hours of scientifically useful integration time was acquired, 
of which 38 hours had seeing better than $1''$. 
The cumulative integration time for the three pointings in the CDFS is 27 hours, with 
a median seeing of $0\farcs 95$. The single pointing in the CL1252 field 
has an integration time of 24 hours, with a median seeing of $0\farcs 65$.
These observations were carried out during five different observing runs 
from September 2002 to November 2003.

The sample described in this paper consists of 38 galaxies 
with velocity dispersions, of which 20 are early-type galaxies at 
$z\sim 1$ , and 18 are early-type galaxies at lower redshift, or late-type 
galaxies.  
100\% of our primary targets yielded velocity dispersions.
The CL1252 observations also yielded four velocity dispersions of cluster galaxies.
The FP of the 1252 cluster is discussed by 
(Holden et al. 2005).

\subsection{Data Reduction}
The spectroscopic data were reduced using standard IRAF tasks. 
Lamp flat fields were taken before or after each night, in sequences
of five exposures.
We used the sequence closest in time to the science observation.
Cosmic rays were removed using the L.A.Cosmic task 
(van Dokkum 2001).
Afterward, all frames were checked manually.
We subtracted a two-dimensional sky spectrum from each exposure, 
obtained by median averaging the four dithered exposures in a sequence, 
masking the target and secondary or serendipitous objects, if present.
The atmospheric emission lines, which are bright and abundant in the observed 
wavelength range, were used to perform the wavelength calibration. 
We corrected for distortion in the spatial direction by tracing the target. 
All individual exposures were optimally weighted to obtain maximum $S/N$.

There are various atmospheric absorption features in the observed wavelength
range. Because the strength and shape of these features 
change with airmass and atmospheric conditions we needed to correct 
each exposure separately. To this end we included a blue star in each of 
our masks, which was reduced along with the 
galaxy spectra. After the final combination, the regions in the galaxy spectra 
with atmospheric features were divided by the normalized spectrum of the blue 
star.
Spectroscopic standard stars were used 
to do a relative flux calibration. One-dimensional spectra were extracted by 
adding those pixel rows with more than 25\% of the flux of the brightest row, 
weighting optimally.

The smoothed one-dimensional spectra are shown in Figure \ref{fig:spec}. 
The coordinates of the objects for which we measured 
velocity dispersions (see Section \ref{sig}) in Table 1. Redshifts, $S/N$ and emission
lines are given in Table 2.

\subsection{Velocity Dispersions}\label{sig}
Velocity dispersions are obtained by fitting template spectra to the observed 
galaxy spectra. 
The fitting method is extensively described by (van Dokkum \& Franx 1996).
The continua of both the observed and the template spectra are filtered 
out in Fourier space and the template spectrum is convolved with a Gaussian to 
match the width 
of the absorption lines in the galaxy spectrum. The part of the galaxy spectrum 
used in the fit is as large as possible. Therefore, our measurements do not 
rely on a few high $S/N$ absorption features.

As templates we use Coud\'e spectra of 132 stars with the appropriate wavelength range 
from the sample constructed by (Valdes et al. 2004),
with a 
spectral range from F0 to M6, including both very low and high 
metallicity stars, and different luminosity classes. These spectra have a FWHM 
resolution of about 1\AA. Each stellar spectrum needs to be smoothed to each 
galaxy spectrum separately before being re-binned.
A second order function is fitted to the width of atmospheric emission lines 
as a function of wavelength to obtain the spectral resolution to which the 
template spectra are smoothed.

When fitting the galaxy spectra, we weight with the inverse of the sky 
brightness, and we mask the region around the atmospheric A band at 7600\AA.
The spectrum above 9300\AA~is omitted because of the strong atmospheric absorption, 
the ever increasing brightness of the sky emission lines and the decreasing 
system throughput. 

After performing the fit for a small number of 
templates, masking and weighting as described, 
we check the residuals from the fit. Regions with emission lines, large 
sky line residuals and remaining data artifacts such as cosmic ray remnants are masked 
if present. 
We then apply fit the galaxy spectrum with all template spectra.
We check whether the
obtained parameters change strongly if one or two strong features are masked out, 
but we conclude that this generally is not the case: 
excluding the strongest features from the 
fit increases the $\chi ^2$-value but does not change the results 
significantly in most cases. For some spectra, however, including Balmer lines 
in the fit leads to different results, probably because unseen emission line
contributions contaminate these features.
For low quality spectra ($S/N \le 10$ per 1.6$\AA~$ pixel in the extracted, 
one-dimensional spectra) the contributions of Balmer 
lines or other strong features can hardly be checked because 
excluding these strong features leaves 
insufficient signal to obtain a proper fit. Therefore, we exclude objects with 
$S/N<12$ spectra from our analysis, but we mention the effect of including 
these.

For all spectra with $S/N\ge 12$ the random errors are below 3\% 
for $\sigma>200$~km~s$^{-1}$, 
and below 5\% for $\sigma<200$~km~s$^{-1}$. Adding 
a systematic uncertainty (including template mismatch and the error on the 
resolution of the galaxy spectra) of about 10\% for 
$S/N=10$ spectra and 2\% for the highest $S/N$ spectra the total errors range 
from 3\% to 17\% with a median of 7.5\% for our sample of early-type galaxies 
with $S/N\ge 12$. Some galaxies have measured velocity dispersions that are
not much larger than the resolution of the spectra. Although these are included
in the analysis, they play no important role in the derivation of our results.

The best fitting stellar spectral type and the measured velocity dispersion  
are listed in Table 1. These
velocity dispersions are aperture corrected to a $3\farcs 4$ diameter circular 
aperture at the distance of Coma as described by 
(J\o rgenson, Franx, \& Kj\ae rgaard 1995).
This correction ranges from $5\%$ to $7\%$.

\section{Photometry}
\subsection{Profile Fitting and Morphologies}\label{sec:profiles}
The ACS provides us with an unprecedented combination of deep and high 
resolution imaging. The spatial resolution (FWHM) at $z=1$ is $0.8$kpc, 
allowing us to accurately measure the effective radii of early-type galaxies at
this redshift, which typically are a few kpc. We use the single, unstacked,  
flat-fielded frames publicly available through the HST MAST archive. 
For the CDFS the number of frames for different positions 
ranges from 8 to 24 (with $530$s exposure time each).
For the CL1252 
field the number of frames ranges from 10 to 
40 (with $1200$s exposure time each), 
but is mostly 10 as only the center of the cluster has 40 overlapping 
images.

For each galaxy each individual $z$ band image is fitted by 
$r^{1/n}$-models (with $n=1,2,3,4$) 
convolved by a position dependent PSF created with TinyTim (Krist 1995), 
measuring $r_{eff}$ (the effective radius), $\mu_{eff}$
(the surface brightness at $r_{eff}$), the position angle and the 
ellipticity.  Each individually derived set of 
model parameters is distortion corrected by calculating the pixel scales in 
the x- and y-directions, using the polynomial distortion coefficients available 
through the WWW\footnote{http://www.stsci.edu/hst/acs/analysis/PAMS}.
We then average the results and compute the measurement error from the scatter.
The error is generally about 6\% in $r_{eff}$, but the combination of
the uncertainty in $r_{eff}$ and $\mu_{eff}$ is such that it is directed almost
parallel
to the local FP. The error relevant to the offset from the FP is typically 2\%.
 Thus, uncertainties in the offset from the local FP are 
dominated by the uncertainty in $\sigma$, of which the error is pointed almost 
perpendicular to the FP.
The effective radii and surface brightnesses are given in Table 1.
For consistency with earlier studies, these are the values obtained from
fitting a de Vaucouleur profile in our analysis.

To transform the observed $z$ band surface brightnesses to the rest-frame $B$
band, we use the technique described by (van Dokkum \& Franx 1996),
using the templates from (Coleman, Wu, \& Weedman 1980)
and observed colors (see Section \ref{sec:color}) to 
interpolate between the pass-bands. The calculated rest-frame
$B$ band surface brightnesses only depend very weakly on the spectral type of the template used. 
The 
typical difference found for using the Sbc template instead of the E template 
is less than 0.02 mag.
The transformations are a function of redshift.
As an example we give the transformation 
(based on the E template) for a galaxy at $z=1$:

\begin{equation}B_z=z+0.165(i-z)+1.398\end{equation}

Physical sizes and rest-frame $B$-band surface brightnesses are given in Table 3.

Figures \ref{fig:mos} and \ref{fig:moslate} 
show the combined residuals of the $r^{1/4}$-fits along with 
the color images of the 38 galaxies with velocity dispersions.
Figure \ref{fig:mos} shows the early-type galaxies, Figure \ref{fig:moslate} 
the late-type galaxies.
Two numbers are used to characterize the magnitude of the residuals. 
At the upper right of each residual image the absolute 
value of the flux in the asymmetric part of the residual
is given as a percentage of the total flux of the galaxy, 
obtained by subtracting the 
residual rotated by 180 degrees from the residual itself.
In the upper left the absolute value of the flux in the symmetric part of the 
residual is given as a percentage of the total flux of the galaxy. 
We find that the asymmetric residual is a good indicator of morphology. 
Our final morphological qualification is a combination of the magnitude of 
the asymmetric residual and visual inspection of the cause of the asymmetry.
Some of the early-type galaxies (Figure \ref{fig:mos}) 
have significant asymmetric residuals, but these are caused by features on a 
very small scale, in the centers of the galaxies, i.e. not by features attributed 
to spiral arms or other large scale irregularities.
The late-type galaxies (Figure \ref{fig:moslate}) have large 
asymmetric residuals, caused by large scale structures like spiral arms.
We note that the Sercic number does not distinguish well between late- and 
early-type galaxies. For example, the most massive galaxy in our sample has 
n=2 and some galaxies that we classify as late-type galaxies have n=4.

\subsection{Colors}\label{sec:color}
We supplement the ACS imaging with ground-based optical and near-IR imaging from 
FORS2 and ISAAC on the VLT and SOFI on the NTT (Vandame et al., in preparation).
GOODS ACS imaging of the CDFS provides photometry in the $b$, $v$, $i$, and $z$
bands (data release version 1.0),
and ESO's imaging survey\footnote{http://www.eso.org/science/eis/} provides 
SOFI and ISAAC imaging in the $J$ and $K$ bands.
Since the CDFS is not entirely 
covered by ISAAC we use the SOFI data for the objects outside the ISAAC pointings. 
All images were smoothed to match the resolution of the $K$ band data with the 
worst seeing, which is $0\farcs 8$ for the ISAAC imaging and 
$1\arcsec$ for the SOFI imaging. 
The photometric differences between the ISAAC and SOFI datasets are small 
($< 0.01$ mag), since the zero-points of the ISAAC data are based on SOFI photometry.
For the CL1252 field we use optical imaging from ACS ($i$ and $z$) and FORS2
($B$, $V$, and $R$) and near-IR imaging from ISAAC 
(Lidman et al. 2004).
Again, all images were smoothed to match the ground-based data with the worst 
seeing, which is $0\farcs 6$ in the $B$ band.

We measure the flux in each band for each of our spectroscopic targets within 
several apertures with different radii ($1/2/3r_{eff}$ and $0\farcs 8$). 
Contaminating objects within the aperture are masked. We choose to use the 
fluxes measured within a radius of $2r_{eff}$, with a minimum of $0\farcs 8$,
trading off between the 
measurement accuracy and the amount of contamination. 
The $i-z$ and $J-K$ 
colors are given in Table 2, as well as total \textit{AUTO} $z$ band and $K$ band 
magnitudes, as obtained with SExtractor (Bertin \& Arnouts 1996).

The availability of near-IR photometry not only allows us to compute rest-frame
$U-B$ colors, but also rest-frame $B-I$ colors. 
To transform the observed colors to rest-frame $U-B$ and $B-I$ we use the same method as 
used to calculate rest-frame $B$ band surface brightnesses 
(see Section \ref{sec:color}).
For $z=1$, using the E template, the transformation is:

\begin{equation}(U-B)_z=0.836(i-z)+0.276(v-i)-1.111\end{equation}
\begin{equation}(B-I)_z=z-K+0.165(i-z)-0.513(J-K)-0.282\end{equation}

Rest-frame colors are given in Table 3.

\begin{figure*}
\figurenum{3}
\epsfxsize=18cm
Images at www.strw.leidenuniv.nl/\~~vdwel/private/FPpaper/
%\epsffile{f3a.ps}
\epsfxsize=18cm
%\epsffile{f3b.ps}
\epsfxsize=18cm
%\epsffile{f3c.ps}
\figcaption{\small 
Color images and $r^{1/4}$-profile fit residuals of the early-type 
galaxies used in the analysis in the subsequent sections.
The color images of the objects in the CL1252 field consist of $i$ and $z$ 
band images, the color images of the objects in the CDFS 
consist of $v$, $i$ and $z$ band images. The residuals are shown in the $z$ 
band for all objects. The boxes are $5\farcs 4$ on a side. 
The residual images 
are not distortion corrected, which causes the small dissimilarities between 
the color and residual images. The numbers at the upper left and right of 
the residual images are, respectively, the symmetric and asymmetric fluxes 
in the absolute residuals within two effective radii, expressed as percentages 
of the total fluxes of the galaxies within the same radius.
}
\label{fig:mos}
\end{figure*}

\begin{figure*}
\figurenum{3}
\epsfxsize=18cm
%\epsffile{f3d.ps}
\epsfxsize=18cm
%\epsffile{f3e.ps}
\begin{center}
\epsfxsize=6cm
%\epsffile{f3f.ps}
\end{center}
\begin{center}
\epsfxsize=6cm
%\epsffile{f3g.ps}
\end{center}
\figcaption{\small Continued}
\label{fig:mos}
\end{figure*}

\begin{figure*}
\figurenum{4}
\epsfxsize=18cm
Images at www.strw.leidenuniv.nl/\~~vdwel/private/FPpaper/
%\epsffile{f4.ps}
\figcaption{\small 
Color images and $r^{1/4}$-profile fit residuals of the galaxies with 
late-type or irregular morphologies. These galaxies have measured velocity 
dispersions but are not included in the analysis in the subsequent sections.
For an explanation of the images and the numbers, see Figure 3.
}
\label{fig:moslate}
\end{figure*}

\subsection{X-ray Data}
For both the CDFS and the CL1252 field deep X-ray data from Chandra are available, 
such that we can check for the presence of AGN in our galaxy sample.
Giacconi et al. (2002) and Alexander et al. (2003)
provide catalogs of the CDFS data. The Chandra data of the CL1252
field is described by (Rosati et al. 2004), who also constructed a point source catalog.
Eight galaxies in our sample of 38 are identified as AGN, based on their large X-ray 
luminosities (typically $> 10^{42}$erg s$^{-1}$). Five of these are early- 
type galaxies, of which two have emission lines in their spectra.
Besides the eight AGN, two galaxies in our sample are identified as extended X-ray 
sources. This X-ray radiation is accounted for by
diffuse halo gas. The X-ray luminosities of CDFS-4 and CDFS-22 are 
$7.15\times 10^{41}$~erg~s$^{-1}$ and $3.42\times 10^{42}$~erg~s$^{-1}$, respectively.
It is not surprising that CDFS-4 and CDFS-22 turn out to be two of the most massive 
galaxies in our sample.
Also, CDFS-22 is one of the galaxies with an AGN. 
The X-ray properties of our sample of galaxies are given in Table 2.

\section{Masses, Mass-to-Light Ratios and Stellar Populations of Early-Type Galaxies at $z=1$}

\subsection{The Fundamental Plane}
J\o rgensen, Franx, \& Kj\ae rgaard (1996)
have shown that the FP for cluster galaxies in the local universe 
can be described by
\begin{equation}\log{R_{eff}}=1.2\log{\sigma}-0.83\log{I_{eff,B}}+\gamma ,
\end{equation}\label{eq:3}
where $R_{eff}$ is the effective radius in kpc, $\sigma$ the central 
velocity dispersion in km~s$^{-1}$, $I_{eff,B}$ the surface 
brightness in the $B$ band ($\log{I_{eff,B}}=-0.4\mu_{B_z}$), 
and $\gamma$ the intercept.
The values of the coefficients are derived from the early-type galaxies 
in ten nearby clusters, 
and imply that mass and $M/L$ scale as $M/L_B\propto M^{0.28}$.
From the sample of Faber et al. (1989) we derived that the intercept of the FP
lies lower by 0.04 in the projection given above.
The offsets of the high-$z$ galaxies are computed using a local FP 
with coefficients from J\o rgensen et al. (1996) but with the intercept derived from the 
Faber et al. (1989) sample.

Besides the form of the local FP, we need a large sample of local
field early-type galaxies, for example to address issues such as selection 
effects.
We construct our local field sample as in Bernardi et al. (2003):
we take early-type galaxies from the Sloan Digital Sky
Survey and include those with less than 10 
neighbors brighter than $M_i=-20.55$ and closer than $1.4Mpc$.
We convert the rest-frame $g$ and $r$ band surface brightnesses to a 
$B$ band surface brightness using the conversion given by J\o rgensen et al. (1996).
As we need the surface brightness 
in the $g$ and $r$ bands at the same radius,
and Bernardi et al. derive effective radii separately for each band,
we compute the surface brightness in the $r$ band at the effective radius 
as measured in the $g$ band.
We note that the FP coefficients as derived by Bernardi et al. are different 
from those from J\o rgensen et al., but this does not lead to different results.

In Figure \ref{fig:FPe} 
we show the FP of the SDSS galaxies and our sample, where the surface 
brightnesses of all galaxies are transformed to the value they would have 
at $z=1$, assuming luminosity evolution found for massive cluster galaxies,
$\Delta \ln {(M/L_B)} = -1.12z$.
This value for the $M/L$ evolution of massive cluster galaxies
is derived from compiling all existing data in the literature for galaxies more massive
than $M=2\times 10^{11}M_{\odot}$
(van Dokkum \& Franx 1996; Kelson et al. 2000, van Dokkum \& Stanford 2003;
Wuyts et al. 2004, Holden et al. 2005).
Our field sample shown in Figure \ref{fig:FPe} 
includes all early-type galaxies with spectra with $S/N\ge 12$.
This is also the sample used in the analysis throughout the rest of the paper, and in the subsequent figures.
As can be seen, the FP already existed at $z=1$ for a large range in size.
At low masses outliers occur, but the interpretation is not straightforward, as 
selection effects play a major role in this regime (see Section 4.3).

\null
\vbox{
\begin{center}
\figurenum{5}
\leavevmode
\hbox{%
\epsfxsize=9cm
\epsffile{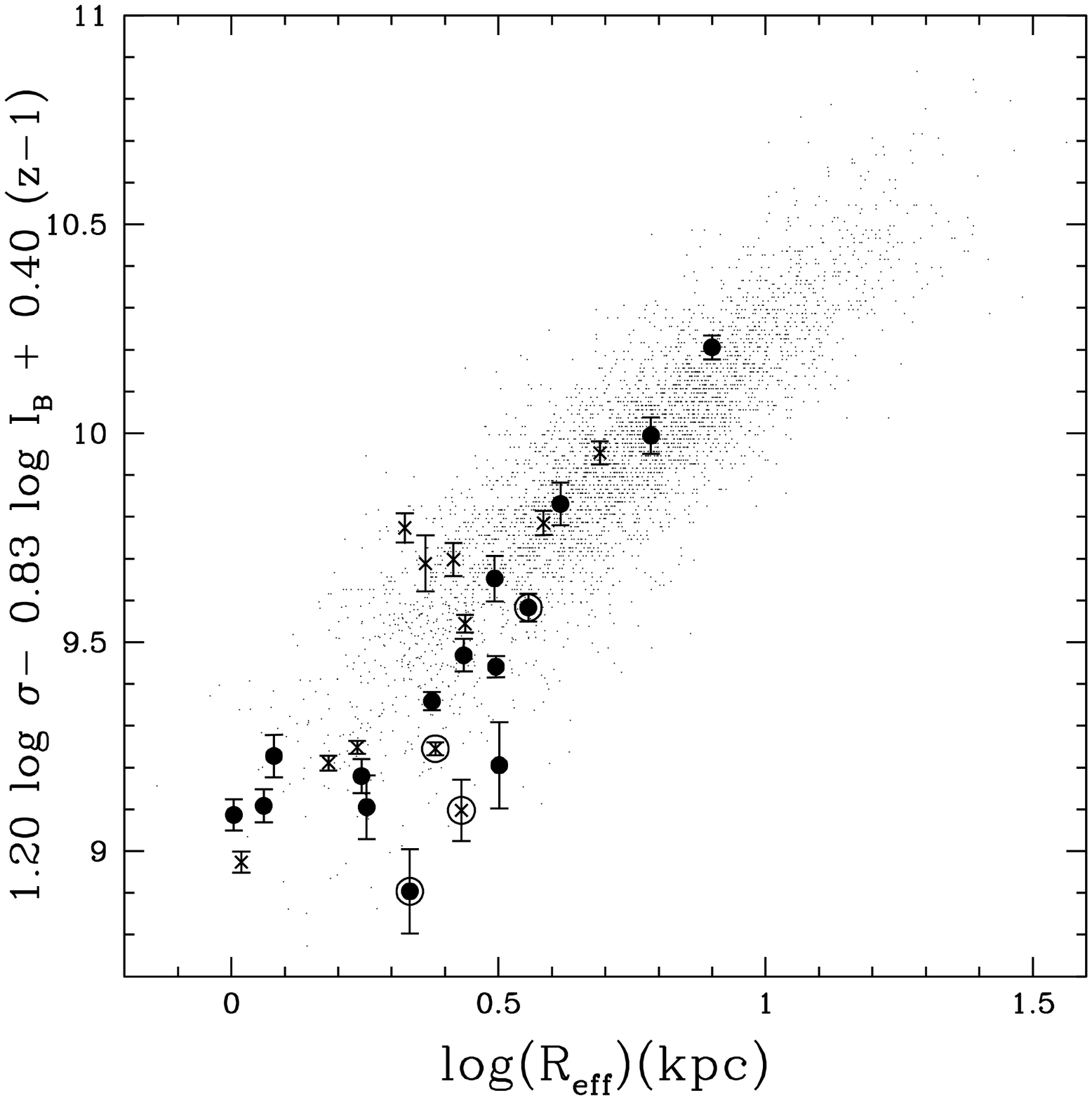}}
\figcaption{
\small 
The Fundamental Plane of our sample of early-type galaxies compared to the  
field early-type galaxy sample from SDSS (small dots, Bernardi et al. 2003). 
The primary sample, early-type galaxies at $z\sim 1$ that satisfy all of our 
selection criteria, are indicated by the filled symbols.
The crosses are fillers, mainly galaxies at redshifts $z\sim 0.7$.
Encircled objects have one or more emission lines in their spectra.
The surface brightness, $I_B$, of every galaxy in this figure is corrected for evolution 
to a value it would have at $z=1$, assuming $\Delta \ln({M/L_B}) =-1.12z$,
which is the evolution of massive cluster galaxies.
The FP already existed at $z=1$ for a large range in size.}
\label{fig:FPe}
\end{center}}

\subsection{Evolution of $M/L$ with redshift}
The offset of high redshift galaxies from the local FP is interpreted as a 
difference in $M/L$ as 
compared to equally massive local galaxies (see, e.g., vam Dokkum \& Franx 1996).
Figure \ref{fig:z_dML} shows the offsets of our field galaxy sample in $\Delta \ln {(M/L_B)}$
as a function of redshift. 
The values are listed in Table 3.
Cluster samples from the literature are also shown.
The galaxies in our sample seem to
evolve faster than the galaxies in the cluster samples, and
the scatter in $\Delta \ln {(M/L_B)}$ is large.
Before we interpret the scatter and 
the apparent difference between field and cluster galaxies,
we need to 
investigate the origin
of the scatter. In Figure \ref{fig:BI_ML} we 
show $M/L_B$ as a function of the rest-frame $B-I$ color.
Galaxies with low
$M/L$ are bluer than galaxies with high $M/L$, as expected from 
the stellar population models shown in the figure.
In Figure \ref{fig:UB_ML} we show a similar 
relation between rest-frame $U-B$ and $M/L_B$, but in this case the correlation
is less clear, due to the fact that the range of $U-B$ colors is much smaller
than the range of $B-I$ colors, and probably also because $U-B$ is more sensitive to 
small variations in the star formation history.
Considering the correlation between color and $M/L$ and the fact that 
$M/L$ evolves with redshift, one expects that color evolves with redshift as well. In Figure 
\ref{fig:z_BI} we show $B-I$ as a function of 
redshift; note the strong similarities between Figures \ref{fig:z_dML} 
and \ref{fig:z_BI}. 
The strong correlation between color and $M/L$, and the similarity in $M/L$ evolution
and color evolution confirm that the observed evolution and scatter of $M/L$ are 
intrinsic, and not due to measurement errors.

We calculate the evolution of $M/L$ of our field sample by performing a linear 
fit and 
minimizing the mean deviation, weighting by the inverse of the error, and 
forcing the fit to go through the $z=0.02$ data point derived from 
the Faber et al. (1989) sample.
We separately consider the evolution of the primary 
sample, which contains the galaxies satisfying all of our selection criteria.
We find that the average 
evolution of our entire early-type galaxy sample is 
$\Delta \ln {(M/L_B)} = (-1.75 \pm 0.16)z$ ($(-1.72 \pm 0.15)z$ for the 
primary sample alone),
which is significantly faster than the evolution found for cluster galaxies,
which is 
$\Delta \ln {(M/L_B)} = (-1.28 \pm 0.08)z$
(van Dokkum \& Franx 1996; Kelson et al. 2000, van Dokkum \& Stanford 2003;
Wuyts et al. 2004, Holden et al. 2005).
The scatter in $\Delta \ln {(M/L_B)}$ is $0.58$ for our field galaxy sample
(0.54 for the primary sample), 
and $0.28$ for the MS1054 cluster sample (Wuyts et al. 2004).

Figure \ref{fig:FPe} suggests that the $M/L$ evolution may depend on galaxy mass, as 
galaxies with small $r_{eff}$ and low $\sigma$ tend to lie lower with respect to the 
local FP as compared to galaxies with large $r_{eff}$ and high $\sigma$.
This was also found for cluster galaxies by Wuyts et al. (2004).
We estimate $M$ and $M/L$ in solar units as described by van Dokkum \& Stanford (2003). 
The values are listed in Table 3.
We explore the mass dependence by color coding the galaxies in Figure \ref{fig:z_dML} according
to their masses. Red points are galaxies with masses larger than $M=2\times 10^{11}M_{\odot}$;
blue points are galaxies that have masses lower than $M=2\times 10^{11}M_{\odot}$.
There is a striking difference between low and high mass galaxies.
For galaxies with masses $M>2\times 10^{11}M_{\odot}$ we measure
$\Delta \ln {(M/L_B)} = (-1.20 \pm 0.18)z$ for our field sample and 
$(-1.12 \pm 0.06)z$ for the cluster samples. For the massive
galaxies in the primary sample alone we find $\Delta \ln {(M/L_B)} = (-1.26 \pm 0.18)z$.
The observed scatter is decreased to $0.34$ for the field sample (0.32 for the primary sample), 
and to $0.28$ for the cluster samples.
When changing this mass cut to $3\times 10^{11}M_{\odot}$, as 
is done by Wuyts et al. (2004),
but thereby limiting the number of galaxies in our sample to four, 
we find $(-1.12 \pm 0.13)z$ and $(-0.99 \pm 0.10)z$ for our sample and the cluster samples, respectively.
We conclude that for high-mass galaxies, there is no difference between the cluster samples and our field sample.
The galaxies with masses $M<2\times 10^{11}M_{\odot}$ in our sample evolve much faster: 
$\Delta \ln {(M/L_B)} = (-1.97 \pm 0.16)z$ ($(-1.90 \pm 0.17)z$ for the primary sample alone).
We verify that
these results do not change if galaxies with spectra with $S/N<12$ are included as well. Therefore,
the accuracy of our results is not limited by the quality of the velocity dispersions.

In Figures \ref{fig:z_dML} and \ref{fig:z_BI} we show evolutionary tracks for a single stellar 
population 
for formation redshifts $z=1$ and 2. This very simple model assumes that 
luminosity evolves with time as $L \propto (t-t_{form})^{\kappa}$, 
where $\kappa$ is derived from stellar population models 
(see van Dokkum et al. (1998) for more details).
These tracks indicate a large spread in formation redshifts. We note, however, 
that $\kappa$ is sensitive to the IMF. Here we use a single stellar population from 
Bruzual \& Charlot (2003)
with a Salpeter IMF and 
solar metallicity, which yields $\kappa=0.97$ in the $B$ band and $0.43$ in
the $I$ band.
According to this model the massive galaxies have 
high luminosity weighted formation redshifts ($z\ge 2$), 
whereas less massive galaxies in our sample have 
lower formation redshifts ($1<z<2$).

\null
\vbox{
\begin{center}
\figurenum{6}
\leavevmode
\hbox{%
\epsfxsize=9cm
\epsffile{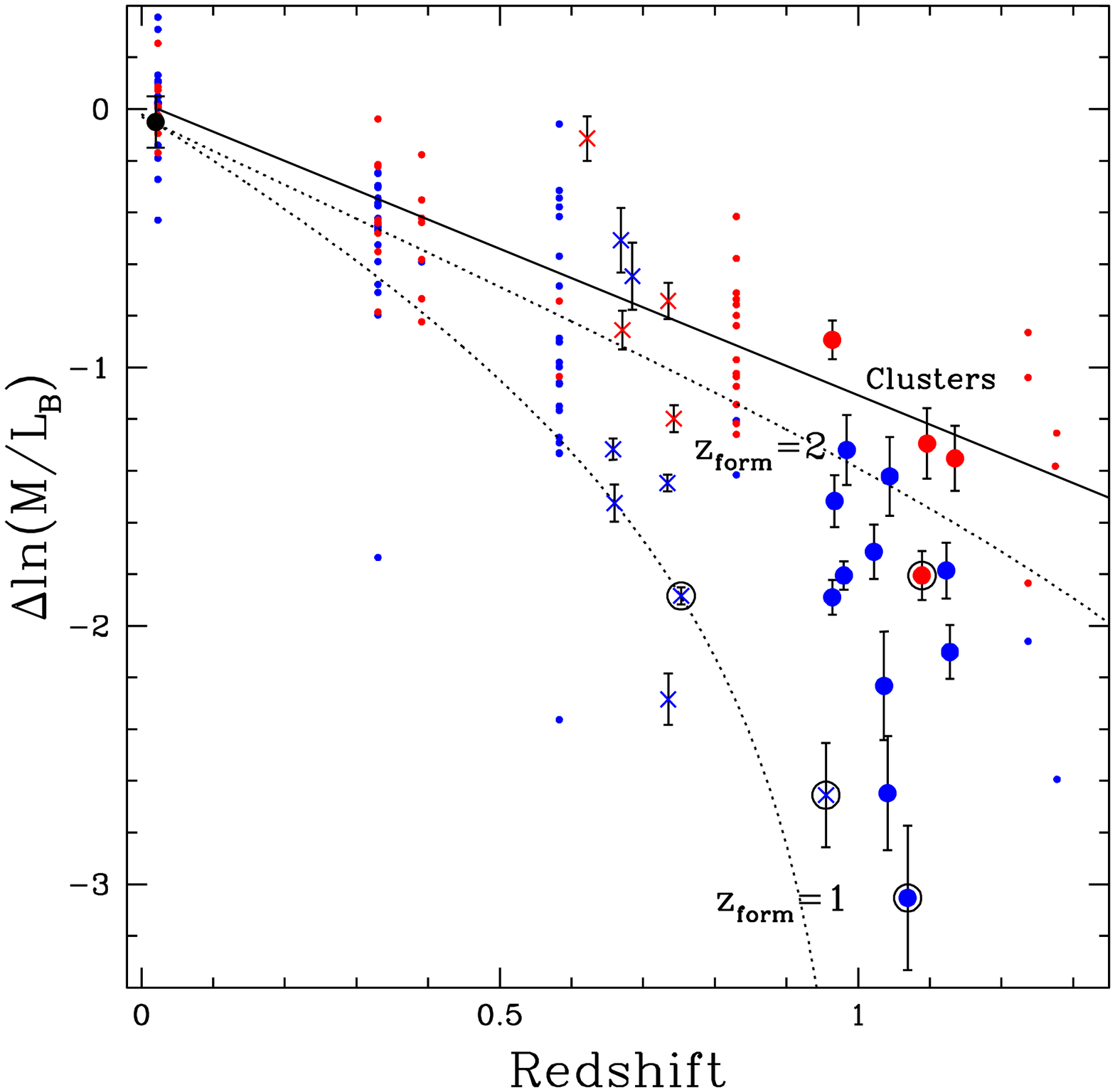}}
\figcaption{
\small 
The offset from the local FP in the rest-frame $B$ 
band for individual cluster galaxies taken from the literature 
(small dots) and the
early-type galaxies with high $S/N$ spectra in our sample.
We distinguish, as in Figure 5, between the primary sample (filled symbols) and the fillers (crosses).
The red symbols are galaxies with masses $M > 2\times10^{11} M_{\odot}$.
The blue symbols are less massive galaxies.
The cluster samples are taken from {J\o rgensen} {et~al.} (1996),
{Kelson} {et~al.} (2000), {van Dokkum \& Franx} (1996), 
{Wuyts} {et~al.} (2004), {Holden} {et~al.} (2004), and 
{van Dokkum \& Stanford} (2003).
The solid line is $\Delta \ln({M/L_B})=1.12z$, which is
the best fitting straight line for the evolution of cluster galaxies with 
masses $M>2\times10^{11} M_{\odot}$. 
The dotted lines are model tracks 
for a single 
stellar population 
with formation redshifts 1 (lower line) and 2 (upper line)
(see text for a more detailed explanation).
These model tracks are forced to go through the black point at $z=0.02$ which 
represents the field galaxies in the sample of Faber et al. (1989).
The range in offsets from the local field FP is large, but there is a strong
correlation with mass.
Massive field galaxies 
evolve as fast as equally massive cluster galaxies, while less
massive galaxies evolve faster.}
\label{fig:z_dML}
\end{center}}

\subsection{The relation between $M$ and $M/L$ and the role of selection effects}
Figure \ref{fig:M_ML} illustrates the tight relation between mass and $M/L_B$.
We also show the SDSS field sample described in Section 4.1.
The $M/L$ of all galaxies have been corrected for evolution as derived from massive cluster galaxies,
$\Delta \ln {(M/L_B)} = -1.12z$. 

At $z=1$, the relation between $M/L$ and $M$ seems much steeper than in the local
universe.
However, we have to take selection effects into account.
Since we selected our sample in the $z$ band, we can transform our 
magnitude limit into a luminosity limit in the rest-frame $B$ band, which is
close to the observed $z$ band at $z\sim 1$ (the median redshift of our
sample of 16 $z\sim 1$ galaxies is $z=1.04$).
Because this luminosity limit only applies to our primary sample,
this discussion does not involve the fillers (mainly galaxies at $z\sim 0.7$). 
Afterward, we will comment on these galaxies.

\null
\vbox{
\begin{center}
\figurenum{7}
\leavevmode
\hbox{%
\epsfxsize=9cm
\epsffile{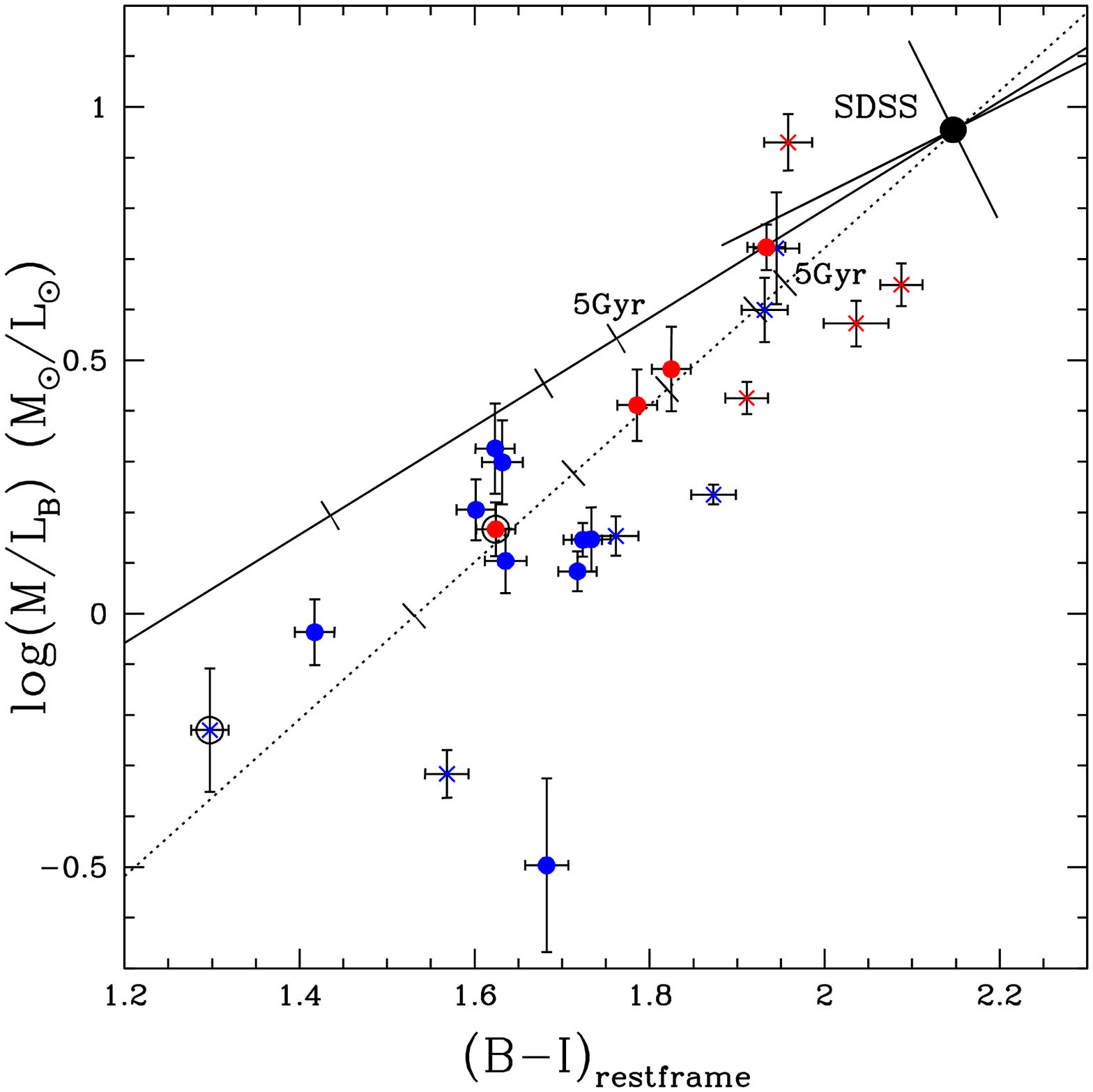}}
\figcaption{
\small 
Rest-frame $B-I$ color versus $M/L_B$ for our early-type galaxy 
sample. For an explanation of the symbols, see Figure 6a.
The large symbol at the upper right indicates the median $B-I$ and $M/L_B$
of the massive galaxies ($M>2\times 10^{11}M_{\odot}$) in the SDSS field early- 
type galaxy sample. The orientation of the distribution around the median values 
and the amount of scatter are indicated by the tilted error bars.
The dotted line is a solar metallicity Bruzual-Charlot model for a 
single stellar population.
The solid line is a model with exponentially declining star formation 
($\tau=1$~Gyr). Both model tracks are shifted vertically to match the SDSS 
data point.
Model ages are indicated by ticks at intervals of 1 Gyr.
The correlation between $M/L$ and color implies that the observed scatter in $M/L$
is real, and can be ascribed to age differences between the stellar populations
of the galaxies.
Our $i-z\ge 0.86$ color selection limit roughly corresponts to $B-I\ge 1.1$ according
to the Bruzual-Charlot models. This shows that our selection criterion only
excludes galaxies with ages less than 1 Gyr, and does not affect our conclusions
regarding the massive, red galaxies.}
\label{fig:BI_ML}
\end{center}}

We first test whether the observed distribution can be fully explained by selection effects,
assuming that the slope and the scatter of the FP do not evolve.
The probability that the observed distribution is drawn from a population with 
the same distribution as the SDSS galaxies is 0.14\%,
if the $M/L$ evolution is the same for all galaxies.
Therefore, the slope evolves, the scatter evolves, or both.
If there is an age difference between high and low mass galaxies, the scatter will most probably 
also evolve differently for high and low mass galaxies. We cannot exclude with high confidence 
that the observed distribution of $M/L$ is due to a larger scatter at high redshift:
if the scatter in $M/L$ at $z=1$ is twice as large as in the local universe, the probability of a non-evolving
slope is 8.0\%.
However, we see no evidence for an increase in the scatter at the high mass end,
where selection effects do not play a role.
Hence, if the scatter evolves, this is only true for galaxies with masses 
$M\sim 10^{11} M_{\odot}$.
Since the increased scatter is most likely caused by young ages, low mass galaxies
would have lower $M/L$, and hence the slope of the relation would also be 
changed.
We consider these findings as strong evidence for mass-dependent
evolution of early-type galaxies. To confirm that we observed a change 
in the slope of the FP, deeper and larger surveys are needed.

\null
\vbox{
\begin{center}
\figurenum{8}
\leavevmode
\hbox{%
\epsfxsize=9cm
\epsffile{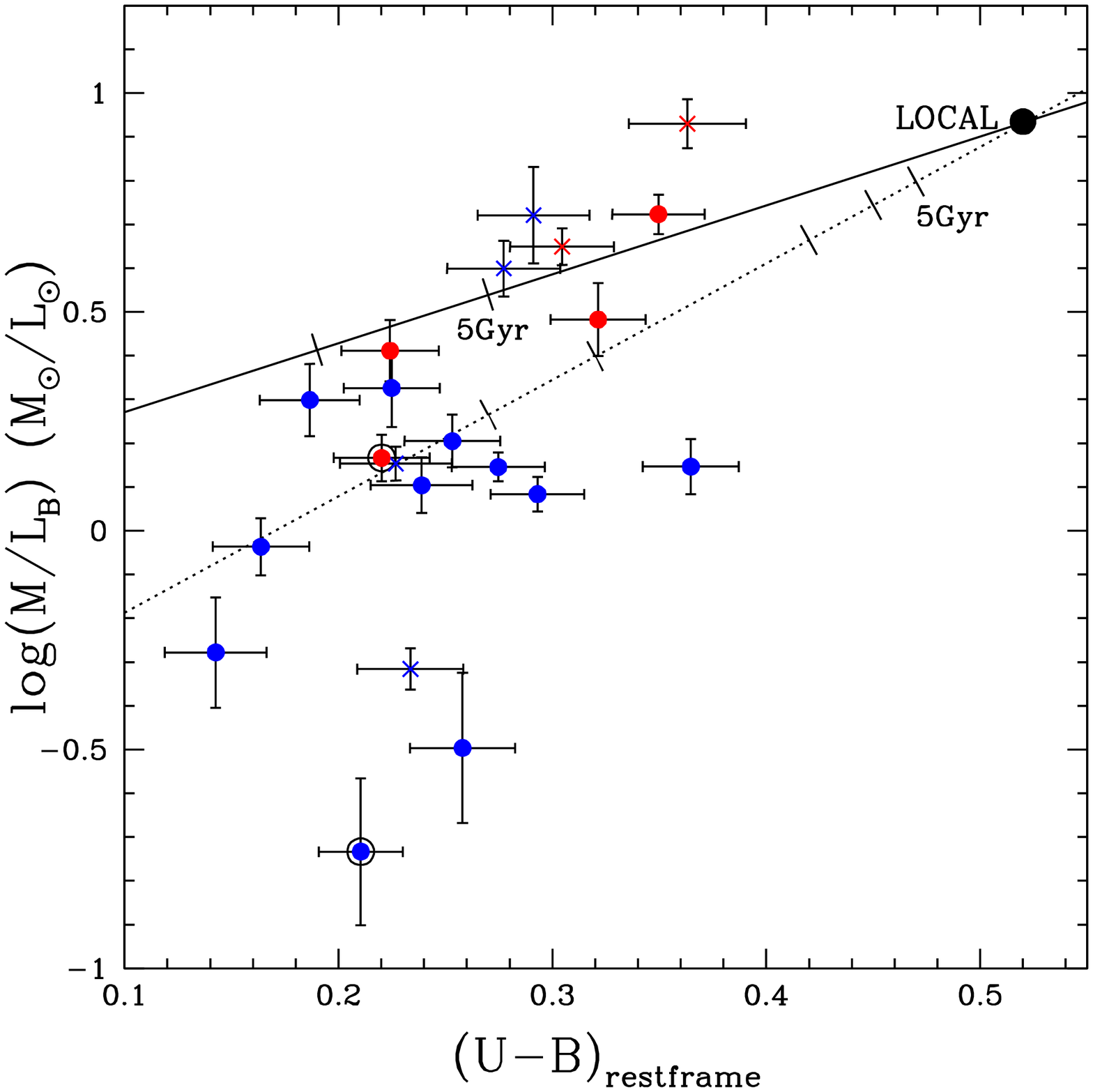}}
\figcaption{\small 
Rest-frame $U-B$ color versus $M/L_B$ for the early-type galaxy 
sample. For an explanation of the symbols, see Figure 6a.
The 'Local' data point is taken from Gebhardt et al. (2003).
There is a similar relation as in Figure 7, but it is less clear because the
range of colors in much smaller in $U-B$ than in $B-I$.
Our $i-z\ge 0.86$ color criterion corresponds to $U-B\ge 0.07$ at $z=1$, which
demonstrates that this criterion would only exclude the most extremely blue galaxies
(with ages well below 1 Gyr).}
\label{fig:UB_ML}
\end{center}}

The location of the luminosity limit at $z=1$ in 
Figure \ref{fig:M_ML} shows 
that our $z\sim 1$
sample is dominated by selection effects for masses $M<6\times10^{10} M_{\odot}$.
The objects which such low masses are only included in the sample because of their 
probably extreme $M/L$. We cannot correct the $M/L$ of that subsample for the bias introduced 
by our luminosity limit.

On the other hand, for the sample of galaxies with higher masses, we can correct for selection
effects, because they are relevant, but not dominant, as can be seen in Figure \ref{fig:M_ML}.
The average evolution of the galaxies in the primary sample 
with masses $M>6\times10^{10} M_{\odot}$
is $\Delta \ln {(M/L_B)} = (-1.55 \pm 0.16)z$.
The median mass of this subsample of 12 galaxies is $M=1.9\times 10^{11} M_{\odot}$. 
We estimate the maximum bias by assuming that the slope is the same 
at $z=1$ and in the local universe, but that
the scatter is a factor two larger at $z=1$ than at $z=0$. 
In that case the observed distribution is expected to follow the short-dashed line in Figure \ref{fig:M_ML}.
At a given mass, the difference between the solid line and the short-dashed line
is the bias introduced by the selection effects in luminosity.
We increase the observed $M/L$ of each galaxy in our primary sample by the difference between 
the solid line and the short-dashed line at the mass of that galaxy.
For galaxies more massive than $M=2\times10^{11} M_{\odot}$ this correction is negligible, but for the
galaxies with masses $M\approx 6\times10^{10} M_{\odot}$ this correction is about 30\%.
Using this method, we find a bias corrected evolution of the 
galaxies with masses $M>6\times10^{10} M_{\odot}$
of $\Delta \ln {(M/L_B)} = (-1.43 \pm 0.16)z$.
Given the uncertainty in the intrinsic scatter, deeper observations 
are necessary to confirm this value.

Besides a bias due to the luminosity limit, errors in the velocity dispersion
produce correlated errors in $M$ and $M/L$, hence underestimating the measured
evolution. Taking the errors in $\sigma$ into account, we find that this
introduces a bias at the level of only 2-3\%, which is several times smaller
than our measurement accuracy.

\null
\vbox{
\begin{center}
\figurenum{9}
\leavevmode
\hbox{
\epsfxsize=9cm
\epsffile{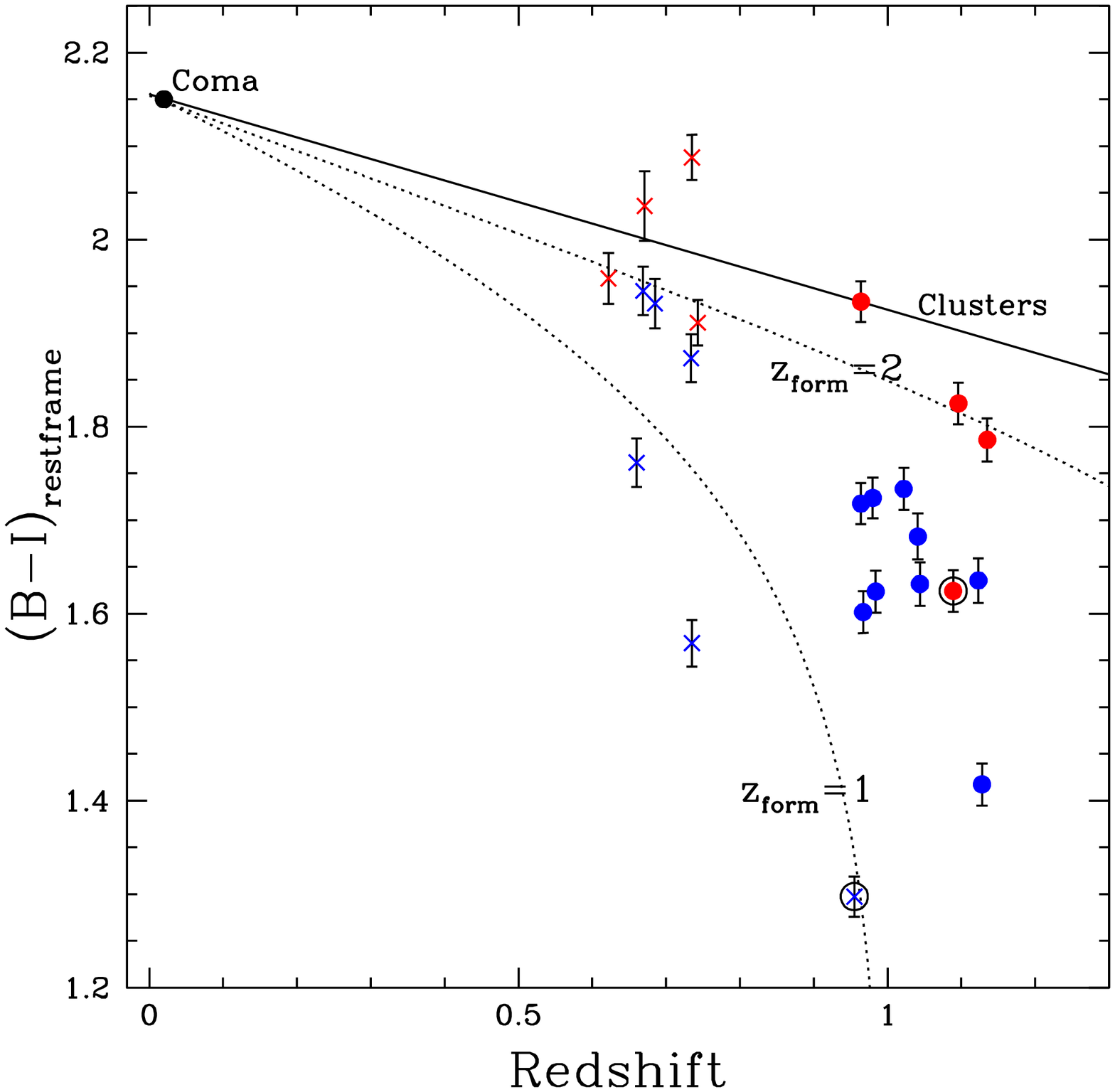}}
\figcaption{\small 
The evolution of the rest-frame $B-I$ color with redshift. For an explanation of the 
symbols, see Figure 6a. As already suggested by the tight relation between $B-I$ and 
$M/L$ in Figure 7, the evolution of color with redshift is very similar to the evolution
of $M/L$ with redshift.
Massive galaxies are the reddest in the local universe, and their color evolves
slower than the color of low mass galaxies.
Also, the color of the most massive field galaxies is very similar to the color
of massive cluster galaxies.}
\label{fig:z_BI}
\end{center}}

The above analysis only involves $z\sim 1$ galaxies satisfying all our selection criteria,
but we note that the relation between $M/L$ and $M$ exists for the 
$z\sim 0.7$ galaxies as well. However, this subsample is selected in an 
inhomogeneous way, therefore it is impossible to correct the observed
evolution. Since all the galaxies roughly lie along lines of constant 
luminosity, the bias toward low $M/L$ galaxies likely explains the 
observed relation between $M$ and $M/L$ for this sub-sample.

Besides by luminosity and morphology, the galaxies in our sample are also 
selected by color.
This potentially introduces an important bias in the measured evolution, 
because of the exclusion of galaxies with blue colors, i.e., low $M/L$.
Our color criterion, however, is quite generous. Even the very blue, low mass
galaxies satisfy this criterion.
For typical Bruzual-Charlot models, our color limit ($i-z=0.86$) corresponds
to $U-V\sim0.67$ at $z=1$, which is 0.2 mag bluer than the limit applied by
Bell et al. (2004b) to select galaxies on the red sequence at $z\sim 1$.
Our color cut is 0.45 mag bluer than the color-magnitude
relation found by the same authors.
From Bruzual-Charlot models we estimate that we only miss galaxies that are
younger than $\sim 1$ Gyr (see Figures 7 and 8). 
Furthermore, Bell et al. (2004b) did not find blue, massive galaxies in the 
entire
COMBO-17 dataset. Hence it is very unlikely that we miss any blue galaxy at the
bright end of our sample because of our color selection criterion.

\null
\vbox{
\begin{center}
\figurenum{10}
\leavevmode
\hbox{%
\epsfxsize=9cm
\epsffile{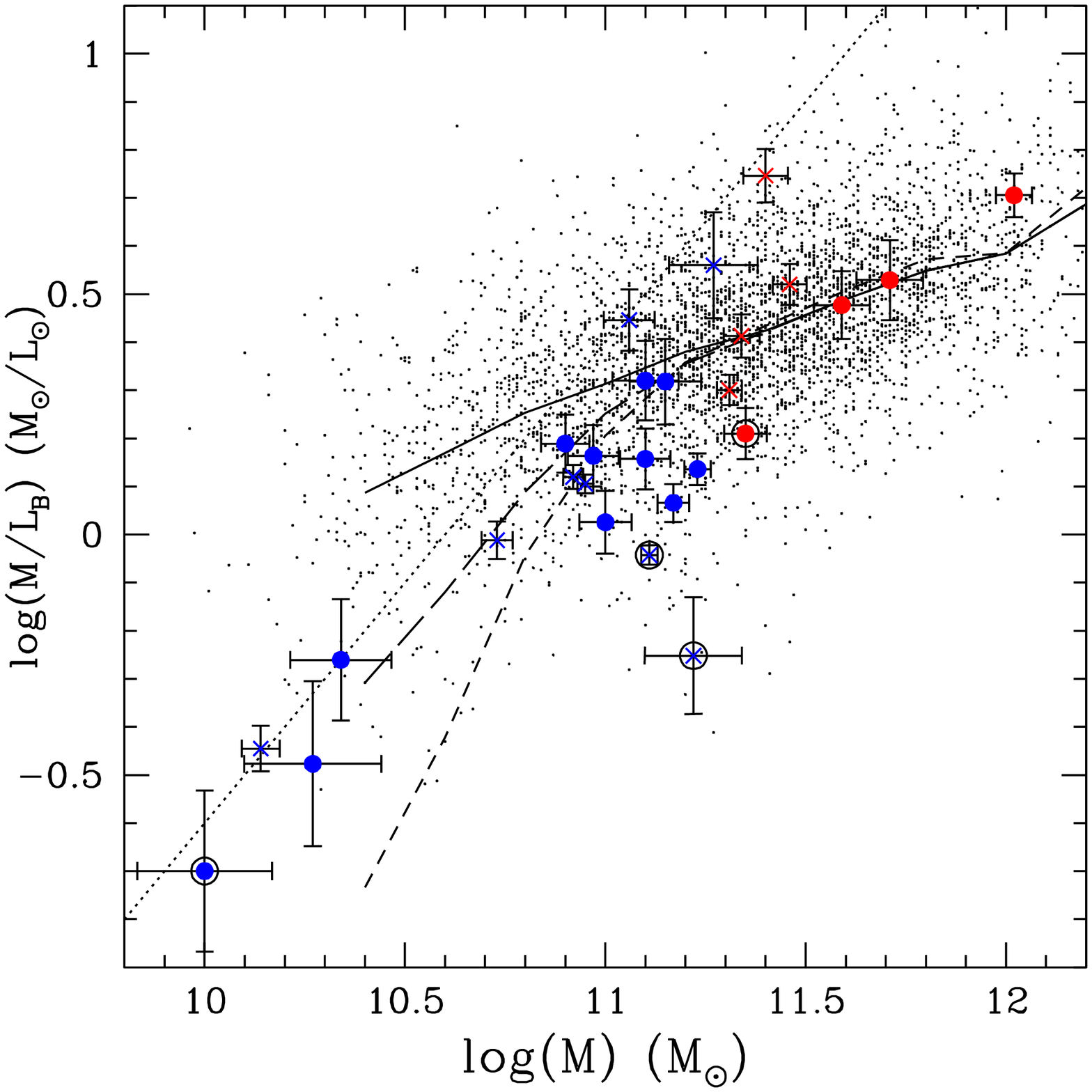}}
\figcaption{
\small 
$M$ versus $M/L_B$ as derived from the FP for the early-type galaxies 
in our sample and for the nearby sample from the SDSS.
For an explanation of the symbols, see Figure 6a.
All data points have been corrected for $M/L$ evolution as found for massive cluster galaxies,
$\Delta \ln({M/L_B}) =-1.12z$, normalizing at $z=1$.
The dotted line indicates our magnitude limit, translated into a luminosity limit at $z=1$.
Therefore, this limit only applies to
the filled circles. The full drawn line indicates the median $M/L_B$ of the 
SDSS early-type galaxies population. The long-dashed line indicates the median 
$M/L_B$ of galaxies that are brighter than our luminosity limit at $z=1$.
Assuming that the scatter in $M/L$ at $z=1$ is a factor 2 larger than in the local universe,
the median of the $M/L$ of galaxies brighter than the luminosity limit follows the short-dashed
line.
It is clear that for the three galaxies in the primary sample 
with the lowest masses selection
effects play such a dominant role that we cannot include these objects
in our efforts to correct for this bias.
Between $6\times 10^{10}M_{\odot}$ and $2\times 10^{11}M_{\odot}$ selection effects are relevant, 
but not dominant. For higher masses, selection effects do not affect our sample.
The difference between galaxies with masses of $\approx 10^{11}M_{\odot}$ and
masses $\approx 10^{12}M_{\odot}$ cannot be explained without an increase in 
the scatter with redshift, or assuming mass-dependent ages.}
\label{fig:M_ML}
\end{center}}

\subsection{Independent evidence for mass-dependent evolution of early-type galaxies}
It is particularly interesting to compare our results to the results from studies involving lensing 
galaxies, since those samples are mass selected. 
Rusin et al. (2003) and van de Ven et al. (2003) find
$\Delta \ln {(M/L_B)} = (-1.29 \pm 0.09)z$ and $(-1.43 \pm 0.30)z$, respectively, using the same
dataset. This seems somewhat
low compared to the evolution of our sample and results found in the literature, 
but the lensing galaxies typically have high masses (see Table 4). The data for individual galaxies published by
van de Ven et al. show that the median mass of the lens sample is $M=2\times 10^{11}M_{\odot}$, whereas
the median mass of our early-type galaxy sample is $M=1.3\times 10^{11}M_{\odot}$.
The galaxies in the lens sample that are more massive than $M=2\times 10^{11}M_{\odot}$ evolve as
$\Delta \ln {(M/L_B)} = (-1.13 \pm 0.31)z$. The galaxies less massive than this evolve much faster:
$(-1.71 \pm 0.30)z$. The similarity between the results from the lensing sample and our sample 
is striking, especially because the lensing sample is mass selected. Hence, it is very hard to
see how a bias toward low $M/L$ galaxies can be responsible for the observed mass dependence in the lensing sample.
It is possible that not all the low mass lenses are genuine early-types.
If we omit the most irregular lenses
(FBQ0951+2635, SBS1520+530, and B1608+656), we still find rapid evolution, $(-1.64 \pm 0.24)z$, 
for low mass lensing galaxies.
Hence, these results provide strong evidence that the observed dependence of $M/L$ evolution on
mass in our sample is real.
One possible complicating factor is that the lensing cross-section of galaxies in groups is 
larger than that of galaxies in the lowest density environments. This may lead to a difference 
between the populations of the lensing sample are our sample.

\null
\vbox{
\begin{center}
\figurenum{11}
\leavevmode
\hbox{%
\epsfxsize=9cm
\epsffile{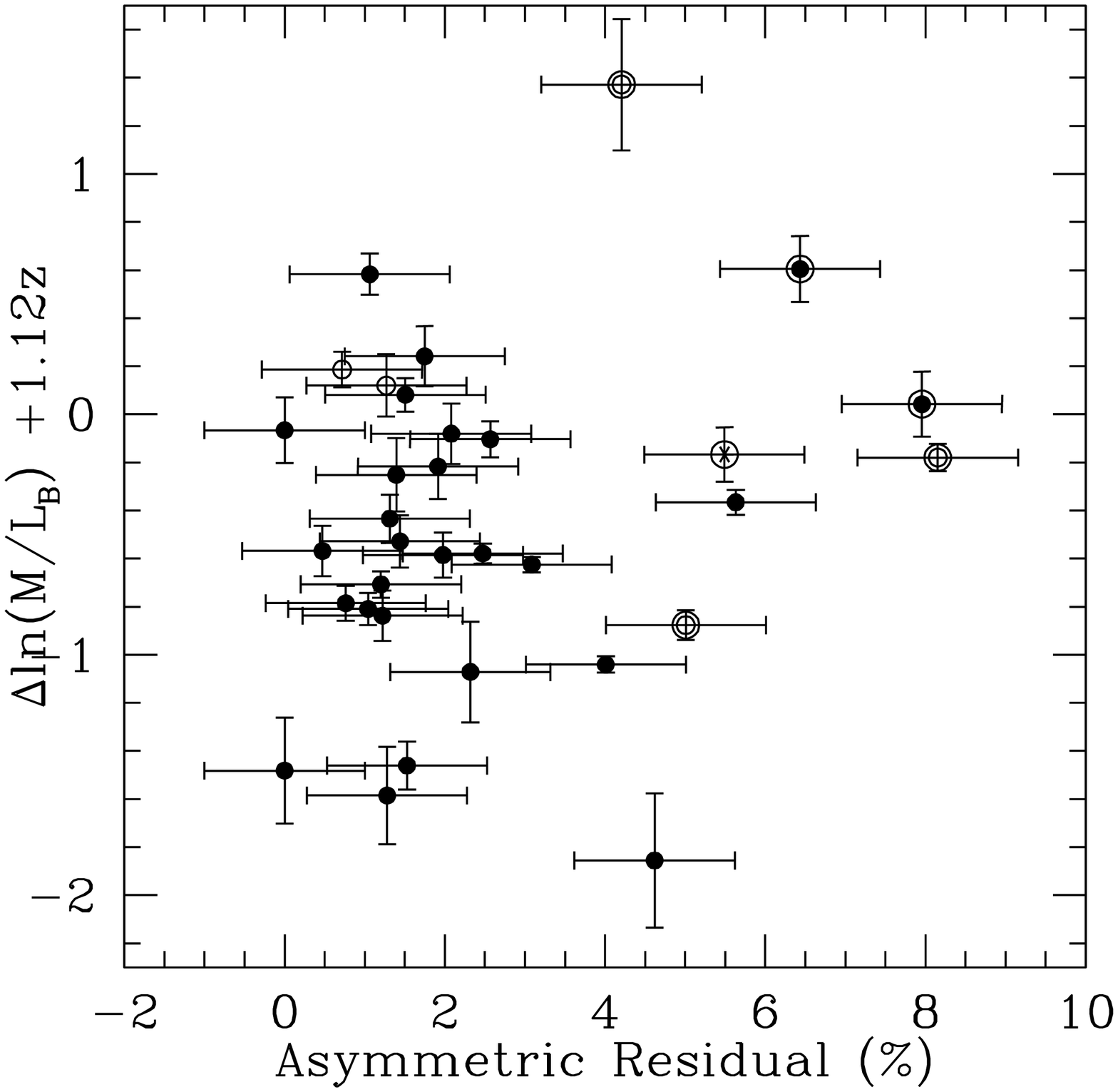}}
\figcaption{
\small 
Asymmetric residuals of the $r^{1/4}$-profile fits (see Figures 3 and 4) versus 
the offset from the local FP of the early-type galaxies in our sample, 
corrected for evolution.
Filled circles are galaxies best fit with $r^{1/4}$-profiles, asterisks with
$r^{1/3}$, and open circles with $r^{1/2}$.
Encircled objects are galaxies with late-type morphologies.
A 2\% or more residual indicates a significant deviation from symmetry in the light 
profile. There is no clear relation between $M/L$ and deviations from smooth 
surface brightness profiles.} 
\label{fig:asym_dML}
\end{center}}

\subsection{Ongoing star formation and morphological deviations}
It is interesting to see that galaxies 
with emission lines have relatively 
low $M/L$, as can be seen from both Figure \ref{fig:z_dML} and \ref{fig:M_ML}. 
Treu et al. (2002)
have already shown that a substantial fraction of the massive
early-type galaxy population at high redshift shows evidence for ongoing star formation
by the presence of emission lines.
Galaxies with emission lines in our sample, however, tend to have low masses.
Figure \ref{fig:M_ML}
suggests that several of the galaxies with masses $\sim 3\times10^{10} M_{\odot}$ are included in the 
magnitude limited 
sample only because they are forming stars.
When excluding galaxies with 
emission lines, the evolution of the sample (without applying a mass cut), is 
$\Delta \ln {(M/L_B)} = (-1.61\pm0.18)z$. The galaxies
with emission lines evolve faster: $\Delta \ln {(M/L_B)} = (-2.41\pm0.45)z$.
On average, the mass of the galaxies without emission lines is 1.5 times larger
than the mass of the galaxies with emission lines.

Mergers or interactions, accompanied with star formation, can lead 
to both deviations from smooth $r^{1/n}$-profiles and low $M/L$ values. 
van Dokkum \& Ellis (2003)
have shown tentative examples of this phenomenon.
Figure \ref{fig:asym_dML} shows the magnitude of the asymmetric residual, 
(described in Section \ref{sec:profiles}), versus the offset from the FP, corrected for luminosity
evolution. 
There is no correlation between deviations from smooth surface profiles and 
$M/L$, and 
we find no evidence for a connection between star formation activity and interactions or mergers.

\subsection{AGN}
Four out of 11 early-type galaxies at $z<0.8$ have AGN, 
and one out of 16 for the $z>0.9$ sample, as determined from X-ray imaging
(see section 3.3).
Of the five early-type galaxies with AGN, three are more massive than 
$M=2\times10^{11} M_{\odot}$ and two have emission lines in their spectra.
The $M/L$ evolution of galaxies in our sample with AGN is $\Delta \ln {(M/L_B)} = (-1.64 \pm 0.31)z$, 
which is not very different from the value for galaxies without AGN: $(-1.76 \pm 0.20)z$.
However, the galaxies with AGN are twice as massive than the galaxies without AGN. 
Considering high mass galaxies only ($M>2\times 10^{11}M_{\odot}$), 
we find $(-1.46\pm 0.24)z$ for galaxies with AGN and $(-1.02\pm 0.41)z$ for
galaxies without AGN.
Now, the trend is reversed, but
only one out of the five galaxies with an AGN has a particularly blue $B-I$ color for its $M/L$.
Therefore, a larger sample is needed to confirm that the epochs of star formation and AGN activity are related. 
Woo et al. (2004) find that up to $z\sim 0.5$ the evolution
of the FP for galaxies selected to have AGN is indistinguishable from otherwise selected galaxies.
Our results provide tentative evidence that a difference may set in at higher redshift.

\section{Comparison with previous results}
Various other authors have measured the $M/L$ evolution of field early-type galaxies.
In this section we compare these previous results with our findings, and 
we comment on the apparent inconsistencies that exist in the literature.
In Table 4 we give the $M/L$ evolution as reported by other papers, as
well as the values we derive using the tabulated datasets given in 
those papers.

Treu et al. (2002)
find $\Delta \ln {(M/L_B)} = (-1.66 \pm 0.31)z$. This number is corrected for selection effects, while 
the value in Table 4 is uncorrected, to make a fair comparison with other results.
The median mass of the sample presented by Treu et al. (2003) is $2.5\times 10^{11}M_{\odot}$ for
a sample extending to $z\sim 0.6$. 
Tabulated data are only available for the lower redshift galaxies presented in Treu et al. (2001).
Only the galaxies with very low redshifts ($z\sim 0.1$) have low masses, therefore the comparison 
between high and low mass galaxies cannot be made.
We note that the evolution of the entire Treu et al. sample is consistent with the evolution
of our entire sample, $(-1.75 \pm 0.16)z$, and that the evolution
of the massive galaxies in our sample is only mildly inconsistent with the evolution 
of the Treu et al. sample.
Several of the high mass galaxies in the Treu et al. sample have emission lines in their spectra,
which contrasts with our sample.

Van Dokkum et al. (2001) find $(-1.35 \pm 0.35)z$ for an average redshift $z=0.42$.
The galaxy masses range from $3\times 10^{10}M_{\odot}$ to $10^{12}M_{\odot}$ and are evenly distributed in $\log{(M)}$.
Remarkably, when taking the tabulated data of the individual galaxies, and applying our fitting method, we find a larger 
value: $(-1.67 \pm 0.23)z$ (see Table 4). Van Dokkum et al. create redshift bins in which they calculate the bi-weight
center of the $M/L$ offset. This method is sensitive to the bin choice, which leads to the difference between the
reported value and the value calculated with our fitting method.
We note that, when using our fitting method, the result of van Dokkum et al.
is very similar to the result presented in this paper.
Splitting the van Dokkum et al. sample into low and high mass galaxies (at $M=2\times 10^{11}M_{\odot}$),
we find $(-1.29 \pm 0.39)z$ for the high mass galaxies, and $(-1.94 \pm 0.34)z$ for the low mass
galaxies.
We verified that this $2.5\sigma$ difference can be explained entirely by the fact that the galaxies are luminosity selected, not mass selected.
Contrary to this study and the work of Treu et al., van Dokkum et al. did not select the galaxies by color, but
by morphology and magnitude only. 

Van Dokkum \& Ellis (2003) find $(-1.25 \pm 0.25)z$. The difference in fitting method, mentioned above, applies 
to this study as well. Additionally, their transformation from $I$-band surface 
brightness to rest-frame $B$-band surface brightness, using $V-I$ colors, is uncertain. Namely, using $V-I$
is an extrapolation for galaxies at $z>0.8$. This leads to a different value 
for the $M/L$ evolution if we use our method to fit to the individual galaxy 
data of the samples of van Dokkum et al. (2001) and van Dokkum \& Ellis (2003):
$(-1.69 \pm 0.13)z$, a value very similar to the result yielded by our sample.
Using the bi-weight center to measure the evolution effectively gives very low weighting factors
to outliers. If the two galaxies with the lowest $M/L$ in the high-$z$ sample of van Dokkum et al.
are omitted, we find $(-1.53 \pm 0.13)z$. The measurement accuracy, together with the uncertain transformation
to rest-frame properties can explain the remaining difference.
The new sample described by van Dokkum \& Ellis (2003) is too small to verify whether there is a trend with 
galaxy mass. There is one galaxy more massive than $M=2\times 10^{11}M_{\odot}$ with
$\Delta \ln {(M/L_B)} = (-1.57 \pm 0.14)z$.
As is the case for the sample of van Dokkum et al. (2001), the sample of van Dokkum \& Ellis (2003)
is not selected by color.
This did not lead to faster evolution due to including blue 
early-type galaxies, which indicates that the color cuts in the other studies
are adequately generous to avoid this bias.

Gebhardt et al. (2003) (hereafter G03) report a brightening in the rest-frame $B$ band of early-type galaxies by 2.4 magnitudes at $z=1$,
which is derived from fitting a cubic spline to the offsets of the galaxies from the local FP.
With our fitting method, we find $(-1.94 \pm 0.20)z$, using the data of the 
individual galaxies in that sample.
This is consistent with the result presented in this paper. 
The difference between the value reported by G03 and our result using their 
data is caused by the three
galaxies at redshifts $z>0.9$. Our linear fit to their data and their fit 
agree well up to $z=0.8$ (1.6 magnitudes brightening).
If we split the G03 sample into high and low mass galaxies,
we find an evolution of $(-1.59 \pm 0.50)z$ for galaxies more massive than $M=2\times 10^{11}M_{\odot}$, and
$(-2.05 \pm 0.17)z$ for less massive galaxies. For low mass galaxies our result, $(-1.97 \pm 0.16)z$, and that
of G03 agree well. The fast evolution for massive galaxies in their sample remains unexplained, but given 
the large uncertainty and the small number of objects the inconsistency is only mild.
We note, however, that contrary to the early-type galaxies in our sample, 
the early-type galaxies in the G03 sample do not show a 
correlation between $M/L$ and rest-frame $U-B$ color.
One of the galaxies in the G03 sample serves as an example of the uncertainty.
HST14176+5226 has been observed spectroscopically before (Ohyama et al.  2002)
and happens to be in the sample of lensing galaxies of Rusin et al. (2003) and van de Ven (2003).
G03 give a stellar velocity dispersion of $\sigma_*=222\pm 8$km~s$^{-1}$ corrected to the same aperture
as our dispersions. 
Ohyama et al. have measured $\sigma_*=245\pm 15$km~s$^{-1}$, also corrected to the same aperture,
which implies a somewhat smaller offset from the local FP. 
van de Ven et al. report a value of $\sigma=292\pm 29$km~s$^{-1}$ as derived from the lensing model.
The latter value leads to a rate of evolution that we find for the massive galaxies in our sample. 
Repeated observation of these massive galaxies may be illuminating.

We conclude that all the results found in the literature are mutually consistent, once differences in 
calculating and presenting the results have been taken into account.

\section{Conclusions}
We obtained ultra-deep spectroscopy for 27 field early-type galaxies with redshifts $0.6<z<1.15$.
The offset of these high redshift galaxies from the local FP is used as a measure of the evolution of $M/L$ and 
as an age estimator. 

The average evolution of the early type galaxies in our sample is $\Delta \ln {(M/L_B)} = (-1.75 \pm 0.16)z$. 
The value we find for galaxies in the primary sample, those galaxies satisfying all our selection criteria, is the same.
The scatter in $\Delta \ln {(M/L_B)}$ is large: 0.58. This shows 
that some galaxies must 
have high luminosity weighted formation redshifts ($z>2$), while others have formed a large fraction of
their stars at redshifts $1<z<2$. Emission lines in the spectra indicate that some galaxies show signs
of ongoing star formation at the epoch of observation.
This is in agreement with the presence of massive early-type
galaxies with emission lines at $z\sim 0.5$ Treu et al. (2002), although the galaxies with emission lines in
our sample tend to have low masses.

We find a tight correlation between $M/L$ and rest-frame color,
which shows that the variation in $M/L$ among the galaxies in the sample is intrinsic,
and due to differences in the stellar populations. 
The galaxies in our sample span a large range of masses. We find that low mass galaxies 
have larger offsets from the local FP than high mass galaxies.
Because luminosity selected samples are biased toward galaxies with low $M/L$ this is a trend 
that is expected. We carefully analyze whether the observed correlation between mass and $M/L$
can entirely be explained by this selection effect or not.
We find that galaxies with masses $M< 6\times 10^{10}M_{\odot}$ are only included in our sample
because they have low $M/L$. For galaxies at $z\sim 1$ with masses larger than $M= 6\times 10^{10}M_{\odot}$, 
our sample is biased, but to a limited amount.
Taking into account the selection effect, we exclude with high confidence that 
the distribution of mass and $M/L$ of our $z\sim 1$ galaxy sample with masses 
$M> 6\times 10^{10}M_{\odot}$ has the same distribution as the low
redshift field early-type galaxy sample taken from Bernardi et al. (2003), corrected for evolution.
We do not claim that we have observed a change of the slope of the FP, because we cannot 
exclude the possibility that our sample is drawn from a distribution
that has the same slope as the Bernardi et al. sample, but with a scatter that is twice as large
at $z=1$. However, the outliers do not occur at the high mass end ($M\sim 10^{12}M_{\odot}$) 
of our galaxy sample, but at lower masses, namely $M\sim 10^{11}M_{\odot}$. 
Therefore, our results show that the evolution of early-type galaxies is mass-dependent,
whether by an increase in the scatter at lower masses, or by systematic faster evolution of lower mass galaxies as compared
to higher mass galaxies, or, which is the most natural explanation, by a combination of these effects.
Assuming the scatter has decreased from $z=1$ to the present day by a factor of 2, we find
that the for bias corrected $M/L$ evolution of $z\sim 1$ early-type galaxies with masses $M> 6\times 10^{10}M_{\odot}$
is $\Delta \ln {(M/L_B)} = (-1.43 \pm 0.16)z$.

Previous studies 
(Treu et al. 2001, 2002; van Dokkum et al. 2001, van Dokkum \& Ellis 2003, Gebhardt et al. 2003),
that claimed to have derived mutually exclusive results, are in fact consistent with our results, 
if the same fitting method is applied to the different datasets.
Particularly interesting
is the consistency of our results with the results from a sample of lensing galaxies
(Rusin et al. 2003; van de Ven 2003), which also
shows the mass-dependence, even though this sample is not biased toward galaxies with low $M/L$:
because of the selection technique, the lens sample contains galaxies with typical $M/L$ at a given mass. 
Our luminosity limited sample is sensitive to outliers, which are present indeed.
The combination of these independent results strengthens the evidence for mass-dependent evolution and 
the combined increase in both slope and scatter with redshift.

Bell et al. (2004)
claim that the mass density of red sequence galaxies increases by at least a factor of 2 from 
$z\sim 1$ to the present. This is partly based on the observation that the luminosity density 
is constant out to $z=1$. The $M/L$ evolution of our galaxy sample implies an increase by a factor of 4 
in the mass density.
In the local universe, most of the mass density 
in early-type galaxies is accounted for by galaxies with 
a velocity dispersion of $\approx 225$km s$^{-1}$ (Kochanek et al. 2000) 
or a mass of $\approx 3\times 10^{11}M_{\odot}$.
If those galaxies, which evolve somewhat slower, dominate the evolution of the mass density, 
the increase is slightly less ($3-3.5$).

The correlation between mass and $M/L$ has been observed in clusters as well (Wuyts et al. 2004), but
this can entirely be explained by selection effects. Both our field sample and the cluster samples
found in the literature are not strongly biased for galaxy masses $M> 2\times 10^{11}M_{\odot}$.
When applying this mass cut, the evolution for cluster galaxies is 
$\Delta \ln {(M/L_B)} = (-1.12 \pm 0.06)z$, and the evolution for field galaxies is 
$\Delta \ln {(M/L_B)} = (-1.20 \pm 0.18)z$ ($-1.26z$ for galaxies in the primary sample).
Galaxies 
with masses comparable to the mass of $L^*$ galaxies in the local universe
($\approx 3\times 10^{11}M_{\odot}$) have luminosity weighted ages that imply formation redshifts
$z\ge 2$, independent of environment.
If progenitor bias is important, the luminosity weighted age of typical early-type galaxies
in the local universe can be considerably lower.

In hierarchical formation models, the predicted difference between the $M/L$ of field and cluster galaxies is 
$\Delta \ln {(M/L_B)} = 0.55$, independent of redshift (van Dokkum et al. 2001). 
This large difference 
is related to the difficulty of constructing isolated galaxies without active star formation, i.e.
to the lack of a mechanism that truncates star formation from within the galaxy.
Our results rule out this prediction at the 99.6\% confidence level, up to $z=1.1$.

Our findings are consistent with down-sizing (Cowie et al. 1996; Kodama 2004).
This idea is independently corroborated by other observations, such as the decrease of the mass
of 'E+A' galaxies with time (Tran et al. 2003), the lack of star formation in massive
galaxies at redshifts $z\le 1$ (De Lucia et al. 2004), the fossil record of star formation
in local early-type galaxies (Thomas et al. 2004), and the claim of mass-dependent evolution
of spiral galaxies (Ziegler et al. 2002, B\"ohm et al. 2004). The lack of age differences between
field and cluster galaxies, and the suggested mass-dependent evolution of 
early-type galaxies show that individual properties of a galaxy, and not environment, 
play an important role in its formation.

We have shown that rest-frame optical colors can be used to measure galaxy masses at high redshift.
This can be regarded as a step toward accurately calibrating SED fitting as a mass estimator.
Certainly including Spitzer photometry in the rest-frame near-infrared will provide 
tight correlations between dynamically derived $M/L$ and $M/L$ derived from SED fitting.
Line-strengths of absorption features in our high $S/N$ spectra will connect low
redshift fossil record studies to evolutionary studies such as these, and constrain
the metallicity range of early-type galaxies at $z\sim 1$, lifting the age-metallicity degeneracy.
Using the mass calibration for high-redshift galaxies, the evolution of the mass density and
the mass function can be determined from volume limited samples.
This will provide strong constraints on formation theories and the importance of 
progenitor bias.

\acknowledgements
We thank the referee for many useful comments, enhancing the quality and the readability of the work.
We thank the ESO staff for their professional and effective assistance 
during the observations. The Lorentz 
center is thanked for its hospitality during various workshops and the Leidsch 
Kerkhoven-Bosscha Fonds for its financial support.

\clearpage
\begin{small}
\begin{table*}[t]
\centering
\caption{Coordinates of the Galaxy Sample}
\begin{tabular}{ccc}
\hline
\hline
ID & $\alpha$ & $\delta$ \\
 & J2000 & J2000\\
\hline
CL1252-1 & 12:52:45.8899 & -29:29:04.5780 \\
CL1252-2 & 12:52:42.3588 & -29:27:47.3112 \\
CL1252-3 & 12:52:42.4793 & -29:27:03.5892 \\
CL1252-4 & 12:52:48.5594 & -29:27:23.2452 \\
CL1252-5 & 12:52:58.5202 & -29:28:39.5256 \\
CL1252-6 & 12:52:56.3846 & -29:26:22.7868 \\
CL1252-7 & 12:53:03.6396 & -29:27:42.5916 \\
CL1252-8 & 12:53:05.1228 & -29:26:29.8680 \\
CL1252-9 & 12:53:05.6213 & -29:26:32.5608 \\
CDFS-1 & 03:32:25.1597 & -27:54:50.1332 \\
CDFS-2 & 03:32:22.9265 & -27:54:34.3429 \\
CDFS-3 & 03:32:26.2940 & -27:54:05.0411 \\
CDFS-4 & 03:32:19.2880 & -27:54:06.1445 \\
CDFS-5 & 03:32:34.8486 & -27:53:50.0705 \\
CDFS-6 & 03:32:42.8569 & -27:53:24.7700 \\
CDFS-7 & 03:32:31.3700 & -27:53:19.1519 \\
CDFS-8 & 03:32:23.6062 & -27:53:06.3463 \\
CDFS-9 & 03:32:17.4796 & -27:52:48.0004 \\
CDFS-10 & 03:32:20.2801 & -27:52:33.0150 \\
CDFS-11 & 03:32:19.3006 & -27:52:19.3400 \\
CDFS-12 & 03:32:45.1488 & -27:49:39.9558 \\
CDFS-13 & 03:32:39.5987 & -27:49:09.6024 \\
CDFS-14 & 03:32:54.2299 & -27:49:03.7722 \\
CDFS-15 & 03:32:41.4049 & -27:47:17.1409 \\
CDFS-16 & 03:32:29.2152 & -27:47:07.5718 \\
CDFS-17 & 03:32:38.4940 & -27:47:02.3640 \\
CDFS-18 & 03:32:37.1944 & -27:46:08.0663 \\
CDFS-19 & 03:32:32.7124 & -27:45:47.4617 \\
CDFS-20 & 03:32:10.0387 & -27:43:33.1237 \\
CDFS-21 & 03:32:19.5922 & -27:43:03.7848 \\
CDFS-22 & 03:32:09.7051 & -27:42:48.1090 \\
CDFS-23 & 03:32:17.9117 & -27:41:22.6795 \\
CDFS-24 & 03:32:32.9855 & -27:41:17.0102 \\
CDFS-25 & 03:32:27.7000 & -27:40:43.6865 \\
CDFS-26 & 03:32:19.1468 & -27:40:40.2190 \\
CDFS-27 & 03:32:21.3600 & -27:40:26.0861 \\
CDFS-28 & 03:32:24.5444 & -27:40:10.4322 \\
CDFS-29 & 03:32:30.1945 & -27:39:30.2407 \\
\hline
\end{tabular}
\end{table*}
\end{small}

\begin{small}
\begin{table*}[t]
\centering
\caption{The galaxy sample}
\begin{tabular}{lccccccccccccc}
\hline
\hline
ID & e & X-ray & $z_{spec}$ & $S/N$ & $\sigma$    & Type & $r_{eff}$ & $\mu_{eff,z}$ & $z_{mag}$ & $i-z$ & $K_{mag}$ & $J-K$ \\
   &   &       &            &       & km~s$^{-1}$ &      & $''$      &               &           &       &           &       \\
\hline
CL1252-1 & \nodata & A & 0.671 & 48 & $219\pm12$ & K2 & 0.46 & 21.69 & 20.72 & 0.53 & 17.27 & 1.95 \\
CL1252-2 & \nodata & \nodata & 0.658 & 75 & $216\pm6$ & G2 & 0.18 & 20.04 & 20.35 & 0.46 & \nodata & \nodata \\
CL1252-3$^*$ & $H\beta\gamma$,[\ion{O}{3}] & A & 0.844 & 83 & $166\pm7$ & F6 & 1.04 & 22.78 & 19.30 & 0.54 & \nodata & \nodata \\
CL1252-4$^*$ & [\ion{O}{3}] & \nodata & 0.743 & 93 & $202\pm8$ & F8 & 1.02 & 23.00 & 19.56 & 0.41 & 17.49 & 1.58 \\
CL1252-5 & \nodata & A & 0.743 & 61 & $251\pm9$ & K1 & 0.36 & 21.07 & 19.91 & 0.53 & 16.98 & 1.81 \\
CL1252-6 & \nodata & \nodata & 0.734 & 123 & $211\pm5$ & G2 & 0.24 & 20.43 & 20.19 & 0.51 & 17.51 & 1.77 \\
CL1252-7 & [\ion{O}{2}] & A & 0.753 & 135 & $213\pm5$ & G8 & 0.33 & 20.53 & 19.56 & 0.47 & \nodata & \nodata \\
CL1252-8 & [\ion{O}{2}] & \nodata & 1.069 & 19 & $63\pm13$ & F4 & 0.27 & 21.92 & 21.41 & 1.00 & \nodata & \nodata \\
CL1252-9 & \nodata & \nodata & 1.036 & 18 & $102\pm16$ & F8 & 0.22 & 21.71 & 21.59 & 0.88 & \nodata & \nodata \\
CDFS-1 & [\ion{O}{2}] & A & 1.089 & 27 & $231\pm15$ & F4 & 0.44 & 21.97 & 20.36 & 1.12 & 17.70 & 1.83 \\
CDFS-2 & \nodata & \nodata & 0.964 & 40 & $200\pm9$ & G0 & 0.39 & 21.47 & 20.10 & 1.04 & 17.15 & 1.74 \\
CDFS-3 & \nodata & \nodata & 1.044 & 17 & $300\pm30$ & G4 & 0.15 & 20.39 & 21.14 & 1.07 & 18.29 & 1.75 \\
CDFS-4 & \nodata & X & 0.964 & 33 & $336\pm18$ & G0 & 1.00 & 22.94 & 19.55 & 1.11 & 16.44 & 1.89 \\
CDFS-5 & \nodata & \nodata & 0.685 & 32 & $194\pm15$ & K1 & 0.37 & 21.64 & 20.42 & 0.66 & 18.27 & 1.81 \\
CDFS-6 & \nodata & \nodata & 0.660 & 49 & $208\pm9$ & G0 & 0.15 & 19.36 & 20.10 & 0.59 & 17.32 & 1.67 \\
CDFS-7 & \nodata & \nodata & 1.135 & 17 & $232\pm19$ & G4 & 0.74 & 23.32 & 20.58 & 1.14 & 17.26 & 1.97 \\
CDFS-8 & \nodata & \nodata & 1.125 & 6 & $253\pm70$ & G4 & 0.20 & 21.53 & 21.66 & 1.19 & 18.70 & 1.84 \\
CDFS-9 & \nodata & \nodata & 1.097 & 8 & $215\pm45$ & G0 & 0.21 & 21.20 & 21.21 & 1.22 & 18.09 & 1.86 \\
CDFS-10 & [\ion{O}{2}] & \nodata & 1.119 & 9 & $275\pm49$ & F6 & 0.091 & 19.68 & 21.50 & 1.10 & 18.72 & 1.70 \\
CDFS-11 & [\ion{O}{2}] & \nodata & 1.096 & 10 & $208\pm33$ & G2 & 0.23 & 21.87 & 21.67 & 1.23 & 17.93 & 1.9 \\
CDFS-12 & \nodata & \nodata & 1.123 & 20 & $262\pm20$ & G4 & 0.14 & 20.42 & 21.29 & 1.17 & 18.00 & 1.87 \\
CDFS-13 & \nodata & \nodata & 0.980 & 46 & $247\pm10$ & G4 & 0.20 & 20.93 & 20.17 & 1.05 & 17.22 & 1.83 \\
CDFS-14 & \nodata & \nodata & 0.984 & 18 & $197\pm21$ & G4 & 0.39 & 22.19 & 20.85 & 1.01 & 17.25 & 1.78 \\
CDFS-15 & \nodata & \nodata & 0.622 & 55 & $317\pm21$ & G8 & 0.31 & 20.94 & 20.09 & 0.62 & 17.41 & 1.73 \\
CDFS-16 & \nodata & \nodata & 0.669 & 31 & $262\pm36$ & G8 & 0.33 & 21.09 & 20.12 & 0.66 & 17.32 & 1.81 \\
CDFS-17$^*$ & \nodata & \nodata & 0.954 & 18 & $305\pm31$ & F4 & 0.83 & 23.33 & 20.35 & 0.89 & 17.84 & 1.81 \\
CDFS-18 & \nodata & \nodata & 1.096 & 14 & $324\pm32$ & G0 & 0.51 & 22.21 & 20.3 & 1.20 & 16.98 & 1.92 \\
CDFS-19 & [\ion{O}{2}] & \nodata & 0.955 & 15 & $229\pm35$ & G4 & 0.34 & 20.35 & 19.30 & 0.49 & 17.53 & 1.56 \\
CDFS-20 & \nodata & \nodata & 1.022 & 25 & $199\pm15$ & G2 & 0.34 & 21.69 & 20.65 & 1.19 & 16.94 & 1.93 \\
CDFS-21 & \nodata & \nodata & 0.735 & 56 & $149\pm8$ & F4 & 0.073 & 18.52 & 20.81 & 0.56 & 18.05 & 1.59 \\
CDFS-22 & \nodata & A,X & 0.735 & 46 & $225\pm11$ & K1 & 0.67 & 22.25 & 19.73 & 0.76 & 16.84 & 1.77 \\
CDFS-23 & \nodata & \nodata & 1.041 & 13 & $70\pm15$ & G4 & 0.39 & 22.59 & 21.23 & 1.09 & 18.51 & 1.77 \\
CDFS-24$^*$ & \nodata & A & 1.042 & 25 & $210\pm16$ & F8 & 1.87 & 24.35 & 19.60 & 0.96 & 17.10 & 2.00 \\
CDFS-25 & \nodata & \nodata & 0.967 & 28 & $258\pm18$ & F6 & 0.13 & 20.02 & 21.10 & 1.02 & 18.13 & 1.65 \\
CDFS-26$^*$ & \nodata & \nodata & 1.129 & 14 & $249\pm25$ & G0 & 1.12 & 23.83 & 20.19 & 1.14 & 17.43 & 1.99 \\
CDFS-27 & [\ion{O}{2}] & \nodata & 1.128 & 8 & $135\pm30$ & F6 & 0.67 & 23.68 & 21.15 & 0.91 & 18.69 & 1.58 \\
CDFS-28$^*$ & [\ion{O}{2}] & A & 0.954 & 12 & $445\pm84$ & G4 & 0.80 & 23.54 & 20.64 & 0.80 & 18.52 & 1.78 \\
CDFS-29 & \nodata & \nodata & 1.128 & 19 & $221\pm17$ & F6 & 0.21 & 20.89 & 20.86 & 1.12 & 17.58 & 1.65 \\
\hline
\tablecomments{
IDs labeled with an asterisk are late-type galaxies. An 'A' in the third column indicates an AGN, an 'X' indicates an extended 
X-ray source. The $S/N$ in column 5 in per \AA~(1.6\AA~per pixel).
$\sigma$ is the aperture corrected central velocity dispersion measured 
with the best fitting stellar template. 'Type' indicates which spectral type fits best to the galaxy spectrum. The error on the 
effective radius may be taken from Table 3, which lists the physical sizes of the galaxies. $\mu_{eff,z}$ is 
the surface brightness in the $z$-band at the effective radius. The errors may be taken from the errors on the surface brightness
in the rest-frame $B$-band, given in Table 3.}
\end{tabular}
\end{table*}
\end{small}

\begin{small}
\begin{table*}[t]
\centering
\caption{Physical parameters of the galaxy sample}
\begin{tabular}{lcccccccc}
\hline
\hline
ID & $z_{spec}$ & $\log{(R_{eff})}$ & $\mu_{eff,B}$  & $\log{(M)}$ & $\log{(M/L_B)}$       & $\Delta \ln{(M/L_B)}$ & $U-B$ & $B-I$ \\
   &            & $kpc$             & $mag/arcsec^2$ & $M_{\odot}$ & $M_{\odot}/L_{\odot}$ &                       &       &       \\
\hline
CL1252-1 &	$0.671$ & $0.58 \pm 0.02$ & $23.64 \pm 0.07$ & $11.34 \pm 0.04$ & $0.57$ & $-0.86 \pm 0.07$ & $0.68$ & $2.04$  \\
CL1252-2 &	$0.658$ & $0.18 \pm 0.02$ & $21.92 \pm 0.10$ & $10.92 \pm 0.02$ & $0.29$ & $-1.32 \pm 0.04$ & $0.61$ & \nodata  \\
CL1252-3$^*$ &	$0.844$ & $0.88 \pm 0.02$ & $24.35 \pm 0.05$ & $11.39 \pm 0.04$ & $0.15$ & $-1.82 \pm 0.06$ & $0.42$ & \nodata  \\
CL1252-4$^*$ &	$0.743$ & $0.86 \pm 0.01$ & $24.61 \pm 0.06$ & $11.54 \pm 0.03$ & $0.54$ & $-1.01 \pm 0.06$ & $0.52$ & $1.67$  \\
CL1252-5 &	$0.743$ & $0.44 \pm 0.02$ & $22.80 \pm 0.10$ & $11.31 \pm 0.03$ & $0.43$ & $-1.20 \pm 0.05$ & $0.68$ & $1.91$  \\
CL1252-6 &	$0.734$ & $0.24 \pm 0.02$ & $22.17 \pm 0.06$ & $10.95 \pm 0.02$ & $0.23$ & $-1.45 \pm 0.03$ & $0.62$ & $1.87$  \\
CL1252-7 &	$0.753$ & $0.38 \pm 0.01$ & $22.17 \pm 0.04$ & $11.11 \pm 0.02$ & $0.08$ & $-1.88 \pm 0.03$ & $0.47$ & \nodata  \\
CL1252-8 &	$1.069$ & $0.33 \pm 0.05$ & $23.39 \pm 0.18$ & $10.00 \pm 0.17$ & $-0.73$ & $-3.05 \pm 0.28$ & $0.21$ & \nodata  \\
CL1252-9 &	$1.036$ & $0.25 \pm 0.04$ & $23.21 \pm 0.16$ & $10.34 \pm 0.13$ & $-0.28$ & $-2.23 \pm 0.21$ & $0.14$ & \nodata  \\
CDFS-1 &	$1.089$ & $0.56 \pm 0.05$ & $23.41 \pm 0.17$ & $11.35 \pm 0.05$ & $0.17$ & $-1.81 \pm 0.09$ & $0.22$ & $1.62$  \\
CDFS-2 &	$0.964$ & $0.50 \pm 0.03$ & $23.10 \pm 0.09$ & $11.17 \pm 0.04$ & $0.08$ & $-1.89 \pm 0.07$ & $0.29$ & $1.72$  \\
CDFS-3 &	$1.044$ & $0.08 \pm 0.05$ & $21.89 \pm 0.22$ & $11.10 \pm 0.08$ & $0.30$ & $-1.42 \pm 0.15$ & $0.19$ & $1.63$  \\
CDFS-4 &	$0.964$ & $0.90 \pm 0.02$ & $24.58 \pm 0.09$ & $12.02 \pm 0.04$ & $0.72$ & $-0.89 \pm 0.07$ & $0.35$ & $1.93$  \\
CDFS-5 &	$0.685$ & $0.42 \pm 0.04$ & $23.59 \pm 0.13$ & $11.06 \pm 0.06$ & $0.60$ & $-0.65 \pm 0.13$ & $0.28$ & $1.93$  \\
CDFS-6 &	$0.660$ & $0.02 \pm 0.02$ & $21.26 \pm 0.10$ & $10.73 \pm 0.04$ & $0.15$ & $-1.53 \pm 0.07$ & $0.23$ & $1.76$  \\
CDFS-7 &	$1.135$ & $0.79 \pm 0.04$ & $24.69 \pm 0.13$ & $11.59 \pm 0.07$ & $0.41$ & $-1.35 \pm 0.13$ & $0.22$ & $1.79$  \\
CDFS-8 &	$1.125$ & $0.21 \pm 0.09$ & $22.93 \pm 0.36$ & $11.08 \pm 0.21$ & $0.37$ & $-1.23 \pm 0.37$ & $0.26$ & $1.62$  \\
CDFS-9 &	$1.097$ & $0.23 \pm 0.05$ & $22.62 \pm 0.21$ & $10.97 \pm 0.16$ & $0.10$ & $-1.76 \pm 0.26$ & $0.33$ & $1.77$  \\
CDFS-10 &	$1.119$ & $-0.13 \pm 0.11$ & $21.09 \pm 0.43$ & $10.82 \pm 0.14$ & $0.05$ & $-1.86 \pm 0.23$ & $0.17$ & $1.52$  \\
CDFS-11 &	$1.096$ & $0.27 \pm 0.07$ & $23.29 \pm 0.29$ & $10.98 \pm 0.13$ & $0.30$ & $-1.30 \pm 0.23$ & $0.35$ & $1.81$  \\
CDFS-12 &	$1.123$ & $0.06 \pm 0.05$ & $21.82 \pm 0.22$ & $10.97 \pm 0.06$ & $0.10$ & $-1.79 \pm 0.11$ & $0.24$ & $1.64$  \\
CDFS-13 &	$0.980$ & $0.38 \pm 0.04$ & $22.54 \pm 0.15$ & $11.23 \pm 0.03$ & $0.15$ & $-1.81 \pm 0.05$ & $0.27$ & $1.72$  \\
CDFS-14 &	$0.984$ & $0.49 \pm 0.04$ & $23.78 \pm 0.22$ & $11.15 \pm 0.09$ & $0.33$ & $-1.32 \pm 0.14$ & $0.22$ & $1.62$  \\
CDFS-15 &	$0.622$ & $0.32 \pm 0.03$ & $22.96 \pm 0.13$ & $11.40 \pm 0.06$ & $0.93$ & $-0.11 \pm 0.09$ & $0.36$ & $1.96$  \\
CDFS-16 &	$0.669$ & $0.36 \pm 0.04$ & $23.07 \pm 0.16$ & $11.27 \pm 0.11$ & $0.72$ & $-0.51 \pm 0.13$ & $0.29$ & $1.95$  \\
CDFS-17$^*$ &	$0.954$ & $0.82 \pm 0.05$ & $24.93 \pm 0.15$ & $11.85 \pm 0.08$ & $0.87$ & $-0.46 \pm 0.14$ & $0.15$ & $1.64$  \\
CDFS-18 &	$1.096$ & $0.62 \pm 0.03$ & $23.63 \pm 0.09$ & $11.71 \pm 0.08$ & $0.48$ & $-1.29 \pm 0.14$ & $0.32$ & $1.83$  \\
CDFS-19 &	$0.955$ & $0.43 \pm 0.03$ & $21.84 \pm 0.14$ & $11.22 \pm 0.12$ & $-0.23$ & $-2.66 \pm 0.20$ & $-0.32$ & $1.30$  \\
CDFS-20 &	$1.022$ & $0.43 \pm 0.05$ & $23.24 \pm 0.20$ & $11.10 \pm 0.06$ & $0.15$ & $-1.71 \pm 0.11$ & $0.36$ & $1.73$  \\
CDFS-21 &	$0.735$ & $-0.28 \pm 0.04$ & $20.27 \pm 0.14$ & $10.14 \pm 0.05$ & $-0.32$ & $-2.28 \pm 0.10$ & $0.23$ & $1.57$  \\
CDFS-22 &	$0.735$ & $0.69 \pm 0.03$ & $24.21 \pm 0.10$ & $11.46 \pm 0.04$ & $0.65$ & $-0.74 \pm 0.07$ & $0.30$ & $2.09$  \\
CDFS-23 &	$1.041$ & $0.50 \pm 0.05$ & $24.10 \pm 0.17$ & $10.27 \pm 0.17$ & $-0.50$ & $-2.65 \pm 0.22$ & $0.26$ & $1.68$  \\
CDFS-24$^*$ &	$1.042$ & $1.18 \pm 0.05$ & $25.86 \pm 0.13$ & $11.90 \pm 0.06$ & $0.48$ & $-1.33 \pm 0.11$ & $0.08$ & $1.47$  \\
CDFS-25 &	$0.967$ & $0.00 \pm 0.08$ & $21.63 \pm 0.36$ & $10.90 \pm 0.06$ & $0.21$ & $-1.52 \pm 0.10$ & $0.25$ & $1.60$  \\
CDFS-26$^*$ &	$1.129$ & $0.96 \pm 0.04$ & $25.24 \pm 0.11$ & $11.83 \pm 0.08$ & $0.52$ & $-1.22 \pm 0.14$ & $0.17$ & $1.59$  \\
CDFS-27 &	$1.128$ & $0.74 \pm 0.06$ & $25.10 \pm 0.17$ & $11.07 \pm 0.18$ & $0.16$ & $-1.62 \pm 0.26$ & $-0.06$ & $1.34$  \\
CDFS-28$^*$ &	$0.954$ & $0.80 \pm 0.06$ & $25.12 \pm 0.18$ & $12.17 \pm 0.15$ & $1.29$ & $0.30 \pm 0.27$ & $-0.03$ & $1.50$  \\
CDFS-29 &	$1.128$ & $0.24 \pm 0.06$ & $22.30 \pm 0.20$ & $11.00 \pm 0.07$ & $-0.04$ & $-2.10 \pm 0.10$ & $0.16$ & $1.42$  \\
\hline
\tablecomments{
IDs labeled with an asterisk are late-type galaxies. The rest-frame $B$-band surface brightness at the effective radius is 
corrected for cosmological surface brightness dimming. The error on $M/L$ in solar units is the same
as the error on $M$. $U-B$ and $B-I$ are rest-frame colors.
}
\end{tabular}
\end{table*}
\end{small}

\begin{small}
\begin{table*}[t]
\centering
\caption{Comparison with previous results}
\begin{tabular}{lccccccc}
\hline
\hline
Ref. & $\Delta \ln{(M/L_B)}/z$ & $\Delta \ln{(M/L_B)}/z$ & $\Delta \ln{(M/L_B)}/z$ & $\Delta \ln{(M/L_B)}/z$ & $<\log (M/M_{\odot})>$ & $<z>$ & $N$ \\
     & reported                & fitted                  & high mass               & low mass                &                        &       &     \\
\hline
Treu et al. 2001 & $-1.64\pm0.12$ & $-1.64\pm0.34$ & $-1.55\pm0.27$ & $-1.47\pm1.89$ & $2.3\times 10^{11}$ & 0.29 & 19 \\
van Dokkum et al. 2001 & $-1.35\pm0.35$ & $-1.67\pm0.23$ & $-1.29\pm0.39$ & $-1.94\pm0.33$ & $1.5\times 10^{11}$ & 0.42 & 18 \\
Treu et al. 2001, 2002 & $-1.84$ & \nodata & \nodata & \nodata & $2.5\times 10^{11}$ & 0.38 & 29\\
van Dokkum et al. 2001, 2003 & $-1.25\pm0.25$ & $-1.68\pm0.13$ & $-1.41\pm0.29$ & $-1.77\pm0.24$ & $1.3\times 10^{11}$ & 0.56 & 27 \\
Gebhardt et al. 2003 & $-2.21$ & $-1.94\pm0.20$ & $-1.59\pm0.49$ & $-2.05\pm0.17$ & $8.5\times 10^{10}$ & 0.64 & 21 \\
This paper & $-1.75\pm0.16$ & $-1.75\pm0.16$ & $-1.20\pm0.18$ & $-1.97\pm0.16$ & $1.6\times10^{11}$ & 0.90 & 27 \\
\hline
Rusin et al. 2003 & $-1.24\pm0.21$ & \nodata & \nodata & \nodata & \nodata & 0.54 & 21 \\ 
van de Ven et al. 2003 & $-1.43\pm0.30$ & $-1.36\pm0.18$ & $-1.13\pm0.31$ & $-1.71\pm0.30$ & $2.3\times 10^{11}$ & 0.54 & 21 \\
\hline
\tablecomments{
Column 2 lists the values of the evolution of $M/L$ as reported in the cited papers. 
Column 3 lists
the values we derive using our fitting technique on the tabulated data 
of the individual galaxies in the cited papers. Column 4 lists the evolution
of galaxies more massive than $M=2\times 10^{11}M_{\odot}$, using our
fitting technique, and column 5 lists the evolution of less massive galaxies.
Average masses, redshifts, and sample sizes ($N$) of the samples 
are also listed.
When two references are given, the results presented in the newest paper are 
based on data of both papers.
Treu et al. (2002) only list redshifts and velocity dispersions of the galaxies.
The reported values are all uncorrected for selection effects. 
Treu et al. (2002) only give an error for the uncorrected value, and
Gebhardt et al. (2003) do not give an error.
Besides the magnitude limited samples, the two studies of the same sample of 
lensing galaxies are also listed. We only fitted the tabulated data from van de Ven et al. (2003),
but the data presented by Rusin et al. (2003) yield the same results.}
\end{tabular}
\end{table*}
\end{small}

\end{document}